\documentclass[fleqn,usenatbib]{mnras}

\usepackage{newtxtext}
\usepackage{mathptmx}

\usepackage[T1]{fontenc}
\usepackage{ae,aecompl}

\usepackage{graphicx}	
\usepackage{amsmath}	
\usepackage{amssymb}	
\usepackage{bm}
\usepackage[usenames,dvipsnames]{color}
\usepackage{verbatim}
\usepackage{multirow}
\usepackage{booktabs}
\usepackage{subfigure}
\usepackage[font=footnotesize,labelfont=bf]{caption}
\usepackage{footnote}
\makesavenoteenv{tabular}
\makesavenoteenv{table}
\usepackage{supertabular}
\usepackage{threeparttable}
\usepackage{longtable}
\usepackage{rotating}
\usepackage{lscape}

\usepackage{lineno}

\usepackage{blindtext}
\makeatletter
\newcommand{\removelatexerror}{\let\@latex@error\@gobble}
\makeatother

\usepackage[]{algorithm2e}
\usepackage{fancyhdr}

\newcommand{\um}[1]{\ \mathrm{#1}}
\newcommand{\mic}[1]{\ \umu\mathrm{#1}}

\newcommand{\selavy}{\textsc{selavy}}
\newcommand{\aegean}{\textsc{aegean}}
\newcommand{\caesar}{\textsc{caesar}}
\newcommand{\pybdsm}{\textsc{pybdsm}}
\newcommand{\scorpio}{\textsc{Scorpio}}

\newcommand{\hii}{H\textsc{ii}}
\newcommand*\hiit{{H\normalfont\textsc{ii}}}

 {\pagebreak[4]\global\pdfpageattr\expandafter{\the\pdfpageattr/Rotate 90}}%
 {\pagebreak[4]\global\pdfpageattr\expandafter{\the\pdfpageattr/Rotate 0}}%

\title[EMU: Compact radio sources in SCORPIO]{Evolutionary Map of the Universe (EMU):\\Compact radio sources in the \textsc{Scorpio} field towards the Galactic plane}

\author[S. Riggi et al.]{
S. Riggi$^{1}$\thanks{E-mail: simone.riggi@inaf.it}, %
G. Umana$^1$, %
C. Trigilio$^1$, %
F. Cavallaro$^{11,1}$, %
A. Ingallinera$^1$, %
P. Leto$^1$, %
F. Bufano$^1$, %
\newauthor%
R.P. Norris$^{2,3}$, %
A.M. Hopkins$^{4,2}$, %
M.D. Filipovi\'c$^2$, %
H. Andernach$^5$, %
J.Th. van Loon$^6$, %
\newauthor%
M. J.~Micha{\l}owski$^{7}$, %
C. Bordiu$^{1,8}$, %
T. An$^9$, %
C. Buemi$^1$, %
E. Carretti$^{10}$, %
J.D. Collier$^{11,2}$, %
\newauthor%
T. Joseph$^{12}$, %
B.S. Koribalski$^{3,2}$, %
R. Kothes$^{13}$, %
S. Loru$^1$, %
D. McConnell$^3$, %
M. Pommier$^{14}$, %
\newauthor%
E. Sciacca$^1$, %
F. Schillir\`o$^1$, %
F. Vitello$^{10}$, %
K. Warhurst$^{15}$ %
and M. Whiting$^{3}$%
\\%
\\%
$^{1}$INAF-Osservatorio Astrofisico di Catania, Via Santa Sofia 78, 95123 Catania, Italy\\%
$^2$School of Science, Western Sydney University, Locked Bag 1797, Penrith, NSW 2751, Australia\\%
$^3$CSIRO Astronomy \& Space Science, P.O. Box 76, Epping, NSW 1710, Australia\\%
$^4$Australian Astronomical Optics, Macquarie University, 105 Delhi Rd, North Ryde, NSW 2113, Australia\\%
$^5$Depto. de Astronomía, DCNE, Universidad de Guanajuato, Cj\'on.\ de Jalisco, Col.\ Valenciana, Guanajuato, CP 36023, Mexico\\%
$^6$Lennard-Jones Laboratories, Keele University, ST5 5BG, UK\\%
$^7$ Astronomical Observatory Institute, Faculty of Physics, Adam
Mickiewicz University, ul.~S{\l}oneczna 36, 60-286 Pozna{\'n}, Poland\\%
$^8$ Centro de Astrobiología (INTA-CSIC), Ctra. M-108, km. 4, 28850 Torrej\'on de Ardoz, Madrid, Spain\\%
$^9$Shanghai Astronomical Observatory, Chinese Academy of Sciences, 80 Nandan Road, Shanghai, 200030, China\\%
$^{10}$ INAF Istituto di Radioastronomia, Via Gobetti 101, 40129 Bologna, Italy\\%
$^{11}$The Inter-University Institute for Data Intensive Astronomy (IDIA), Department of Astronomy, University of Cape Town, Rondebosch, 7701, South Africa\\%
$^{12}$School of Physics and Astronomy, University of Manchester, Oxford Road, Manchester, M13 9PL, UK\\%
$^{13}$ Dominion Radio Astrophysical Observatory, Herzberg Astronomy and Astrophysics, National Research Council Canada, PO Box 248, Penticton, BC V2A 6J9, Canada\\%
$^{14}$Univ Lyon, Univ Lyon1, Ens de Lyon, CNRS, Centre de Recherche Astrophysique de Lyon UMR5574, 9 av Charles André F- 69230, Saint-Genis-Laval, France\\%
$^{15}$ CSIRO Astronomy \& Space Science, 33 Onslow St, Geraldton, WA 6530. Australia%
}

\date{Accepted 2020 December 28. Received 2020 December 23; in original form 2020 September 30}

\begin{document}
\label{firstpage}
\pagerange{\pageref{firstpage}--\pageref{lastpage}}
\maketitle

\begin{abstract}
We present observations of a region of the Galactic plane taken during the Early Science Program of the Australian Square Kilometre Array Pathfinder (ASKAP).
In this context, we observed the \textsc{Scorpio} field at 912 MHz with an uncompleted array consisting of 15 commissioned antennas. The resulting map covers a square region of $\sim$40 deg$^{2}$, centred on $(l,b)$=(343.5$^{\circ}$,0.75$^{\circ}$), with a synthesized beam of 24"$\times$21" and a background rms noise of 150-200 $\mu$Jy/beam, increasing to 500-600 $\mu$Jy/beam close to the Galactic plane.
A total of 3963 radio sources were detected and characterized in the field using the \caesar{} source finder. We obtained differential source counts in agreement with previously published data after correction for source extraction and characterization uncertainties, estimated from simulated data. The ASKAP positional and flux density scale accuracy were also investigated through comparison with previous surveys (MGPS, NVSS) and additional observations of the \textsc{Scorpio} field, carried out with ATCA at 2.1 GHz and 10" spatial resolution. These allowed us to obtain a measurement of the spectral index for a subset of the catalogued sources and an estimated fraction of (at least) 8\% of resolved sources in the reported catalogue.
We cross-matched our catalogued sources with different astronomical databases to search for possible counterparts, finding $\sim$150 associations to known Galactic objects. Finally, we explored a multiparametric approach for classifying previously unreported Galactic sources based on their radio-infrared colors.
\end{abstract}




\begin{figure*}
\centering%
\includegraphics[scale=0.9]{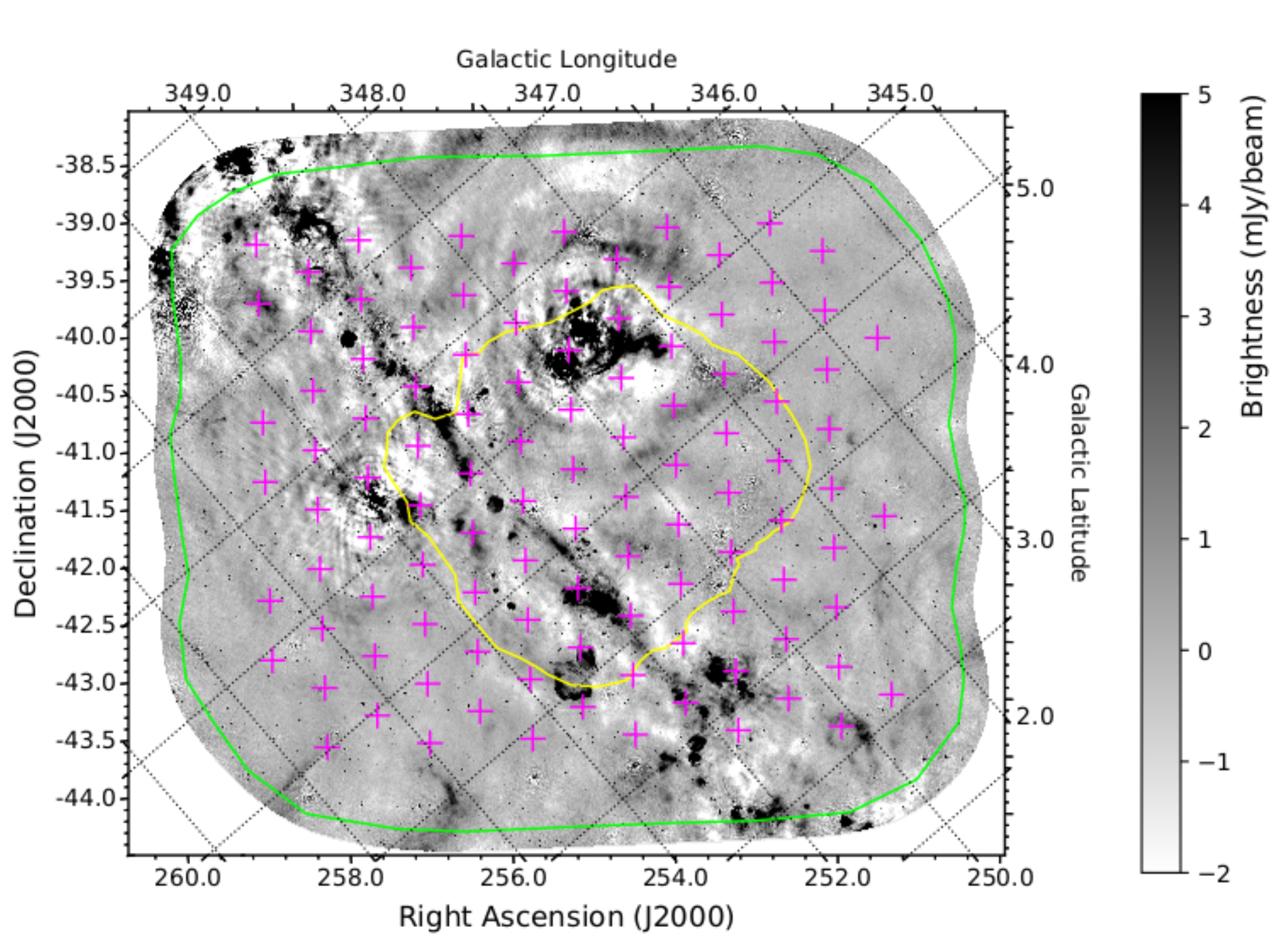}
\caption{912 MHz mosaic of the \scorpio{} region observed with ASKAP. The purple crosses indicate the beam centre positions for the three interleaved pointings. The green contour delimits the mosaic area considered for source catalogue extraction with the \caesar{} source finder. The yellow contour denotes the \scorpio{} region observed with the ATCA telescope at 2.1 GHz (see text).}
\label{fig:mosaic}
\end{figure*}

\section{Introduction}
\label{sec:intro}
The next-generation deep radio continuum surveys, such as the Evolutionary Map of the Universe (EMU) \citep{Norris2011}, planned at the Australian SKA Pathfinder (ASKAP) telescope \citep{Johnston2008} will open a new era in radio astronomy with potential new discoveries expected in several fields, from Galaxy evolution and characterization to Galactic Science.\\In the full-operational mode, the ASKAP telescope is made of 36 12-metre antennas installed in Western Australia, each one equipped with a Phased Array Feed (PAF) receiver \citep{Schinckel2012}, operating with a bandwidth of 288 MHz over the frequency range 700-1800 MHz. The PAF system forms 36 beams and provides an instantaneous field of view of $\sim$30 square degrees, allowing ASKAP to survey the southern sky with unprecedented speed ($\sim$220 deg$^{2}$ per hour at a target 1$\sigma$ rms of 100 $\mu$Jy/beam) and higher resolution ($\sim$10" at 950 MHz) compared to existing surveys.\\
The array has a maximum baseline of 6 km and was completed in mid 2019. During the commissioning phase, the ASKAP EMU Early Science Program (ESP) was launched (October 2017), in which several target fields were observed with the aim of validating the array operations, the observation strategy and the data reduction pipeline. During this preparatory phase, it became evident that the imaging performance exceeded those of past observations, enabling valid scientific results even with an incomplete array.
The \textsc{Scorpio} field was the first Galactic field observed during the ASKAP ESP using 15 commissioned antennas \citep{Umana2020}.\\%
The \scorpio{} survey \citep{Umana2015} started in 2011 with multiple scientific goals. The original objectives were the study and characterization of different types of Galactic radio sources, with a focus on radio stars and circumstellar regions (e.g. \hii{} regions). Recently, the study and characterization of stellar relics, such as Galactic supernova remnants (SNRs) in connection with observations at different wavelengths (infrared and gamma-ray primarily) has become an additional target of interest \citep{Ingallinera2017}. The survey also represents an important testbench for the ASKAP data reduction pipeline in the Galactic plane and for the analysis methods designed for the upcoming EMU survey.\\%
This paper is the second of a series of works planned with ASKAP \textsc{Scorpio} Early Science data. In the first paper \citep{Umana2020} 
we discuss the ASKAP's potential for the discovery of different classes of Galactic objects in comparison with original ATCA observations, and described the data reduction strategies adopted to produce the final mosaic.
The goal of this paper is to report a first catalogue of the compact sources present in \textsc{Scorpio}. This work will also serve as a validation on real data of the designed source extraction algorithms tested so far with simulated data \citep{Riggi2016,Riggi2019}.\\
The paper is organized as follows. The ASKAP \textsc{Scorpio} radio observations and data reduction are briefly described in Section~\ref{sec:observations}. In Section~\ref{sec:compactsources} we describe the source extraction methodologies used to build the source catalogue, and discuss the typical performance achieved in source detection and characterization (e.g. completeness, reliability, positional and flux density accuracy). In Section~\ref{sec:analysis} we present the analysis conducted on the resulting source catalogue, from source counts to spectral indices, while in Section~\ref{sec:sourceclassification} we report a comparison with existing astronomical databases and a preliminary study of unclassified sources. Finally, we report in Section~\ref{sec:summary} a summary of the results obtained and future prospects.

\section{Observational data of the Scorpio field}
\label{sec:observations}

\subsection{ASKAP 912 MHz observations and data reduction}
\label{subsec:observations}
\begin{figure}
\centering%
\includegraphics[scale=0.35]{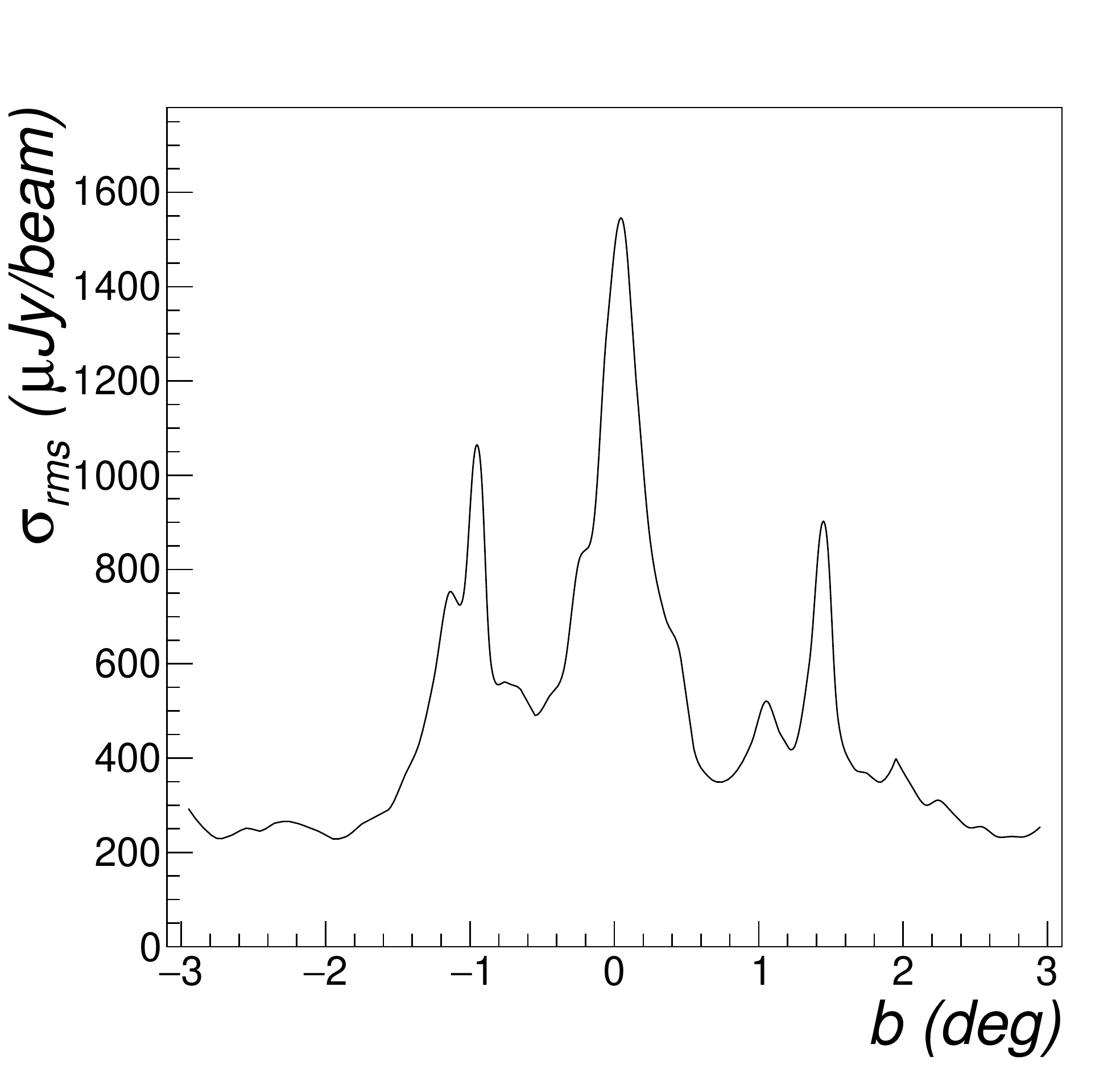}
\caption{Estimated background noise of \scorpio{} mosaic in $\mu$Jy/beam as a function of the Galactic latitude $b$ and averaged over Galactic longitude $l$.}
\label{fig:bkgrms}
\end{figure}

The \textsc{Scorpio} field was observed in January 2018 with 15 antennas equipped with the new PAF system version (Mk II) in band 1 (from 792 to $1032\um{MHz}$). In this array configuration, the minimum and maximum baselines were respectively $22.4\um{m}$ and $2.3\um{km}$. The former corresponds to a maximum theoretical largest angular scale (LAS) around 50 arcmin at 912 MHz.
The total surveyed area covers $\sim\!40$ square degrees centred on $l$=343.5$^{\circ}$, $b$=0.75$^{\circ}$, extending by a factor of $\sim$4.8 the area surveyed with past \scorpio{} observations done with the Australian Telescope Compact Array (ATCA) \citep{Umana2015,Ingallinera2019}.\\
The calibration and imaging procedures adopted to produce the final mosaic (shown in Fig.~\ref{fig:mosaic} and referred to as the \scorpio{} ASKAP map in the rest of the paper) are described in detail in the \scorpio{} paper 1 \citep{Umana2020}. 
The green contour in Fig.~\ref{fig:mosaic} delimits the field region considered for source extraction (see Section~\ref{sec:compactsources}), while the yellow contour denotes the \scorpio{} region observed with the ATCA telescope at 2.1 GHz \citep{Umana2015} (see Section~\ref{subsec:atcaobservations} for details).
The synthesized beam of the final map in J2000 coordinates is 24"$\times$21" at a position angle of 89$^{\circ}$.\\
The background level and rms noise were estimated with the \caesar{} finder using parameter values reported in Table~\ref{tab:caesar-pars}.
The background rms noise was obtained by interpolating the median absolute deviation (MAD) of pixel fluxes computed over moving sampling boxes of size 10 times the area of the synthesized beam. The background level varies considerably across the surveyed area. In Fig.~\ref{fig:bkgrms} we report the estimated background noise in $\mu$Jy/beam as a function of the Galactic latitude coordinate $b$, averaged over the Galactic longitude coordinate $l$. We observe a noise level $\sim$200 $\mu$Jy/beam in regions far from the Galactic plane and without bright sources. Close to the Galactic plane, the background noise increases due to the Galactic diffuse emission and the bright emission from extended sources, filling the beam of the telescopes and increasing the system temperature. In regions free of extended sources we observe a background noise around 500-600 $\mu$Jy/beam. Only 20\% of the field area has a 5$\sigma$ noise level smaller than 1 mJy/beam, while for $\sim$70\% of the field the 5$\sigma$ noise is smaller than 2 mJy/beam. 

\subsection{ATCA 2.1 GHz observations and data reduction}
\label{subsec:atcaobservations}
The \scorpio{} field was observed with the Australia Telescope Compact Array (ATCA) in the 6A and 6D configurations at the reference frequency of 2.1 GHz, using the 16-cm CABB receiver (observing band from 1.1 to 3.1 GHz) \citep{Wilson2011}. In this array configuration, the theoretical upper limits for the LAS ranges from $\sim$4.3 to $\sim$12.2 arcmin. The observations, conducted in different runs from 2011 to 2012, the data reduction strategy and the scientific results are extensively described elsewhere \citep{Umana2015, Riggi2016,Cavallaro2018,Ingallinera2019}. ATCA observations cover only a small portion of the \textsc{Scorpio} field observed with ASKAP, equivalent to 8.4 square degrees (see Fig.\ref{fig:mosaic} for a comparison of the surveyed area size). In Fig.~\ref{fig:atcamosaic} we present the \textsc{Scorpio} ATCA mosaic. We refer to this as the \scorpio{} ATCA map. The achieved rms is $\sim$30-40 $\mu$Jy/beam and the synthesized beam in J2000 coordinates is 9.8"$\times$5.8" (position angle of -3$^{\circ}$).\\
The ATCA data, obtained with a bandwidth of $\sim$1.7 GHz, were
divided into 7 sub-bands ($\langle\nu/\text{GHz}\rangle$=1.449, 1.681, 1.844, 2.065, 2.337, 2.614, 2.895) and imaging was independently performed on each of them to produce additional mosaics. The 2.895-GHz channel map was not considered as it was significantly affected by noise and imaging artefacts. The remaining sub-band mosaics (1-6) along with the full band mosaic are used throughout the paper as ancillary data to complement the ASKAP catalogue with value-added information, such as the source spectral indices (see Section~\ref{sec:spectralindex}), or to estimate the expected fraction of extended sources (see Section~\ref{sec:extendedsources}).

\subsection{MOST 843 MHz observations}
The \emph{Molonglo Galactic Plane Survey 2nd Epoch (MGPS-2)} \citep{Murphy2007}, carried out with the Molonglo Observatory Synthesis Telescope (MOST) at a frequency of 843 MHz, completely covers the \scorpio{} field observed with ASKAP with a lower spatial resolution (45"$\times$45" cosec|$\delta$|) and a source detection threshold of $\sim$10 mJy. 799 MGPS sources fall in the \scorpio{} region. Their position uncertainty is considered better than 1"-2" \citep{Murphy2007}.

\subsection{NVSS 1.4 GHz observations}
The \emph{NRAO VLA Sky Survey} (NVSS) \citep{Condon1998} covers the \scorpio{} region north of DEC=-40$^{\circ}$ at a frequency of 1.4 GHz with an angular resolution of 45". The detection threshold is $\sim$2.5 mJy. A number of 853 NVSS sources fall in the \scorpio{} region.

\subsection{TGSS 150 MHz observations}
The \emph{TIFR GMRT Sky Survey} (TGSS) \citep{Intema2017} fully covers the \scorpio{} field at the reference frequency of 150 MHz and with an angular resolution of 25"$\times$25"/$\cos(DEC-19^{\circ})$ and a median rms noise of 3.5 mJy/beam. 249 sources from the first alternative data release (ADR) fall in the \scorpio{} region.

\subsection{GLEAM 200 MHz observations}
The \emph{GaLactic and Extragalactic All-sky Murchison
Widefield Array} (GLEAM) survey \citep{Hurley-Walker2017} partially covers the \scorpio{} field in (1$^{\circ}\le|b|\le$10$^{\circ}$, 345$^{\circ}<l<$67$^{\circ}$) at the reference frequency of 200 MHz (bandwidth 60 MHz) with an angular resolution of $\sim$2 arcmin and an rms noise of 10-20 mJy/beam. 51 sources from the GLEAM Galactic plane catalogue \citep{Hurley-Walker2019} fall in the \scorpio{} region.

\subsection{Supplementary surveys}
In this work we will also make use of the following infrared surveys for source classification studies (see Section~\ref{sec:unclassifiedsources}):
\begin{itemize}
\item \emph{AllWISE} \citep{Cutri2013} of the \emph{Wide-field Infrared Survey Explorer} (WISE, \citealt{Wright2010}): The survey is fully covering the \scorpio{} mosaic region with $\sim$9.3$\times$10$^{5}$ sources detected with S/N>5 in at least one of the four bands at 3.4$\mic{m}$ (W1), 4.6$\mic{m}$ (W2), 12$\mic{m}$ (W3) and 22$\mic{m}$ (W4). The angular resolutions are 6.1", 6.4", 6.5" and 12" and the 5$\sigma$ flux sensitivities for point sources are 0.08 mJy, 0.11 mJy, 1 mJy and 6 mJy respectively.
\item \emph{GLIMPSE} (Galactic Legacy Infrared MidPlane Survey Extraordinaire) 8.0$\mic{m}$ surveys \citep{Churchwell2009} of the \emph{Spitzer Space Telescope} \citep{Werner2004}: The surveys (GLIMPSE-I and GLIPMSE-3D) are partially covering ($\sim$74\%) the \scorpio{} mosaic region with 8.2$\times$10$^{5}$ sources detected with S/N>5. The angular resolution is 2" and the 5$\sigma$ flux sensitivity $\sim$0.4 mJy.
\item \emph{Hi-GAL} (Herschel infrared Galactic plane Survey) 70$\mic{m}$ survey \citep{Molinari2016} of the \emph{Herschel Space Observatory} \citep{Pilbratt2010}: The survey is partially covering ($\sim$50\%) the \scorpio{} mosaic region with 5654 sources detected with S/N>5. The angular resolution is $\sim$8.5" and the 1$\sigma$ flux sensitivity $\sim$20 MJy/sr.
\end{itemize}

\section{Compact sources in the \textsc{Scorpio} field}
\label{sec:compactsources}

\subsection{Source finding}
\label{sec:sfinding}
Compact sources were extracted from the \scorpio{} ASKAP map with the \caesar{} source finder \citep{Riggi2016,Riggi2019} using
an iterative flood-fill algorithm in which the detection threshold is initially set to 5$\sigma$ and the flooding threshold to 2.5$\sigma$. At each iteration the background and noise maps are recomputed, excluding the sources found in the previous iteration and the detection threshold is lowered in steps of 0.5$\sigma$. Two iterations were used in this work, corresponding to a final 4.5$\sigma$ seed threshold level. Algorithm parameter values, reported in Table~\ref{tab:caesar-pars}, resulted from a fine-tuning procedure carried out both on the simulated data sample described in \cite{Riggi2019} and on the simulated maps described in Section~\ref{sec:catalog-uncertainties}.\\We detected 5663 islands\footnote{By "island" (or "blob") we denote a group of connected pixels with brightness above a merge threshold and around a seed pixel with brightness above a detection threshold. Island counts include also islands nested in other islands.} in the map. 5413 of these were selected as compact source candidates and successfully fitted with a mixture of Gaussian components.\\
To reject imaging artefacts and sources with a poor quality characterization, we applied these selection criteria to the extracted source sample:
\begin{itemize}
\item Source fit converged with a $\tilde{\chi}^{2}<$10;
\item Positive fitted component peak flux;
\item Fitted component centroid inside the source island and the mosaic boundary region;
\item Separation between any pair of source components larger than 8" (or 2 pixels)
\end{itemize}
After the selection, 4262 source islands and 4813 fitted source components are left in the preliminary catalogue.
\\These source counts still include spurious sources, mainly due to imaging artefacts and over-deblending of extended/diffuse emission, surviving the selection criteria. To produce the final catalogue we visually inspected the entire field labelling each source component as "real" or "spurious" in case of a clear unambiguous identification.
4144 fitted source components were tagged as "real" in the final catalogue. Taking into account that the selected survey region is 37.7 square degrees, a density of $\sim$110 selected compact sources per square degree is obtained.\\
In Table~\ref{tab:finderresults} we report a summary of the number of sources extracted at different selection stages. Source numbers labelled with "SEL+VIS SEL" refer to the number of sources passing both the selection criteria and the visual selection. The number of detected sources classified per number of fitted components is also reported.\\
Using the same procedure and source selection adopted for the ASKAP catalogue, we extracted 2227 sources and 2369 fitted components from the \scorpio{} ATCA mosaics. The source numbers obtained at different selection stages on the full bandwidth map are reported in Table~\ref{tab:atcafinderresults}. 

\begin{table}
\caption{Number of sources extracted from the Scorpio ASKAP mosaic at 912 MHz with the \caesar{} finder at different selection stages and per number of fitted components. The \textsc{no sel} column reports the number of sources extracted by the source finder without any quality cuts applied. The \textsc{sel} column reports the number of sources obtained after applying the selection criteria described in the text. The \textsc{sel+vis sel} column reports the number of sources passing both the quality selection and the visual inspection.}
\begin{tabular}{c|ccc|}
\hline%
\hline%
\multirow{2}{*}{\# components} & \multicolumn{3}{|c|}{Selection}\\%
\cmidrule(lr){2-4}%
& \textsc{no sel} & \textsc{sel} & \textsc{sel+vis sel}\\%
\hline%
0 & 250 & 0 & 0\\
1 & 4754 & 3857 & 3786\\
2 & 440 & 308 & 173\\
3 & 131 & 59 & 4\\
$>$3  & 88 & 38 & 0\\
\hline%
All & 5663 & 4262 & 3963\\
\hline%
\end{tabular}
\label{tab:finderresults}
\end{table}

\begin{table}
\caption{Number of sources extracted from the Scorpio ATCA mosaic at 2.1 GHz with the \caesar{} finder at different selection stages and per number of fitted components. The \textsc{no sel} column reports the number of sources extracted by the source finder without any quality cuts applied. The \textsc{sel} column reports the number of sources obtained after applying the selection criteria described in the text. The \textsc{sel+vis sel} column reports the number of sources passing both the quality selection and the visual inspection.}
\begin{tabular}{c|ccc|}
\hline%
\hline%
\multirow{2}{*}{\# components} & \multicolumn{3}{|c|}{Selection}\\%
\cmidrule(lr){2-4}%
& \textsc{no sel} & \textsc{sel} & \textsc{sel+vis sel}\\%
\hline%
0 & 146 & 0 & 0\\
1 & 3021 & 2104 & 2096\\
2 & 465 & 188 & 120\\
3 & 158 & 53 & 11\\
$>$3 & 102 & 30 & 0\\
\hline%
All & 3892 & 2375 & 2227\\
\hline%
\end{tabular}
\label{tab:atcafinderresults}
\end{table}

\subsection{Source cross-matching}
\label{sec:crossmatching}
To validate and complement the ASKAP catalogue, we cross-matched it with the ATCA catalogue and with the catalogues of supplementary radio (MGPS, NVSS, TGSS, GLEAM) and infrared (AllWISE, GLIMPSE, Hi-GAL) surveys listed in Section~\ref{sec:observations}. A summary of the match results is reported in Table~\ref{tab:radioIRmatch} and details are provided in the following sections.

\subsubsection{Cross-matching with ATCA source catalogue}
\label{sec:atca-crossmatches}
731 ASKAP source components (out of 856 components in the catalogue falling in the ATCA mosaic region) match with at least one source component from the ATCA catalogue within a search radius of 24". We visually inspected each cross match rejecting spurious and unclear associations, e.g. cases in which the ASKAP source flux density measurement is potentially affected by close background sources visible in the ATCA map but not in ASKAP. We finally selected 648 matches. The majority are one-to-one matches (596), while the remaining (52) are one-to-many matches (up to three components). The number of matches purely arising by chance was estimated by averaging the number of matches found between ASKAP catalogue and multiple random ATCA catalogues in which the measured source positions were uniformly randomized inside the ATCA mosaic. With this procedure we found 3.8$\pm$0.3 matches between both catalogues. We thus concluded that $\sim$99.5 percent of the matches found are likely to be real.

\subsubsection{Cross-matching with supplementary radio surveys}
We found 688 MGPS sources potentially associated to ASKAP sources within a match radius of 45". 546 of them were finally selected for further analysis after excluding multi-matches and sources with potentially unreliable flux densities, e.g. affected by imaging artefacts or closely located to very extended sources.\\
Within a match radius of 45" we found 226 NVSS sources matching to one or more ASKAP source components. 189 single matches were finally selected after a visual inspection.\\192 out of 217 TGSS source matches can be associated to a single ASKAP source within a match radius of 24". Similarly, 32 out of 40 GLEAM source matches found within a radius of 45" were selected for further analysis after excluding multi-matches or ambiguous associations.

\subsubsection{Cross-matching with supplementary infrared surveys}
\label{subsec:ir-crossmatches}
For the cross-match we considered the AllWISE survey as the primary dataset reference due to its full coverage. To limit the number of spurious associations, we selected sources with S/N>5 in both the W3 and W4 bands with a fraction of saturated pixels smaller than 10\%, reducing the number of sources within the ASKAP field to $\sim$1.2$\times$10$^{5}$. 
A crossmatch with ASKAP source catalogue using a match radius equal to the ASKAP beam size (24") yields 1319 matches. About 2900 catalogued sources do not therefore have IR counterparts with fluxes above W3 and W4 detection thresholds. The number of source associations arising by chance, estimated using artificial catalogues with random offsets applied, is however compatible with the number of obtained matches, indicating that the vast majority of the matches are spurious.\\Selecting a smaller radius (e.g. 8" as in the Galactic object search analysis) and requiring flux information in all bands, leads to 384 associations allowing us to reduce the number of potential spurious matches to $\sim$40\%. The number of associations decreases to 225 and 41, respectively if we require triple and quadruple matches with GLIMPSE and Hi-GAL catalogues.\\The impact of the 5$\sigma$ limits requested in W3 and W4 bands was investigated on the star/galaxy/quasar dataset provided by \cite{Clarke2020}\footnote{\url{10.5281/zenodo.3768398}}. The magnitude selection cuts W3$<$11.32 and $W4$<8.0\footnote{See the WISE Explanatory Supplement at \url{http://wise2.ipac.
caltech.edu/docs/release/allsky/expsup/}} were found to remove a very large fraction ($\sim$98\%) of extragalactic sources from the sample but, unfortunately, also potential radio stars. We thus expect the unmatched sources to be mostly extragalactic, with a smaller percentage of "IR-quiet" (e.g. pulsars) or faint mid-infrared Galactic objects.
\begin{table}
\caption{Number of cross-matches found between ASKAP source component catalogue and different radio and infrared surveys. Column 4 reports the total number of sources falling in the \scorpio{} mosaic for each survey catalogue. In the last two rows we summed up the number of sources from different catalogues.}
\begin{threeparttable}
\begin{tabular}{l|c|c|c}
\hline%
\hline%
Surveys & radius (") & \#matches & total\\%
\hline%
ATCA & 24 & 648 & 2369\\%
MGPS & 45 & 546 & 799\\%
NVSS & 45 & 189 & 853\\%
TGSS & 24 & 192 & 249\\%
GLEAM & 45 & 32 & 51\\%
AllWISE & 8 & 384 & 123374\\%
AllWISE + GLIMPSE & 8 & 225 & -\tnote{*}\\%
AllWISE + GLIMPSE + Hi-GAL & 8 & 41 & -\tnote{\dag}\\%
\hline%
\end{tabular}
\begin{tablenotes}
\footnotesize%
\item[*] GLIMPSE catalogue has 820112 sources in \scorpio{}
\item[\dag] Hi-GAL catalogue has 5654 sources in \scorpio{}
\end{tablenotes}
\end{threeparttable}
\label{tab:radioIRmatch}
\end{table}

\subsection{Catalogue selection effects and uncertainties}
\label{sec:catalog-uncertainties}
To evaluate the expected source extraction and characterization accuracy of the produced catalogue, we made use of both simulated data and cross-matches found with other surveys.\\
Simulated samples were drawn from the \scorpio{} mosaics (both ASKAP and ATCA) using the following approach\footnote{Simulation tasks (compact source subtraction, point-source generation) can be performed using application scripts provided along with \caesar{}.}. First, a residual map was obtained by subtracting all compact sources down to a very low detection threshold (2$\sigma$). Imaging artefacts, extended and diffuse sources were not removed.
A sample of 50 simulated maps was then generated by artificially adding uniformly spaced point sources to the residual mosaic, with a density of 100 deg$^{-2}$ and exponential flux density distribution ($\propto$e$^{-\lambda S}$ with $\lambda$=1.6, as for real data above the detection threshold).
\caesar{} was finally run on the simulated maps and the same set of quality criteria used for the \scorpio{} mosaics were applied to the simulated catalogue. Source extraction and characterization metrics were estimated by cross-matching injected sources with extracted sources.

\begin{figure}
\centering%
\includegraphics[scale=0.35]{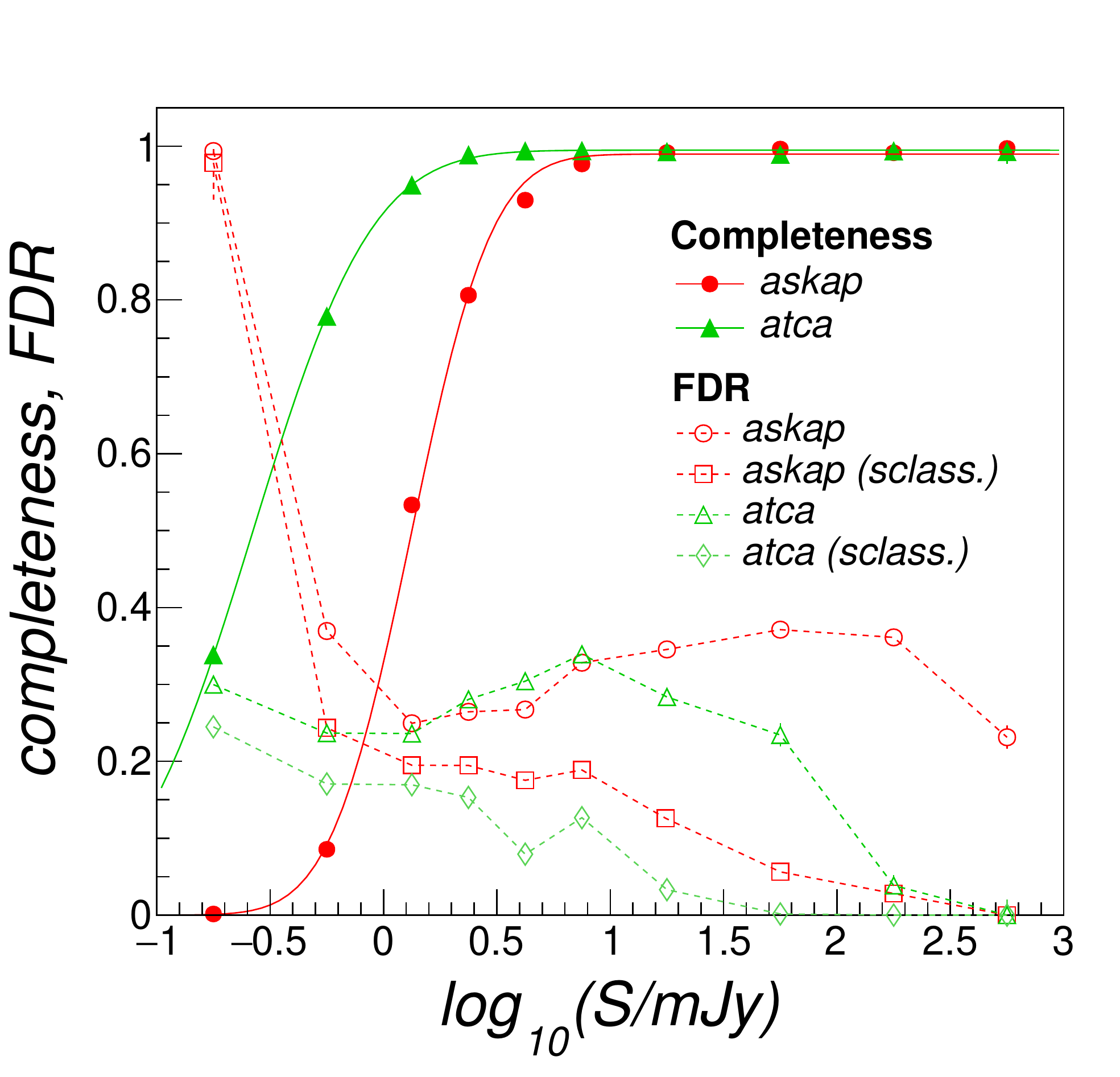}
\caption{Completeness (filled markers) and false detection rate (FDR) (open markers) as a function of the injected and measured source flux density respectively obtained on the simulated ASKAP (red markers) and ATCA (green markers) source catalogues. Solid lines represent the fitted completeness model of equation~\ref{eq:completeness-model}. Open squares and diamonds (labelled as "sclass") represent the false detection rate obtained after applying a neural network classifier, trained to identify spurious fit components in the catalogue.}
\label{fig:completeness-reliability}
\end{figure}

\subsubsection{Completeness and reliability}
\label{sec:completeness-reliability}
Catalogue completeness and reliability were studied with simulated data as a function of the source flux density. Sources injected in the simulated maps are considered as the true catalogue. Extracted sources that do not crossmatch to any injected source are thus counted as false detections and contribute to increase the false detection rate (or decrease the reliability). Results are reported in Fig.~\ref{fig:completeness-reliability} for both ASKAP (red markers) and ATCA (green markers) simulated data. The completeness $C$ is shown with filled markers and can be parametrized as a function of the flux density $S$ as:
\begin{equation}
\label{eq:completeness-model}
C=\frac{C_{0}}{2}\left(1 + \text{erf}\left\{\frac{\log_{10}(S)-\log_{10}(S_{0})}{k}\right\}\right)
\end{equation}
with parameters: $C_{0}$=0.99, $\log_{10}(S_{0})$=$-$2.88, $k$=0.40 (ASKAP) and $C_{0}$=0.99, $\log_{10}(S_{0})$=$-$3.58, $k$=0.59 (ATCA). For ASKAP (ATCA) observations we expect the catalogue to be $>$90\% complete above 5 mJy (1 mJy). 
\\The false detection rate, reported in Fig.~\ref{fig:completeness-reliability} with open dots (ASKAP) and open triangles (ATCA), is found to be of the order of 20-30\%, slightly decreasing by a few percent when moving away from the Galactic plane. False detections are largely due to sidelobes around bright sources (compact or extended) and overdeblending of extended sources. Reliability can be improved by $\sim$10\% after applying a neural network classifier previously trained to identify "good" sources from the data (see \citealt{Riggi2019} for more details). Results are shown in Fig.~\ref{fig:completeness-reliability} with open squares (ASKAP) and open diamonds (ATCA). 
To further reduce the false detection rate to acceptable levels (below 1\%) we visually inspected the produced catalogue removing spurious sources that were not identified in the automated procedure (see Section~\ref{sec:sfinding}).\\ 
For the future we expect this can be substantially improved in multiple ways, e.g. increasing the number of classifier parameters, tuning of classifier hyperparameters, introducing specialized classifiers trained to identify imaging artefacts (e.g. around bright sources) which cannot be efficiently
removed with the classifier adopted in this work.

\subsubsection{Source position accuracy}
\label{subsec:position-accuracy}
Positional uncertainties in RA and Dec can be expressed as:
\begin{equation}
\left.
\begin{aligned}
&\sigma_{\alpha}=\sqrt{\sigma_{\alpha,fit}^{2}+\sigma_{\alpha,cal}^{2}}\\%
&\sigma_{\delta}=\sqrt{\sigma_{\delta,fit}^{2}+\sigma_{\delta,cal}^{2}}
\end{aligned}    
\right.
\end{equation}
where $\sigma_{fit}$ are the position uncertainties due to the fitting process (including noise and uncertainties in fit parameters) while $\sigma_{cal}$ are the uncertainties due to the calibration process.\\
$\sigma_{fit}$ is provided by the fitting routine for each source and can be compared with values obtained from simulated data as a function of the source signal-to-noise ratio (S/N). 
We found that the fit uncertainties are largely underestimated with respect to the expected values found in the simulations. The latter are a factor 3-4 larger than the semi-analytical estimates provided by \cite{Condon1997}\footnote{A discrepancy (a factor $\sim$2) between analytical and measured uncertainties was observed also in other analysis carried out with simulated data using \caesar{} \citep{Riggi2019} or alternative finders \citep{Hopkins2015}.}. This is partly expected given that the semi-analytical estimate and the estimate obtained from the fitting routine are neglecting the effect of the correlated noise in the map. A correction for this effect, for example using equation 41 of \cite{Condon1997} in the maximal-smoothing limit, would yield an expected analytical uncertainty $\sim$2.6 larger. We have finally set $\sigma_{fit}$ to the values obtained from the simulations and parametrized it as a function of S/N as:
\begin{equation}
\left.
\begin{aligned}
&\sigma_{\alpha,fit}^{\text{ASKAP}}=\frac{9.86}{S/N} ~\rmn{arcsec},\;\;\;\sigma_{\delta,fit}^{\text{ASKAP}}=\frac{6.01}{S/N}~\rmn{arcsec}\\%
&\sigma_{\alpha,fit}^{\text{ATCA}}=\frac{2.28}{S/N}~\rmn{arcsec},\;\;\;\sigma_{\delta,fit}^{\text{ATCA}}=\frac{2.98}{S/N}~\rmn{arcsec}%
\end{aligned}    
\right.
\end{equation}
Position fit uncertainties are of the order of 1.5"-1.8" (0.5"-0.6") at the detection threshold and better than 0.5" (0.15") with S/N>20 for ASKAP (ATCA) data.\\Calibration uncertainties $\sigma_{cal}$ can be inferred from the position spread observed in bright sources with respect to a reference catalogue. The Radio Fundamental Catalogue (RFC, version rfc\_2020b)\footnote{\url{http://astrogeo.org/rfc/}}, for instance, contains $\sim$17000 radio sources measured in multiple VLBI observations with milliarcsecond accuracy. 5 RFC sources, detected in the X band, cross-match with ASKAP bright sources (four with $S/N>$200 and one with $S/N>$20) and have position uncertainty smaller than 0.02". The standard deviations ($s_{\alpha}$=0.5", $s_{\delta}$=0.4") of the observed position offset provide a first measure of $\sigma_{cal}$ since both RFC and ASKAP fit position uncertainties are negligible at high S/N. Additionally, we inferred ASKAP calibration errors using 344 and 80 sources with S/N>50 matching to MGPS and NVSS catalogues, respectively. The observed standard deviations ($s_{\alpha}$=1.3"-1.5", $s_{\delta}$=1.3"-1.4") suggest a slightly larger calibration uncertainty of $\sigma_{cal,\alpha}$=1.0"-1.1" and $\sigma_{cal,\delta}$=0.8"-1.1", after subtracting (in quadrature) the ASKAP fit errors and the MGPS ($\sim$0.9" from \citealt{Murphy2007}) or NVSS ($\sim$1" from \citealt{Condon1998}) uncertainties.\\
As no ATCA-RFC matches were found, to infer calibration errors for ATCA, we considered 38 sources detected with S/N>50 and matched to MGPS sources. In this case, we did not use NVSS data as no matches were found with ATCA above the considered significance level. From the offset standard deviations ($s_{\alpha}$=2.1", $s_{\delta}$=2.5"), we obtained a calibration uncertainty of $\sigma_{cal,\alpha}$=1.9" and $\sigma_{cal,\delta}$=2.4", after subtracting (in quadrature) the ATCA fit errors and the MGPS uncertainties. Such estimates should be regarded as upper limits if MGPS positional uncertainties are underestimated.\\
Systematic offsets possibly introduced by the fitting procedure were investigated with simulated data for both ASKAP and ATCA. No bias (e.g. median offsets smaller than 0.04") was found and, hence, no corrections were applied to the catalogues.\\
To assess the absolute positional accuracy of the ASKAP ESP data, we considered the median offsets found with respect to the matched sources in RFC, ATCA, MGPS and NVSS catalogues. We considered only the ASKAP sources with S/N>50. From the comparison, we found that the ASKAP Dec offsets are negligible (smaller than 0.05"), while the ASKAP RA is systematically higher than that of the reference catalogues: $\langle\alpha\rangle$=1.4" (RFC), $\langle\alpha\rangle$=1.4" (ATCA), $\langle\alpha\rangle$=1.6" (NVSS), $\langle\alpha\rangle$=1.7" (MGPS)\footnote{The median offset was found to be 2.7" but we corrected for the systematic offset of -1" reported in \citealt{Murphy2007} with respect to NVSS data.}. The RA offset is consistent as a function of the ASKAP source flux density above S/N>20 (e.g. varying by $\sim$0.2 at most from bin to bin) and slightly larger than the total position uncertainties estimated above. 
The astrometric offset, affecting also other ASKAP Early Science observations, is due to a known bug in the ASKAPSoft software version used to process \scorpio{} ASKAP15 data. Recent analysis carried out on newer observations done with the full ASKAP array and improved versions of the reduction software and calibration procedure, do not report significant astrometric offsets. Indeed, we were not able to detect comparable offsets from the preliminary \scorpio{} ASKAP36 maps. For this analysis, we have decided to account for the observed systematic offset by increasing the calibration uncertainty to 1.5".


\subsubsection{Source flux density accuracy}
\label{subsec:flux-accuracy}
\begin{figure}
\centering%
\includegraphics[scale=0.42]{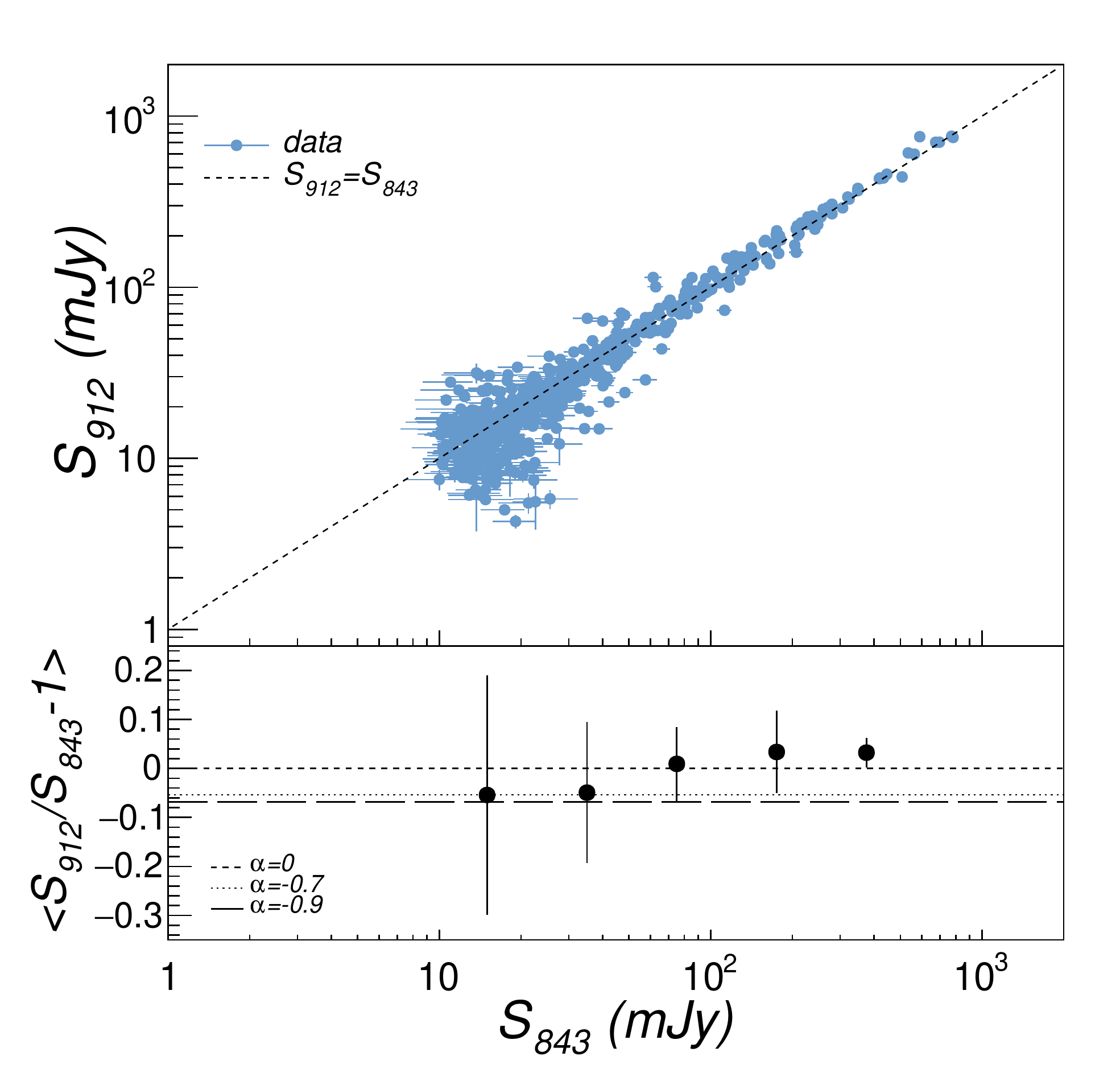}%
\caption{Top: Correlation between flux densities $S_{912}$ and $S_{843}$ of cross-matched sources found in ASKAP (912 MHz) and MGPS (843 MHz) catalogues. The dashed black line corresponds to a one-to-one relation ($S_{912}$=$S_{843}$). Bottom: Relative difference between ASKAP and MGPS flux densities. Black dashed and dotted lines represent the difference expected from $S\propto\nu^{\alpha}$ for three different spectral indices (0,-0.7,-0.9).}
\label{fig:fluxscalecomparison}
\end{figure}

Uncertainties on the measured flux density $S$ are mainly due to the uncertainties of the source fitting process in the presence of noise ($\sigma_{fit}$, expressed here as a percentage of the flux density) and the flux scale calibration uncertainties ($\sigma_{cal}$, usually given as a percentage of the flux density). Total relative uncertainties $\sigma$ can be therefore expressed as:
\begin{equation}
\sigma=\sqrt{\sigma_{fit}^{2}+\sigma_{cal}^{2}}
\end{equation}
$\sigma_{fit}$ values for each source are obtained by error propagation using source parameter errors provided by the fitting routine. These were found a factor $\sim$3 smaller with respect to the expected values obtained from simulated data. Similarly to what was done in Section~\ref{subsec:position-accuracy}, we have set the fit uncertainties for both ASKAP and ATCA data to the following parametrized values, derived from the simulations as a function of the source signal-to-noise ratio (S/N):
\begin{equation}
\left.
\begin{aligned}
&\sigma_{fit}^{ASKAP}= 0.14\times\left[\log_{10}{(S/N)}\right]^{-2.72}\\%
&\sigma_{fit}^{ATCA}= 0.14\times\left[\log_{10}{(S/N)}\right]^{-2.74}
\end{aligned}    
\right.
\end{equation}
Fit uncertainties are $\sim$35\% at the detection threshold and better than 5\% with S/N>30.\\It is now known from simulation studies \citep{Hopkins2015,Riggi2019} that many source finders produce a biased flux density measurement as they approach the detection threshold. \caesar{}, like other widely used finders (e.g. \selavy{}, \aegean{}, \pybdsm{}), tends to overestimate the flux density at low S/N. As discussed in \cite{Hopkins2015}, this is likely due to a poorly constrained Gaussian fit. Systematic flux density offsets $\langle\Delta S/S\rangle_{fit}$ were therefore characterized using simulated data and parametrized as a function of the source S/N as follows:
\begin{equation}
\left.
\begin{aligned}
&\langle\Delta S/S\rangle_{fit}^{ASKAP}=0.05\times\left[\log_{10}{(S/N)}\right]^{-4.97}\\%
&\langle\Delta S/S\rangle_{fit}^{ATCA}=0.06\times\left[\log_{10}{(S/N)}\right]^{-4.22}
\end{aligned}    
\right.
\end{equation}
Systematic fit biases are $\sim$25-30\% at the detection threshold and smaller than $\sim$1\% with S/N>25. Measured flux densities in both ASKAP and ATCA catalogues were corrected using the above parametrizations.\\
To assess the absolute flux density scale reliability, we carried out a comparison of the measured ASKAP source flux densities with those reported in previous catalogues at a nearby frequency, such as the MGPS source catalogue. The correlation between ASKAP ($S_{912}$) and MGPS ($S_{843}$) source flux densities is reported in the top panel of Fig.~\ref{fig:fluxscalecomparison}.
The dashed black line corresponds to a perfect correlation ($S_{912}$=$S_{843}$).
The bottom panel represents the median relative flux density difference observed as a function of the MGPS flux density. Error bars correspond to the semi-interquartile range of the data in each bin. If all sources were to have a single power-law spectrum $S\propto\nu^{\alpha}$, we would ideally expect to observe a relative difference of (912/843)$^{\alpha}-$1 between the two catalogues. Expected offsets are shown in Fig.~\ref{fig:fluxscalecomparison} with black dashed and dotted lines for three choices of expected spectral indices: $\alpha$=$-$0.7 for extragalactic sources, $\alpha$=0 for Galactic sources with thermal-dominated emission and an average $\alpha$=$-$0.9 as observed in Section~\ref{sec:spectralindex} or in other radio surveys in the Galactic plane \citep{Cavallaro2018,Wang2018}. For sources brighter than $\sim$70 mJy, less affected by possible flux density reconstruction bias on both catalogues, we observe that ASKAP flux densities are on average larger by $\sim$3\% with respect to MGPS fluxes. Provided that the source sample is not completely dominated by Galactic thermal sources, we would infer that ASKAP fluxes are overestimated by 9-10\% compared to those expected from a mixed population of sources with an average spectral index of $-$0.9.\\
We also compared the ASKAP flux densities of selected sources with those expected from a power-law fit obtained with at least four spectral points using MGPS, NVSS and ATCA data. 45 sources (28 with S/N>50) are well-modelled by a power-law ($\tilde{\chi}^{2}$<2, spectral indices ranging from $-$0.7 to $-$0.9) over the entire frequency range and, thus, can be used as additional flux calibrators. ASKAP flux densities for these sources were on average found in excess by $\sim$10\% (S/N<50) and $\sim$5\% (S/N>50) with respect to the predicted values. The standard deviation of the observed offset of bright sources (S/N>50) is $\sim$5\% and can be used as a measure of $\sigma_{cal}$. These results suggest a flux density scale inconsistency in the ASKAP data calibration or imaging process that has to be investigated with the full array and improved releases of the data reduction pipeline. The origin of the observed shift is in fact not fully understood at this early stage stage of ASKAP observations in which the calibration process is not yet optimized. For this reason, we have decided not to correct for it and assume a larger calibration uncertainty (10\%) to carry out the analysis described in Section~\ref{sec:analysis}. For the ATCA data we have instead assumed $\sigma_{cal}$=3\% following previous works on the \scorpio{} field \citep{Cavallaro2018}.

\section{Analysis}
\label{sec:analysis}

\begin{figure}
\centering%
\subtable[Integrated to peak ratio vs S/N]{\includegraphics[scale=0.35]{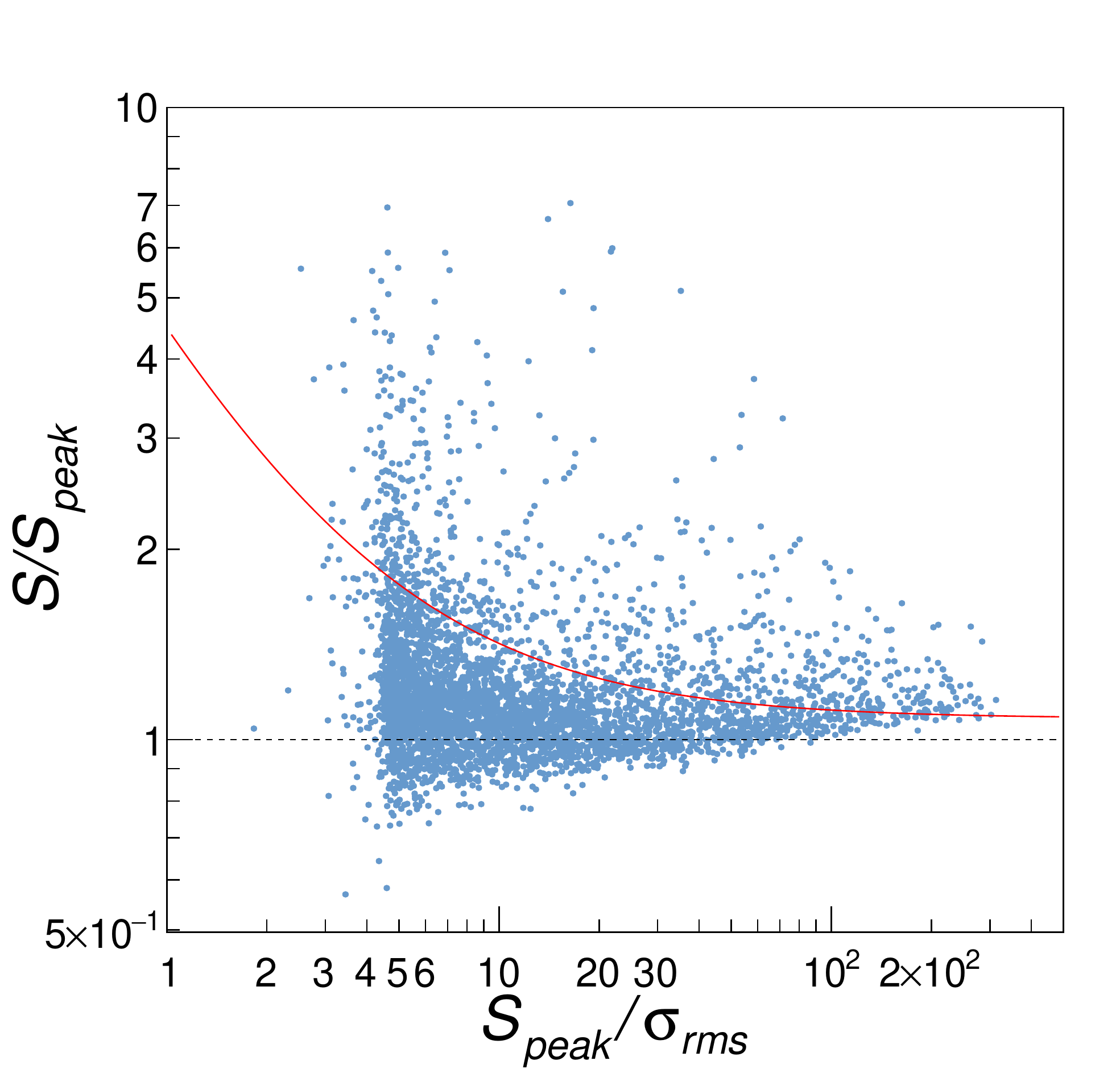}\label{fig:extsourcefraction_1}}\\%
\vspace{-0.35cm}%
\subtable[Resolved source fraction]{\includegraphics[scale=0.35]{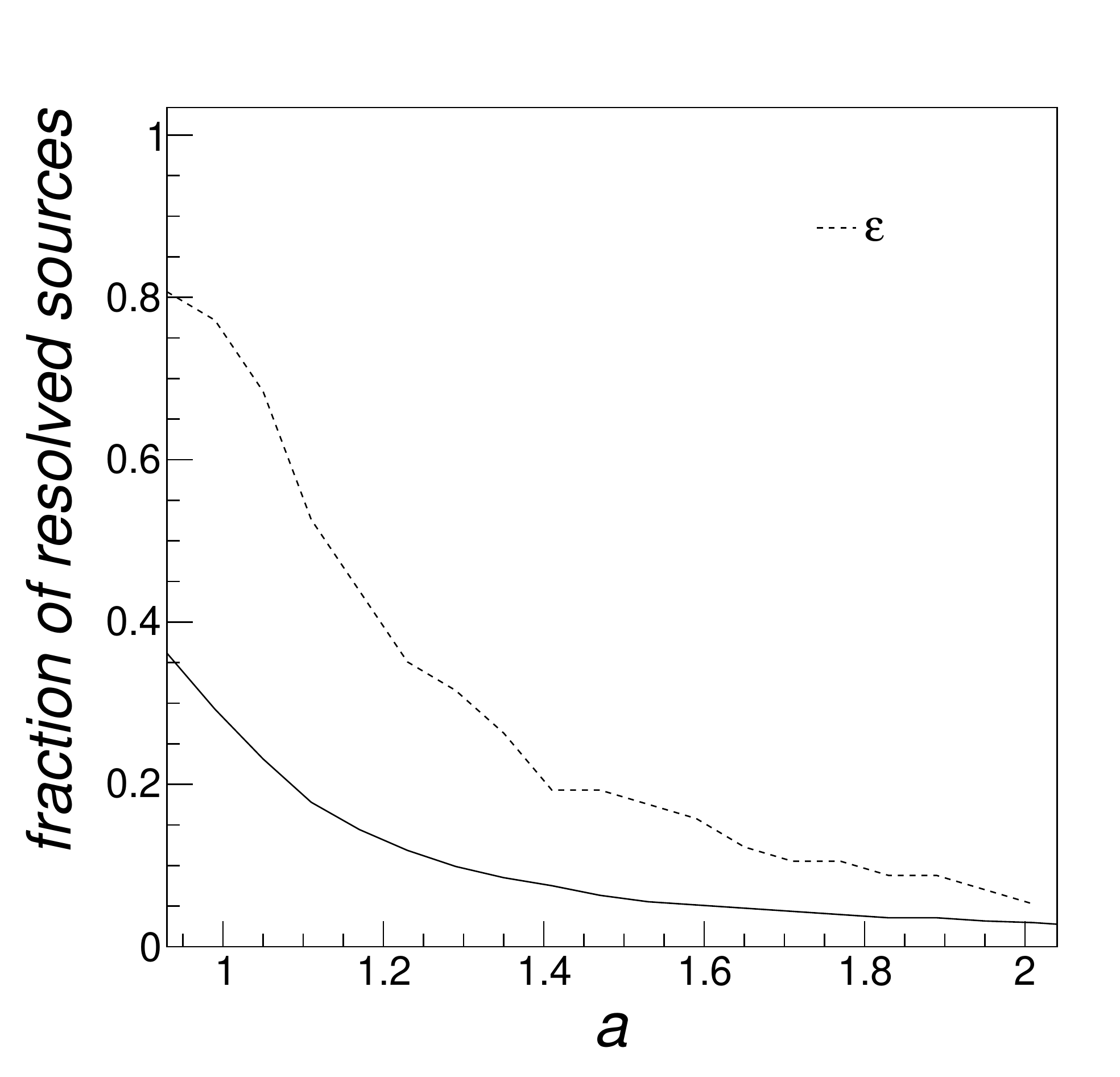}\label{fig:extsourcefraction_2}}%
\caption{Upper panel: Ratio between integrated $S$ and peak $S_{peak}$ flux densities for catalogued sources as a function of their signal to noise ratio $S/N$. The red line indicates the curve $S/S_{peak}$=$a$+$\frac{b}{S/N}$ ($a$=1.08, $b$=3.4), used in the XXL survey, above which sources are labelled as resolved.
Lower panel: Fraction of catalogued sources that would be classified as resolved as a function of the applied threshold on the parameter $a$ (with $b$ fixed to 3.4), shown with a solid black line. The dashed black line represents the fraction of truly resolved sources $\varepsilon$ (found by comparison with ATCA data) that would be classified as resolved as a function of the applied threshold on the parameter $a$.}%
\label{fig:extsourcefraction}
\end{figure}

\subsection{Estimation of the fraction of resolved sources}
\label{sec:extendedsources}
A widely used method \citep{Franzen2015,Franzen2019} to identify non point-like sources in the catalogue is based on the ratio $S$/$S_{peak}$ between the integrated and peak flux density which is expected to be larger than 1 for extended sources. In the XXL survey by \cite{Butler2018}, sources were classified as resolved if they fulfil the following empirical relation:
\begin{equation}
S/S_{peak}> a + \frac{b}{S/N}    
\end{equation}
with $a=$1.08 to account for possible calibration errors and $b$=2.03, defined so as to keep 90\% of the data having $S$/$S_{peak}$<1 above the curve $S$/$S_{peak}$= $a$ - $\frac{b}{S/N}$. We tuned the value of $b$ for ASKAP \scorpio{} survey using simulated data (see Section~\ref{sec:catalog-uncertainties}). To have less than 5\% of misclassified simulated point-sources, the value of $b$ has to be increased to $\sim$3.4.
In Fig.~\ref{fig:extsourcefraction_1} we report the $S$/$S_{peak}$ as a function of S/N for \scorpio{} source components. The red line represents the empirical function used in the XXL survey with parameters: $a$=1.08, $b$=3.4. We report in Fig.~\ref{fig:extsourcefraction_2} the fraction of catalogued sources that would be classified as extended if thresholded in $S$/$S_{peak}$ as a function of the applied threshold on the parameter $a$ (solid black line), assuming $b$ fixed to 3.4. Adopting the same criterion as in the XXL survey ($a$=1.08) we would classify $\sim$20\% of the catalogued sources as resolved. This is comparable to values reported in other surveys carried out in the Galactic plane (e.g. $\sim$20\% in the THOR survey \citep{Wang2018}) or far from the Galactic plane and with different spatial resolutions (e.g. see \citealt{Butler2018} and references therein)\footnote{For comparison, the angular resolution of XXL and THOR survey are 4.8" and 25", respectively.}.\\Using the \textsc{Scorpio} ATCA higher resolution image, it is possible to estimate the number of truly resolved sources at least for a portion of the \textsc{Scorpio} field. In Section~\ref{sec:atca-crossmatches} we found that 52/648 ASKAP sources match to more than one ATCA source, e.g. $\sim$8\% of the ASKAP catalogued sources have a genuine extended nature. Using this limited source sample, we can compute the fraction of truly resolved sources that would be classified as extended, $\varepsilon$, according to the $S$/$S_{peak}$ criterion described above. We report the results in Fig.~\ref{fig:extsourcefraction_2} with a dashed black line. As can be seen, the $S$/$S_{peak}$ parameter is not sensitive enough to allow a perfect identification of known resolved sources (e.g. for $a$=1.08 we obtained an identification efficiency of $\sim$60\%).\\We therefore repeated the analysis including additional parameters. Following \cite{Riggi2019}, we considered in particular the ratio $E$/$E_{beam}$ between the source fitted ellipse eccentricity $E$ and the beam ellipse eccentricity $E_{beam}$. We found no significant improvements in the extended source identification capabilities compared to the case of a single discriminant parameter. In light of these results, we will conservatively assume $\sim$8\% as a lower limit in the fraction of extended sources expected in the reported catalogue. Future observations of the \scorpio{} field with the complete array and at higher ASKAP frequencies will provide an improved angular resolution (by a factor of 3-4), enabling a direct identification of additional extended sources over the entire field. 

\begin{table}
\caption{Number of ASKAP sources detected at one or multiple radio frequencies after cross-matching them with MGPS, ATCA (also including sub-band data) and NVSS catalogues.}
\begin{tabular}{c|c|c}
\hline%
\hline%
\# frequencies & \# sources & percentage(\%)\\%
\hline%
1 & 2986 & 72.1\\
2 & 504 & 12.2\\
>2 & 654 & 15.8\\
3 & 119 & 2.9\\
4 & 25 & 0.6\\
5 & 52 & 1.3\\
6 & 63 & 1.5\\
7 & 222 & 5.4\\
8 & 152 & 3.7\\
9 & 21 & 0.5\\%
\hline%
\end{tabular}
\label{tab:frequency_multiplicity}
\end{table}

\begin{figure}
\centering%
\includegraphics[scale=0.4]{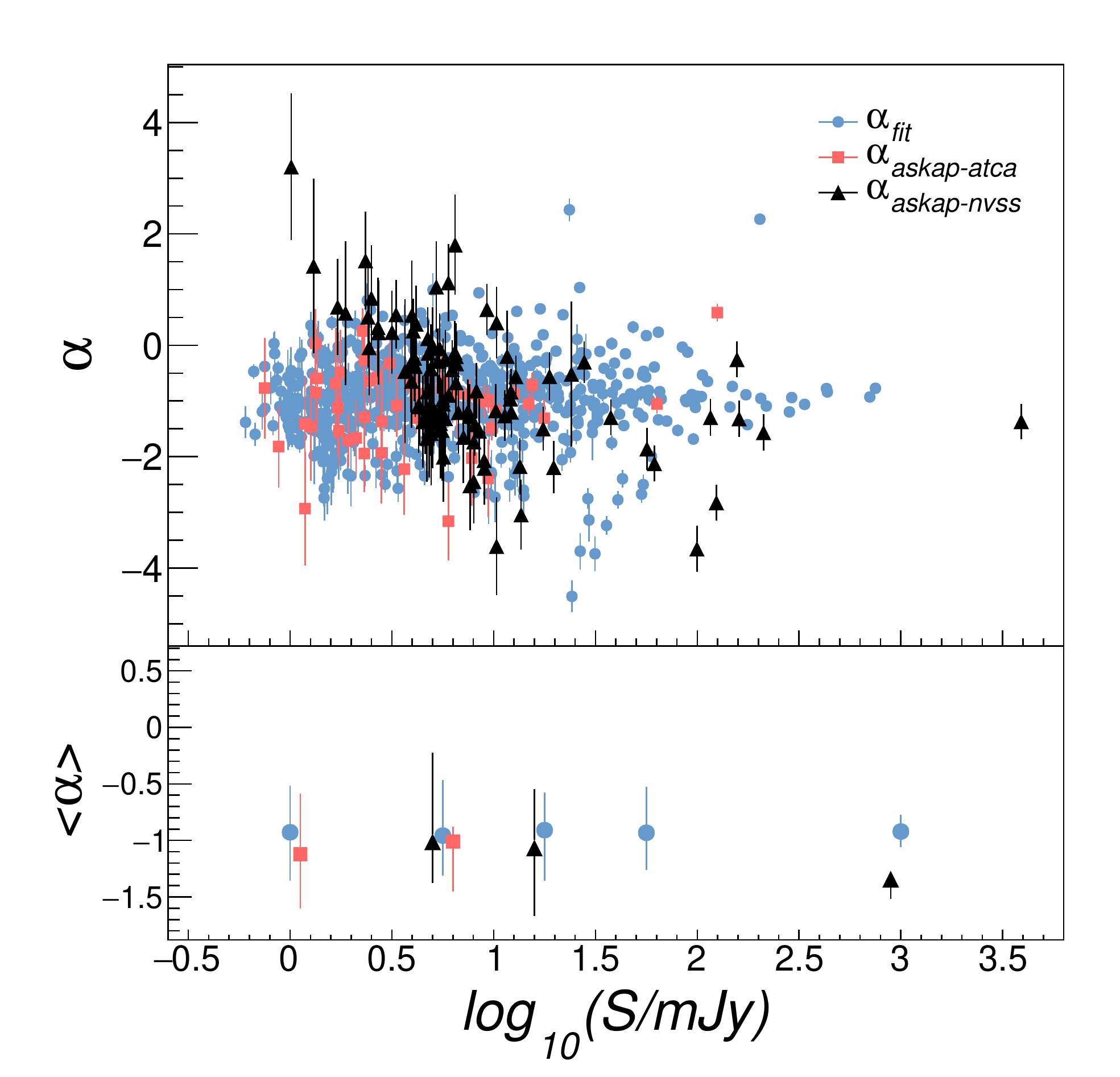}
\caption{Top panel: Spectral indices of ASKAP 912 MHz source components as a function of ASKAP flux density. Blue dots represent the spectral indices obtained from a power-law fit of available ASKAP-ATCA-NVSS-MGPS data, while the red squares and black triangles indicate the spectral indices obtained from only ASKAP-ATCA and ASKAP-NVSS frequencies, respectively. Bottom panel: Median spectral index for the three samples as a function of ASKAP flux density. Error bars are the interquartile range (IQR) in each flux density bin.}%
\label{fig:spectralindex}
\end{figure}

\subsection{Spectral indices}
\label{sec:spectralindex}
Following the convention $S\propto\nu^{\alpha}$ (where $S$ is the integrated source flux), we estimated the spectral index $\alpha$ of ASKAP catalogued sources using the cross-matches found in Section~\ref{sec:crossmatching} with the ATCA (including also sub-band data), MGPS and NVSS source catalogues. The number of ASKAP sources detected at one or multiple radio frequencies are reported in Table~\ref{tab:frequency_multiplicity}. More than 70\% of the sources do not cross-match with any of the considered catalogues and thus no spectral index information can be reported for them. 12\% of the sources have flux information only at two frequencies. For them only a two-point spectral index can be reported. These are to be considered as first-order estimates and might not represent a good estimate particularly for sources in which a turnover is present between the two frequencies. For example this is the case for some of the Gigahertz Peaked Spectrum (GPS) and Compact Steep Spectrum (CSS) radio sources (about 40, or $\sim$20\%, in the compilation presented in \citealt{Jeyakumar2016}). For spectral indices obtained only from ASKAP and ATCA data, taking into account the source extraction performance in both catalogues, we expect to obtain an unbiased estimate of $\alpha$ above 10 mJy at 912 MHz with uncertainties smaller than $\sim$0.2.\\For $\sim$16\% of the sources, we have flux information in at least three different frequencies. These data were fitted with a power-law model to determine the spectral indices. When the fit does not converge or does not pass a minimum quality criterion ($\tilde{\chi}^{2}$<10), e.g. due to the presence of outliers in one or more frequency bands, a robust linear regression is performed excluding data points with larger fit residuals. Without improvement (e.g. the $\tilde{\chi}^{2}$ of the robust fit is still larger than the quality threshold), the reported source spectral index is finally set to the value found using only two frequencies, ASKAP-ATCA or ASKAP-NVSS (if no ATCA information is available).\\
In Fig.~\ref{fig:spectralindex} (upper panel) we report the spectral indices obtained from the radio spectrum fitting procedure (635 sources, blue dots) on available ASKAP-ATCA-NVSS-MGPS matches, and those obtained using only ASKAP-ATCA (49 sources, red squares) and ASKAP-NVSS (96, black triangles) matches. Some sources ($\sim$1\% of the sample) have a rather extreme spectral index ($\alpha$<$-$3 or $\alpha$>2.5). Although there is a chance they are pulsars ($\alpha$<$-$3) or hyper-/ultra-compact \hii{} regions with an optically thick free-free emission ($\alpha$>2.5), their measured indices are somewhat questionable.
We note that the majority of the potentially unreliable indices are indeed obtained from two nearby frequencies (ASKAP-NVSS), for which a change in flux density (e.g. due to errors) of $\sim$10\% would lead to a $\sim$0.5 variation in the measured spectral index. These spectral indices should therefore be treated with caution and be re-estimated with data at additional frequencies. Five sources, in particular, present a very steep spectrum ($\alpha$<$-$3), obtained by fitting ASKAP, MGPS and NVSS data. Spectral index values are in this case mainly determined by NVSS measurements. From a visual inspection we were not able to spot possible issues in the ASKAP or MGPS flux estimate, e.g. due to a complex background or imaging artefacts. NVSS data for two sources (\texttt{J171435-392543}, \texttt{J171448-394756}) are instead considerably affected by artefacts and thus the reported flux densities may be not accurate. Future ASKAP \scorpio{} observations, bridging the frequency gap between MGPS and NVSS measurements, will allow us to determine the spectral slope more reliably.\\Median spectral indices are reported in the bottom panel of Fig.~\ref{fig:spectralindex} as a function of the ASKAP source flux density. The median value of fitted spectral indices for brighter sources (>10 mJy) is $\langle\alpha\rangle=-$0.92 (IQR=0.68). A robust least squares linear fit to the data of Fig.~\ref{fig:spectralindex} (top panel) yields similar values. These estimates are consistent within the quoted uncertainties with those obtained in a completely independent analysis \citep{Cavallaro2018} carried out on a portion of the \textsc{Scorpio} ATCA field using only ATCA sub-band data.
\begin{figure}
\centering%
\subtable[Response matrix]{\includegraphics[scale=0.32]{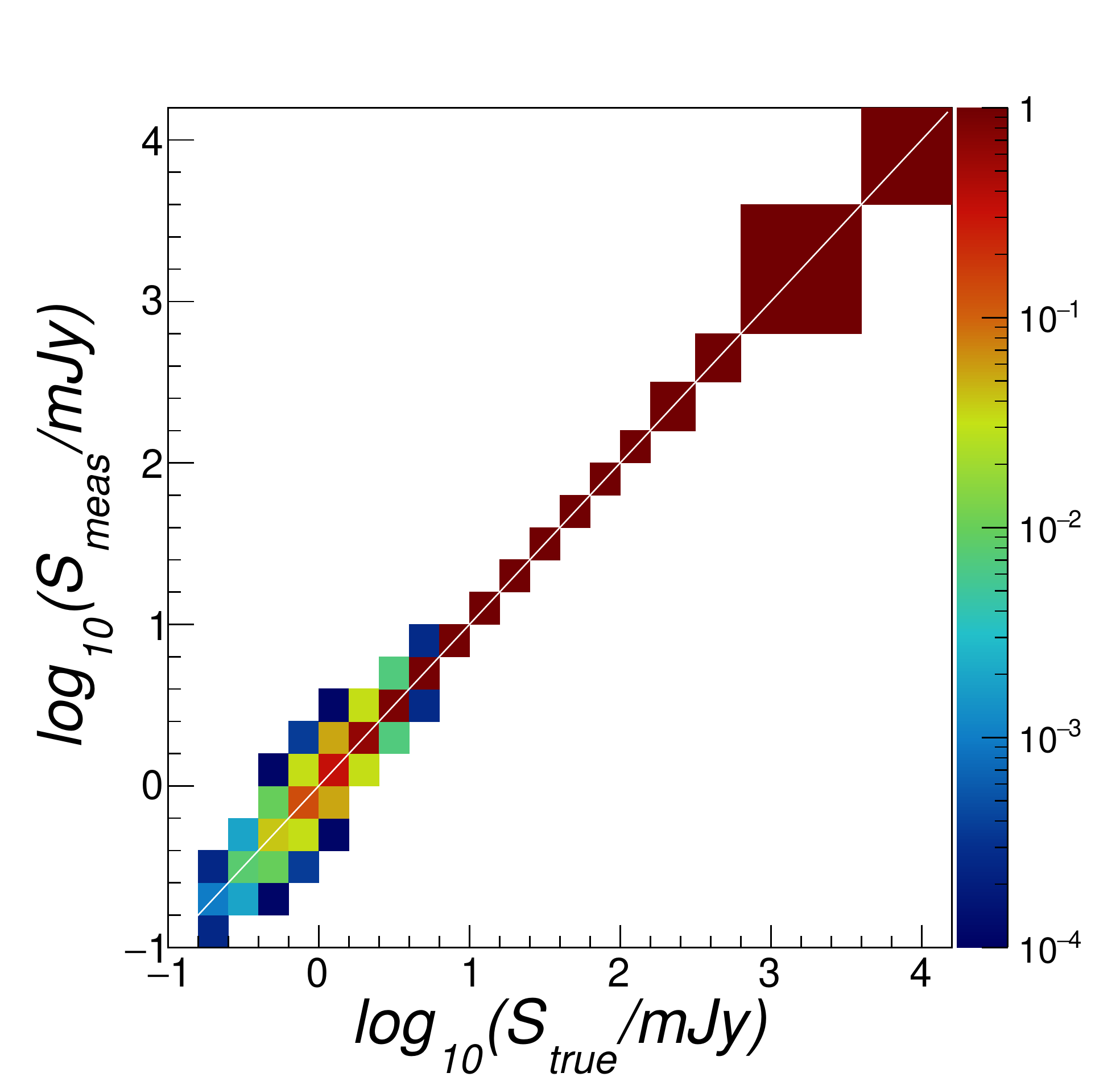}\label{fig:sourcecounts_1}}\\%
\vspace{-0.4cm}%
\subtable[Differential source counts]{\includegraphics[scale=0.38]{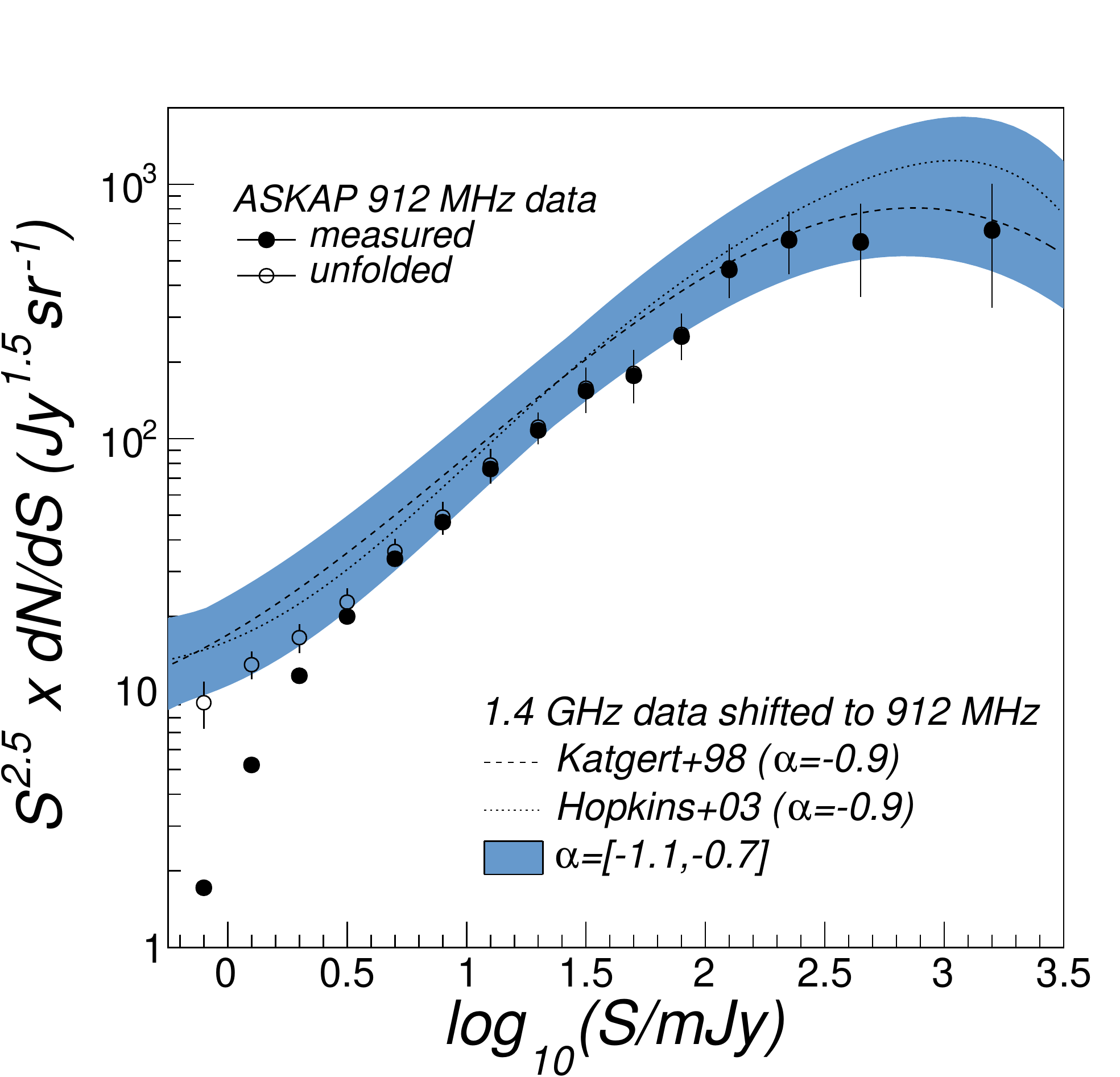}\label{fig:sourcecounts_2}}%
\caption{Top: Response matrix used to unfold source detection effects (detection efficiency, flux bias and uncertainty) from raw source counts. 
The color represents the probability $r_{ji}$ for a source with true flux density in bin $i$ to be detected with measured flux density in bin $j$; 
Bottom: Differential uncorrected source counts of the \textsc{Scorpio} survey catalogue normalized to standard Euclidean counts ($\propto$ $S^{-2.5}$) shown with filled black dots. Open black dots represents the unfolded differential source counts obtained with the procedure described in the text and using the response matrix above. The shaded area represents the existing extragalactic source count data at 1.4 GHz, parametrized by \protect\cite{Katgert1988,Hopkins2003} and shifted to a frequency of 912 MHz assuming a spectral index $\alpha$ ranging from $-$1.1 to $-$0.7 (central value $-$0.9). Values relative to the median spectral index $-$0.9 are shown as dashed and dotted lines, respectively.}%
\label{fig:sourcecounts}
\end{figure}

\subsection{Source counts}
\label{sec:sourcecounts}
In Fig.~\ref{fig:sourcecounts_2} (filled black dots) we report the differential source counts $dN$/$dS$ (number of sources $N$ per unit area per unit flux density $S$) obtained by dividing the number of sources found in each flux density bin by the survey area (37.7 deg$^{2}\sim$11.5$\times$10$^{-3}$ sr) and flux density bin width. Counts were not corrected for expected source detection efficiency and flux density reconstruction accuracy (bias and uncertainty) and were normalized by $S^{-2.5}$ (the Euclidean slope) \citep{Condon1984} as conventionally done in other studies of source counts. Error bars denote the statistical Poissonian uncertainties on the obtained counts.\\
To compare it with other surveys we need to unfold the source detection effects from the measured counts.
The former have been modeled through a response matrix $\mathbf{R}=[r_{ji}]$ expressing the probability for a source with true flux density $x\equiv\log_{10}\left(S_{true}\right)$ in bin $i$ to be detected with measured flux density $y\equiv\log_{10}\left(S_{meas}\right)$ in bin $j$. $r_{ji}$ is computed for each flux density bin ($i$,$j$) as:
\begin{equation}
\label{eq:response-matrix}
r_{ji}=\int_{y_{j}-\Delta y_{j}/2}^{y_{j}+\Delta y_{j}/2}\int_{x_i-\Delta x_i/2}^{x_i+\Delta x_i/2} \frac{\varepsilon(x)\varepsilon^{reso}(x)}{\sqrt{2\pi}\sigma(x)}
\exp\left\{ -\frac{\left(y-x\right)^{2}}{2\sigma^{2}(x)}\right\}
dx\;dy
\end{equation}
where $x_i$, $\Delta x_i$ and $y_j$, $\Delta y_j$ are the true and measured flux density bin centres and widths. $\sigma$ and $\varepsilon$ are the flux density uncertainty and the source detection efficiency, respectively\footnote{No bias term is included in equation~\ref{eq:response-matrix}, as we already corrected our data for measurement bias.}. Both have been parametrized as a function of the true flux density using simulated data (see Sections~\ref{subsec:flux-accuracy} and \ref{sec:completeness-reliability}). $\varepsilon^{reso}$ is introduced to correct for the catalogue incompleteness effect related to the source angular size and known as the "resolution bias". Resolved sources with the same integrated flux density of point sources have in fact lower peak flux densities and thus a higher chance to be missed in the catalogue as falling below the source detection threshold. The response matrix was finally normalized such that the probabilities for a true flux density bin $i$ sum up to the overall efficiency in bin $i$, e.g. $\varepsilon_{i}\varepsilon_{i}^{reso}$.\\
$\varepsilon^{reso}$ was set equal to $1-h(>\phi_{lim})$, where $h(>\phi_{lim})$ is the fraction of sources above a maximum angular size $\phi_{lim}$ being missed as resolved. $h(>\phi_{lim})$ was estimated as a function of the flux density as described in \cite{Windhorst1990,Prandoni2001,Thorat2013,Retana-Montenegro2018}:
\begin{align}
\varepsilon^{reso}&= 1-h(>\phi_{lim})\\%
h(>\phi_{lim})&=\exp{[-\ln{2}(\phi_{lim}/\phi_{median})^{0.62}]}
\end{align}
\begin{align}
\phi_{median}&= \left\{
\begin{array}{ll}
2" & S_{1.4\,\text{GHz}}\leq1\,\text{mJy}\\
2"\times S_{1.4\,\text{GHz}}^{0.3} & S_{1.4\,\text{GHz}}>1\,\text{mJy}\\
\end{array} 
\right.\\
\phi_{lim}&=max(\phi_{1},\phi_{2})\\%
\phi_{1}&=\sqrt{b_{maj}b_{min}\times(S/S_{thr}-1)}\\%
\phi_{2}&=\sqrt{b_{maj}b_{min}\times(a+\frac{b}{S/\sigma_{rms}})}%
\end{align}
where $b_{maj}$=24" and $b_{min}$=21" are the beam sizes, $\sigma_{rms}$=300 $\mu$Jy/beam is the minimum noise value in the map, $S_{thr}$=1.5 mJy is the source minimum detection threshold\footnote{The minimum detection threshold value corresponds here to the 5$\sigma$ threshold with respect to the minimum noise ($\sigma_{rms}$=300 $\mu$Jy/beam) in the map, not to the "effective" detection threshold, averaged over the mosaic, which amounts to $\sim$5 mJy.}, $S_{1.4\,\text{GHz}}$ is the integrated flux density at 1.4 GHz computed using a spectral index of $\alpha$=$-$0.9 according to the results obtained in Section~\ref{sec:spectralindex} and in \cite{Cavallaro2018}. $a$=1.08 and $b$=2.03 are the parameters used in Section~\ref{sec:extendedsources} for the identification of resolved sources.
\\The resulting matrix values are reported in Fig.~\ref{fig:sourcecounts_1}. 
Above 5 mJy, the correction applied to the source counts is negligible as the catalogue is $>$98\% complete, as shown in Fig.~\ref{fig:completeness-reliability}, and the flux uncertainty is typically below 1\%. Below 5 mJy, the flux uncertainty rapidly increases (up to 30\% at 1 mJy) while the completeness rapidly degrades to zero.
Source detection inefficiency, thus, causes true counts to be underestimated by a factor $\varepsilon$, while finite flux density uncertainty cause sources to migrate from true to measured flux density bins. As a result, the observed counts in a given bin is contaminated by upward fluctuations from the adjacent lower bins and downward fluctuations from the adjacent upper bins. If the true source counts is steeply falling with flux density, fluctuations from the lower bins dominate and the net effect is an overestimation of the counts.\\Resolution bias also causes an underestimation of the true counts. Computed correction factors were found of the order of 5\% at 5 mJy and smaller than 1\% above 50 mJy.\\To derive the corrected counts $\bm{\mu}$ from uncorrected counts $\bm{\nu}$, knowing the response matrix $\mathbf{R}$, we need to solve the following system of equations:

\begin{equation}
\arraycolsep=1.4pt\def\arraystretch{2.2}
\begin{array}{llllll}
\underbrace{
\left.
\begin{array}{ll}
\left(
\begin{matrix}
  \nu_{1}\\%
  \vdots\\%
  \nu_{M}%
\end{matrix}
\right)
\end{array}
\right.}_{\large{\bm{\nu}}}%
&
\mbox{=}
&
\underbrace{
\left.
\begin{array}{ll}
\left(
\begin{matrix}%
r_{11} & r_{12} \cdots & r_{1N}\\%
\vdots & \vdots & \ddots\\%
r_{M1} & r_{M2} \cdots & r_{MN}%
\end{matrix}
\right)
\end{array}
\right.}_{\large{\textbf{R}}}%
&
\times
&
\underbrace{
\left.
\begin{array}{ll}
\left(
\begin{matrix}
\mu_{1}\\%
\vdots\\%
\mu_{N}\\%
\end{matrix}
\right)
\end{array}
\right.}_{\large{\bm{\mu}}}%
&
\;\;\;M\geq N
\end{array}
\end{equation}
where $M$ is the number of flux bins for the uncorrected counts and $N$ is the number of flux bins for the corrected counts.
To determine the corrected counts $\bm{\widehat{\mu}}$ we made use of a  forward-folding procedure. First we considered a model $f\left(\log_{10}S_{true};\Theta\right)$ with parameters $\Theta$ describing the true source counts $\bm{\mu_{model}}$ as a function of $\log_{10}S_{true}$. 
The model folded with the response matrix $\mathbf{R}$ was fitted to the observed counts $\bm{\nu}$ using a maximum likelihood approach. The best fit provides the expected true model counts $\bm{\widehat{\mu}}_{model}$ and the corresponding observed model counts $\bm{\widehat{\nu}}_{model}$. Corrected counts are finally computed as:
\begin{equation}
\bm{\widehat{\mu}}= \bm{\nu}*\frac{\bm{\widehat{\mu}}_{model}}{\bm{\widehat{\nu}}_{model}}    
\end{equation}
We modeled the true source counts with a power-law functional form. This was found to provide a good fit ($\chi^{2}$/ndf=44.5/45) to our source flux density data above the detection threshold.\\The unfolded differential source counts, normalized by $S^{-2.5}$, are reported in Fig.~\ref{fig:sourcecounts_2} with open black dots while numerical values are reported in Table~\ref{tab:sourcecounts}. 
The error bars represent the total uncertainties $\sigma_{\text{tot}}$=$\sqrt{\sigma_{\text{stat}}^2 + \sigma_{\text{syst}}^2}$ on the unfolded counts, where $\sigma_{\text{stat}}$ are the statistical uncertainties, obtained from the Poissonian errors on the uncorrected counts, and $\sigma_{\text{syst}}$=$\sqrt{\sigma_{\text{fit}}^2+\sigma_{\text{matrix}}^{2}+\sigma_{\text{cal}}^{2}}$ are the systematic uncertainties.
$\sigma_{\text{fit}}$ and $\sigma_{\text{matrix}}$ are obtained by propagating the likelihood fit parameter errors and the response matrix uncertainties in the unfolded counts, respectively. Matrix uncertainties have been computed by taking the variance of several response matrices randomized around parametrization errors for $\varepsilon$ and $\sigma$. Additionally, following \cite{Windhorst1990}, a 10\% uncertainty is considered in the resolution bias correction $\varepsilon_{reso}$. To estimate $\sigma_{\text{cal}}$, we shifted the measured flux densities by $\pm$10\% (the calibration uncertainty quoted in Section~\ref{subsec:flux-accuracy}), repeated the unfolding procedure on the resulting "shifted" spectra and compared the obtained counts with "unshifted" counts.\\
Existing source count data at 1.4 GHz, as parametrized by \cite{Katgert1988,Hopkins2003} and shifted to a frequency of 912 MHz (assuming a spectral index $\alpha$ ranging from $-$1.1 to $-$0.7), are shown for comparison as a shaded blue area. The shaded area also accounts for the count fluctuations ($\sim$20\%) reported in past works comparing source counts from different extragalactic fields \citep{Windhorst1990,Hopkins2003} or from different regions of the same survey \citep{Prandoni2001,Retana-Montenegro2018}. The observed spread can be due to either systematic uncertainties, different correction factors applied, or to the large scale structure of the universe (i.e. cosmic variance).\\
As can be seen, the measured source counts as a function of the source flux density are consistent with the trend reported in other surveys carried out far from the Galactic plane (e.g. see \citealt{Katgert1988,Hopkins2003} and references therein). The discrepancies observed assuming a spectral index $\alpha$=$-$0.7, ranging from 15\% to 20\%, are within the quoted systematic uncertainties. The presence of our Galaxy is also expected to play a role in this comparison. Galactic sources indeed contribute to the overall source counts, although with a smaller fraction ($\sim$5\% at least from the analysis presented in Section~\ref{sec:galsources}). 
An overdensity effect ($\sim$10-15\%), reported for example by \cite{Cavallaro2018}, can not however be clearly identified, being of the same order of the reported uncertainties.\\
For the sake of completeness we derived the source counts in a region outside the Galactic plane (|b|>2) following the same procedure described above and using an updated response matrix, parametrized on the considered mosaic region. No significant differences (within few percent) were found with respect to the source counts obtained over the full mosaic, suggesting that the majority of the unclassified sources have an extragalactic origin.

\setlength{\tabcolsep}{4pt}
\begin{table}
\caption{912 MHz source counts in \scorpio{} field. The columns represent: (1) flux density bin interval $\Delta$S in mJy, (2) flux density bin centre $S$ in mJy, (3) number of
catalogued sources $N$ in each flux density bin, (4) corrected number of sources $N_{corr}$, (5) uncorrected normalized differential source counts ($S^{2.5}dN$/$dS$), (6) corrected normalized differential source counts ($S^{2.5}dN$/$dS|_{corr}$) with statistical (7) and total (8) uncertainties.}
\setlength\tabcolsep{1.5pt}
\begin{tabular}{llllllll}
\hline%
\hline%
$\Delta$S & $S$ & $N$ & $N_{corr}$ & $S^{2.5}dN$/$dS$ & $S^{2.5}dN$/$dS|_{corr}$ & $\sigma_{stat}$ & $\sigma_{tot}$\\%
(mJy) & (mJy) & & & (Jy$^{1.5}$sr$^{-1}$) & (Jy$^{1.5}$sr$^{-1}$) &  & \\%
\hline%
0.23 & 0.50 & 111 & 1640.15 & 0.23 & 3.45 & 0.33 & 0.58\\%
0.37 & 0.79 & 409 & 2183.64 & 1.72 & 9.16 & 0.43 & 1.10\\%
0.58 & 1.26 & 623 & 1544.04 & 5.22 & 12.93 & 0.50 & 1.47\\%
0.93 & 2.00 & 700 & 988.56 & 11.69 & 16.51 & 0.56 & 1.83\\%
1.47 & 3.16 & 600 & 683.18 & 20.00 & 22.77 & 0.87 & 2.59\\%
2.33 & 5.01 & 507 & 541.79 & 33.72 & 36.03 & 1.55 & 4.22\\%
3.69 & 7.94 & 354 & 370.61 & 46.97 & 49.17 & 2.51 & 6.07\\%
5.85 & 12.59 & 287 & 297.43 & 75.98 & 78.74 & 4.46 & 10.09\\%
9.27 & 19.95 & 204 & 210.18 & 107.76 & 111.03 & 7.82 & 15.67\\%
14.69 & 31.62 & 146 & 149.81 & 153.88 & 157.89 & 12.48 & 23.68\\%
23.29 & 50.12 & 84 & 85.92 & 176.65 & 180.68 & 19.50 & 32.96\\%
36.90 & 79.43 & 60 & 61.22 & 251.76 & 256.89 & 33.38 & 53.75\\%
58.49 & 125.89 & 55 & 56.02 & 460.46 & 468.98 & 63.84 & 101.74\\%
157.74 & 223.87 & 46 & 46.77 & 602.18 & 612.27 & 90.02 & 141.26\\%
314.73 & 446.68 & 16 & 16.25 & 590.32 & 599.37 & 149.02 & 219.10\\%
3350.11 & 1584.89 & 8 & 8.11 & 657.56 & 666.73 & 241.78 & 348.37\\%
\hline%
\end{tabular}
\label{tab:sourcecounts}
\end{table}

\section{Source classification}
\label{sec:sourceclassification}

\subsection{Search for known Galactic objects}
\label{sec:galsources}
We cross-matched the ASKAP \textsc{Scorpio} source catalogue with different
astronomical databases to search for possible associations with known Galactic objects. We restricted the search to the following types of objects:
\begin{itemize}
\item \emph{Stars}: associations searched in the SIMBAD Astronomical Database\footnote{\url{http://simbad.u-strasbg.fr/simbad/}}, in the Galactic Wolf-Rayet Star Catalogue\footnote{\url{http://pacrowther.staff.shef.ac.uk/WRcat/index.php}} \citep{Rosslowe2015} and in the Gaia data release 2 catalogue \citep{Brown2018};
\item \emph{Pulsars}: associations searched in the ATNF Pulsar Catalogue\footnote{\url{https://www.atnf.csiro.au/research/pulsar/psrcat/}} \citep{Manchester2005} (version 1.63); 
\item \emph{Planetary Nebulae (PNe)}: associations searched in the Hong Kong/AAO/Strasbourg H-alpha (HASH) Planetary Nebula Database\footnote{\url{http://202.189.117.101:8999/gpne/dbMainPage.php}} \citep{Parker2016};
\item \emph{\hii{} regions}: associations searched in the WISE Catalogue of Galactic \hii{} regions\footnote{\url{http://astro.phys.wvu.edu/wise/}} \citep{Anderson2014}.
\end{itemize}

Extended objects, such as the supernova remnants (SNRs) listed in the Galactic SNR catalogue\footnote{\url{http://www.mrao.cam.ac.uk/surveys/snrs/}} by Green \citep{Green2019}, were not considered in this analysis. Possible associations and new detections will be reported in a future work dedicated to \scorpio{} extended sources.\\We summarize the cross-match results in Table~\ref{tab:catcrossmatches}.
The number of associations found for each catalogue is reported in column 4 while the expected number of false matches is given in column 5. These were estimated with the same procedure described in Section~\ref{sec:crossmatching}, i.e. evaluating the number of matches found within the chosen radius in several "randomized" catalogues. The search radius considered for the matching, reported in column 3, corresponds to the maximum statistical significance of the match signal above the background. The number of objects labelled as "confirmed" in the astronomical databases is reported in column 6.\\
Only 146 objects were associated to known classes of Galactic objects, corresponding to $\sim$4\% of the total number of \scorpio{} catalogued sources. The vast majority of the catalogued sources are thus labelled as not classified.
A 2D map showing the positions of both classified and unclassified sources is reported in Fig.~\ref{fig:catcrossmatches}. 
\begin{table}
\caption{Number of sources from different astronomical catalogues (see text) associated to ASKAP \textsc{Scorpio} radio sources.}
\begin{threeparttable}
\begin{tabular}{l|l|c|c|c|c}
\hline%
\hline%
Obj. Type & Catalogue & $r_{match}$ & $n_{matches}$ &  $n_{matches}^{random}$  & Confirmed\\%
& & (") & & & objects\\%
\hline%
\multirow{3}{*}{Star} & SIMBAD & 4 & 20 & 2.3$\pm$0.3 & 13\\%
 & WR Cat. & 4 & 0 & 0 & 0\\%
 & GAIA DR2 & 2 & 933 & 874.2$\pm$9.6 & 0 \\%
\hline%
Pulsar & ATNF Cat. & 8 & 21 & 0 & 21\\%
\hline%
PNe & HASH & 8 & 38 & 0 & 27\\%
\hline%
\hii{} regions & WISE & 32 & 67 & 0 & 35\\%
\hline%
All & & & 146\tnote{*} & & 96\\%
\hline%
\hline%
\end{tabular}
\begin{tablenotes}
\footnotesize%
\item[*] GAIA matches not included in the final count.
\end{tablenotes}
\end{threeparttable}
\label{tab:catcrossmatches}
\end{table}

\subsubsection{Stars}
\label{sec:stars}
\begin{figure}
\centering%
\includegraphics[scale=0.37]{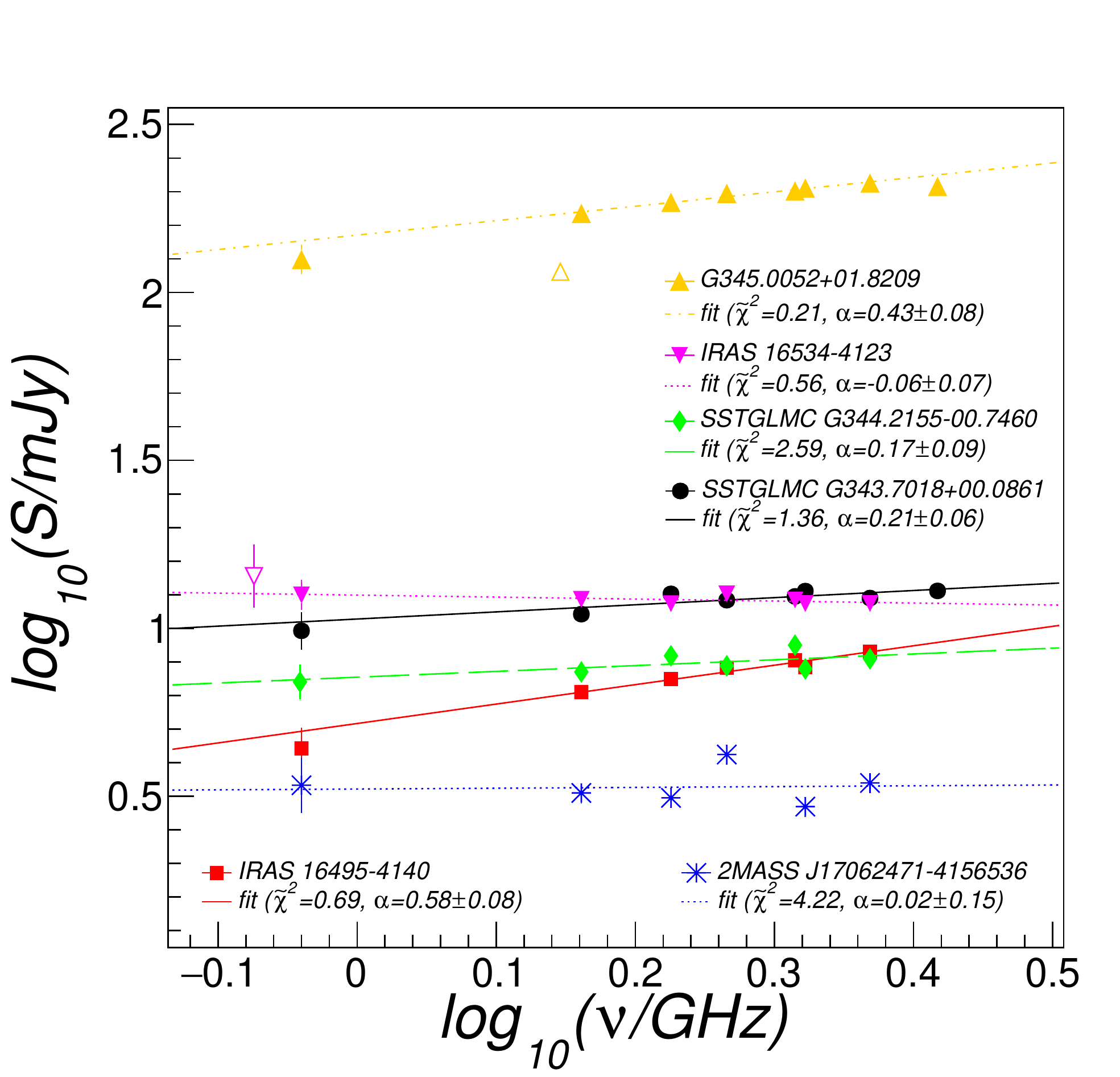}%
\caption{Spectral data of Scorpio sources associated to stars in the SIMBAD database, obtained using ASKAP and ATCA data (filled markers) and previous MGPS and NVSS observations (open markers). The power-law fits are reported with coloured lines.}
\label{fig:star_sed}
\end{figure}

Inside the \scorpio{} ASKAP region, we selected 10628 stars in the SIMBAD database and 19 Wolf-Rayet stars in the \cite{Rosslowe2015} catalogue. These were cross-matched to ASKAP source components within a match radius of 4" in sky coordinates. Since the Gaia catalogue is densely populated (more than 9$\times$10$^{6}$ entries found in the \scorpio{} region) we lowered the search radius to 2", comparable to the positional uncertainties obtained in ASKAP. We found 20 associations in SIMBAD, among them 7 YSO (1 confirmed, 6 candidates), and no associations with Wolf-Rayet stars up to a matching radius of 32". SIMBAD matches are expected to be real as the estimated number of chance matches is 2.3$\pm$0.3 (see Table~\ref{tab:catcrossmatches}). The associations found with Gaia DR2 (933) are instead dominated by random matches and will not be further considered. A list of the associated objects found is reported in Table~\ref{tab:StarMatches}.
The reported classification was investigated with infrared and optical data. Details are reported below:
\begin{enumerate}
\item \texttt{SSTGLMC G343.7018+00.0861}: invisible in optical and near-IR; clear but not particularly red in the mid-IR. Not immersed in a region of star formation. YSO candidate classification is premature. Could well be extragalactic.
\item \texttt{IRAS 16495-4140}: likely a YSO. It is a small clump of stars of which two dominate in the near-IR (2MASS). Together they are bright and red in the mid-IR (WISE; not covered by Spitzer).
\item \texttt{2MASS J17062471-4156536}: classified as a YSO candidate with star-forming activity (and dark cloud) nearby. Could be extragalactic. Invisible in the optical but bright (and red) in 2MASS. Not particularly red in Spitzer and WISE.
\item \texttt{SSTGLMC G344.2155-00.7460}: only detected in the mid-IR, faintly. There is nothing to suggest otherwise that this is a background AGN. It is not near star-forming complexes and it is next to darker areas of extinction, i.e.
viewed through a relatively more transparent region. The YSO classification is questionable.
\item \texttt{IRAS 16534-4123}: likely extragalactic. It is exceedingly faint in optical and very bright in WISE.
\item \texttt{[MHL2007] G345.0052+01.8209 1}: Class I protocluster. Clump of
red stars in 2MASS, no detection in optical, embedded within a diffuse background emission seen with Spitzer (GLIMPSE) and saturating in WISE.
\item \texttt{HD 326586}: F8 star, bright GALEX (UV) source.
\item \texttt{SSTGLMC G342.6544-00.3827}: faint, red. Detected in optical and near-IR, not particularly bright or red in mid-IR. Not necessarily a YSO, and no
indications that it should be radio-loud. Could well be a background AGN.
\item \texttt{IRAS 16472-4401}: related by SIMBAD with an IRAS source, undetected in 2MASS. Close to the source's sky position ($\approx$6 arcsec), a YSO was detected by the ATLASGAL survey at 870 $\mu m$ \citep{Urquhart2018}.
\item \texttt{2MASS J16504054-4328122} (\texttt{IRAS 16470-4323}): while SIMBAD classes it as an AGB star, there is no direct evidence for this. It is strangely yellow in the DECaPS (five-band optical and near-infrared survey of the southern Galactic plane with the Dark Energy Camera (DECam) at Cerro Tololo), unlike any other object around it. It is red in 2MASS but again yellow in Spitzer
and WISE. The Gaia data are marginal, so we cannot completely exclude an extragalactic nature.
\item \texttt{HD 326392}: classified as B8 star $\approx$830 pc far \citep{Brown2018}. At the stellar distance, the measured flux of HD 326392 corresponds to a radio luminosity of $\approx$6$\times$10$^{17}$ ergs s$^{-1}$ Hz$^{-1}$, more than one order of magnitude higher than CU Vir ($\approx$3$\times$ 10$^{16}$ ergs s$^{-1}$ Hz$^{-1}$, \citealt{Leto2006}), a magnetic late B type star well studied at radio regime. The radio luminosity of HD 326392 is instead more comparable to those of strong magnetic stars of B2 spectral type,
level $\approx$10 kG and luminosity close to 10$^{18}$ ergs s$^{-1}$ Hz$^{-1}$
\citep{Leto2017,Leto2018}. This suggests HD 326392 as a possible strong magnetic star.
\item \texttt{2MASS J17122205-4230414}: fairly red but not too faint in the optical, bright in the mid-IR. As such it is isolated, the
post-AGB classification seems more likely than a YSO (as listed in SIMBAD).
It is found to be slightly extended and classified as a planetary nebula by
\citep{Suarez2006} with high confidence, though their optical spectrum shows no trace of [OIII].
\item \texttt{TYC 7872-1355-1}: Anonymous star, but in a reflection nebula.
\item \texttt{MSX6C G346.4809+00.1320}: YSO, Class 0 (invisible in near-IR) with
a 6.7-GHz methanol maser detection by \cite{Gaylard1993} and a single
100\% polarized 1665-MHz OH maser peak by \cite{Caswell1983}.
\item \texttt{2MASS J17074166-4031240}: likely a YSO. It is invisible in the
optical but clear and red in 2MASS; however, the mid-IR emission comes mainly from the bright rim (photo-dissociation region) of a cloud/HII region immediately to its side, most likely the origin of the radio emission.
\item \texttt{CD-38 11343}: Classified as "double star" (er* (M3Ve+M4Ve)) in SIMBAD. Eruptive pair of M dwarfs at 15 pc, with likely expected variability also in radio.
\item \texttt{IRAS 17056-3930}: very faint red optical, but bright(ish) red near-IR and mid-IR source. However, it is very isolated, there is nothing like it within at least 5 arcminute radius in a field of stars with the nearest
sign of star formation activity about 8 arcminutes away. Further investigations are needed to address its nature.
\item \texttt{IRAS 17056-3916}: no detection in optical or near-IR, but very red mid-IR source; possibly unresolved. It is located right at the edge of a dark cloud, which could suggest it is a YSO or explain why it is not seen in the optical. Further investigations are needed to confirm if of extragalactic origin.
\item \texttt{Cl* NGC 6318 PCA 7229}: Classified as "star in cluster" in SIMBAD. This might in fact be the \texttt{AGAL G347.919-00.762} YSO, detected in Spitzer data.
\item \texttt{TYC 7873-953-1}: radio source generically classified by the SIMBAD database as star. This is associated with a visible source about 1.2 kpc far \citep{Brown2018}. Further, this was discovered as a variable radio source at 1.4 GHz from NVSS data \citep{OfekFrail2011}.
\end{enumerate}
We compared the matches found with those obtained in pilot observations of the \scorpio{} field with ATCA at 2.1 GHz. Only 2 of the 10 star associations reported in \cite{Umana2015} are retrieved also in the ASKAP map. Due to a lack of sensitivity, no radio sources are detected in the direction of the two Wolf-Rayet stars (\texttt{HD 151932}, \texttt{HD 152270}) previously detected in ATCA.\\For 6 associated objects, shown in Fig.~\ref{fig:star_sed} and reported as top entries in 
Table ~\ref{tab:StarMatches}, we were able to derive a spectral index measurement (column 14) using \scorpio{} ASKAP and ATCA, MGPS and NVSS data (see Section~\ref{sec:spectralindex}).
Among them we have 4 Young Stellar Object (YSO) candidates with fitted spectral index $\alpha$: 0.21$\pm$0.06 (\texttt{SSTGLMC G343.7018+00.0861}), 0.02$\pm$0.15 (\texttt{2MASS J17062471-4156536}), 0.17$\pm$0.09 (\texttt{SSTGLMC G344.2155-00.7460}), 0.43$\pm$0.08 (\texttt{[MHL2007] G345.0052+01.8209 1}). Taking into account the reported uncertainties, these values are generally consistent with the spectral indices expected in optically thin free-free emission processes from ionized gas. According to \cite{Scaife2012,Ainsworth2012,Anglada2018}, the free-free radio emission at centimetre wavelengths is ascribed to outflow processes (e.g. thermal jets) causing the required gas ionization, particularly in the earliest protostellar stages (Class 0 and Class I). The resulting YSO radio spectral indices $\alpha$ are expected in the range $-$0.1<$\alpha$<1.1 but their values depend on the protostar evolution. For example in collimated outflows, typical of early protostar stages, a radio spectral index $\alpha\sim$0.25 is favoured, while, for standard conical jets, spectral indices around 0.6 are expected \citep{Reynolds1986,Anglada1998}. Existing measurements mostly fall in the above range although in some cases the observed spectral indices ($<-$0.5) suggest a contribution from non-thermal processes \citep{Ainsworth2012}. Our spectral index measurements are found within the range expected from free-free emission. Additional radio data at different frequencies are however needed to further constrain the dominant emission mechanism.

\subsubsection{Pulsars}
\label{sec:pulsars}
\begin{figure}
\centering%
\includegraphics[scale=0.37]{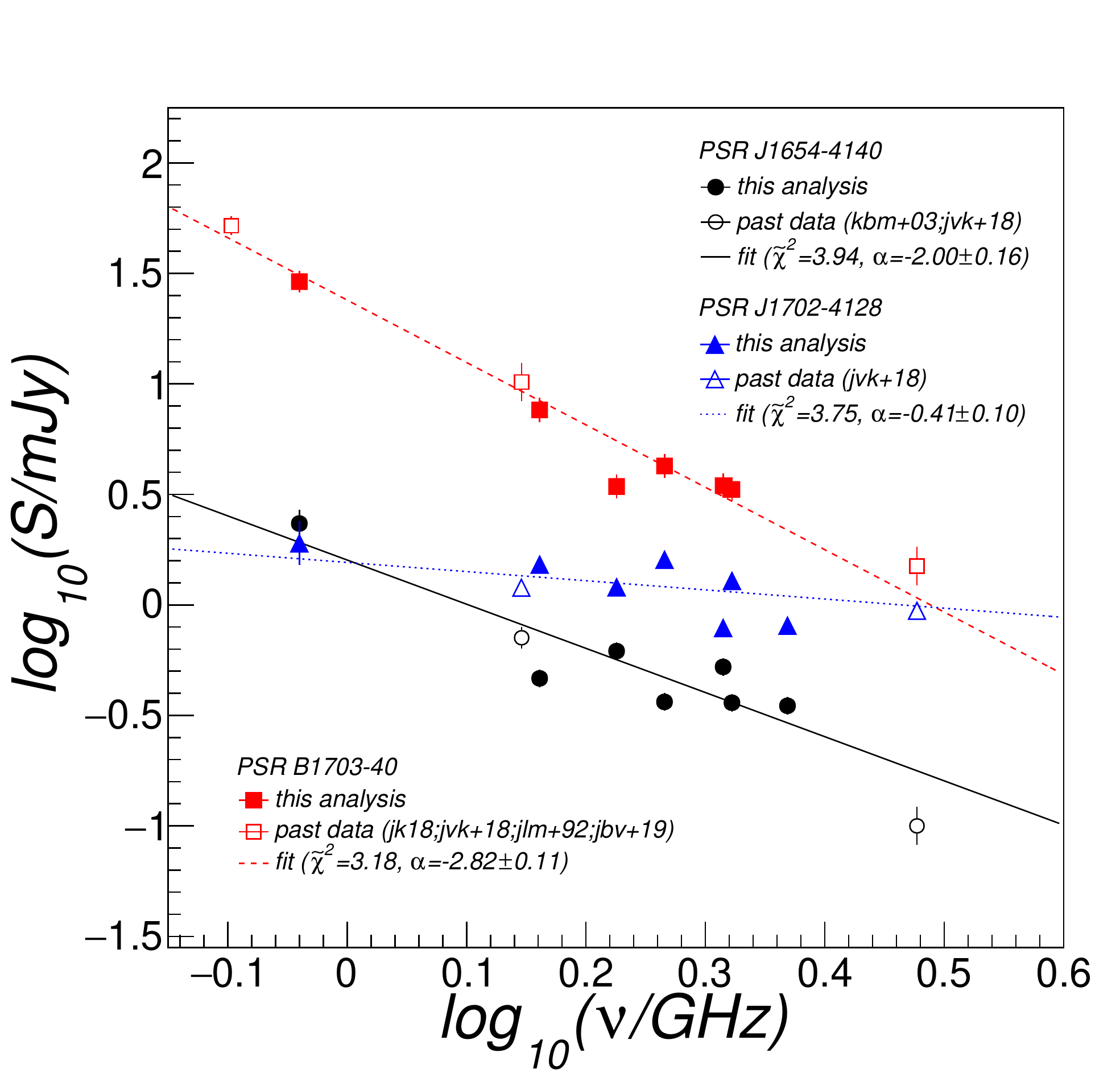}%
\caption{Spectral data for \texttt{J1654-4140} (open black dots), \texttt{B1703-40} (open red squares) and \texttt{J1702-4128} (open blue triangles) pulsars reported in the ATNF pulsar catalogue: kbm+03 \citep{Kramer2003}, jvk+18  \citep{Jankowski2018}, jk18 \citep{Johnston2018}, jlm+92 \citep{Johnston1992}, jbv+19 \citep{Jankowski2019}. ASKAP and ATCA spectral data obtained in this work are shown with filled markers. Single power-law fits are reported with solid, dashed and dotted lines for the three sample pulsars.}
\label{fig:pulsar_sed}
\end{figure}

The ATNF Pulsar database has 58 catalogued pulsars inside the \scorpio{} region. We carried out a search for possible associations with \scorpio{} source components, finding 21 associations within a matching radius of 8" and a number of chance matches compatible with zero.\\
In Table~\ref{tab:ATNFMatches} we report the full list of matched objects. Four of these (\texttt{J1654-4140}, \texttt{B1703-40}, \texttt{J1702-4128}, \texttt{J1702-4217}), shown as top entries in the table, have spectral index information (column 14) that can be obtained from ASKAP and ATCA data only. Column 12 represents the spectral index reported in the ATNF catalogue from measurements at 0.8, 1.4 and 3.0 GHz \citep{Jankowski2018,Jankowski2019,Kramer2003,Johnston2018,Johnston1992}. Flux density measurements, available in the literature for three of them (\texttt{J1654-4140}, \texttt{B1703-40}, \texttt{J1702-4128}), are reported in Fig.~\ref{fig:pulsar_sed} with open markers. 
New measurements, obtained in this work, are shown with filled markers. Both ASKAP and ATCA fluxes compare well to existing data, allowing us to derive a new spectral index measurement for all three pulsars. Fit results are shown in the figure with solid, dashed and dotted lines. The steep spectral index values found for \texttt{J1654-4140} and \texttt{B1703-40}, reported in Table~\ref{tab:ATNFMatches}  (column 14), are compatible with the distribution of pulsar spectral indices compiled in \cite{Maron2000} 
($\langle\alpha\rangle$=$-$1.8$\pm$0.2). The flat spectrum of \texttt{J1702-4128}, confirmed by our ASKAP and ATCA observations, supports the hypothesis of a pulsar wind nebula (PWN) as the origin of the radio emission. This was investigated by \cite{Chang2008} to explain the nature of the X-ray emission from Chandra CXOU J170252.4-412848 but no conclusive evidence was reached. Future ASKAP EMU and POSSUM data might give further evidence for a PWN if this source is linearly polarized.

\subsubsection{Planetary Nebulae}
\label{sec:pn}
\begin{figure}
\centering%
\includegraphics[scale=0.37]{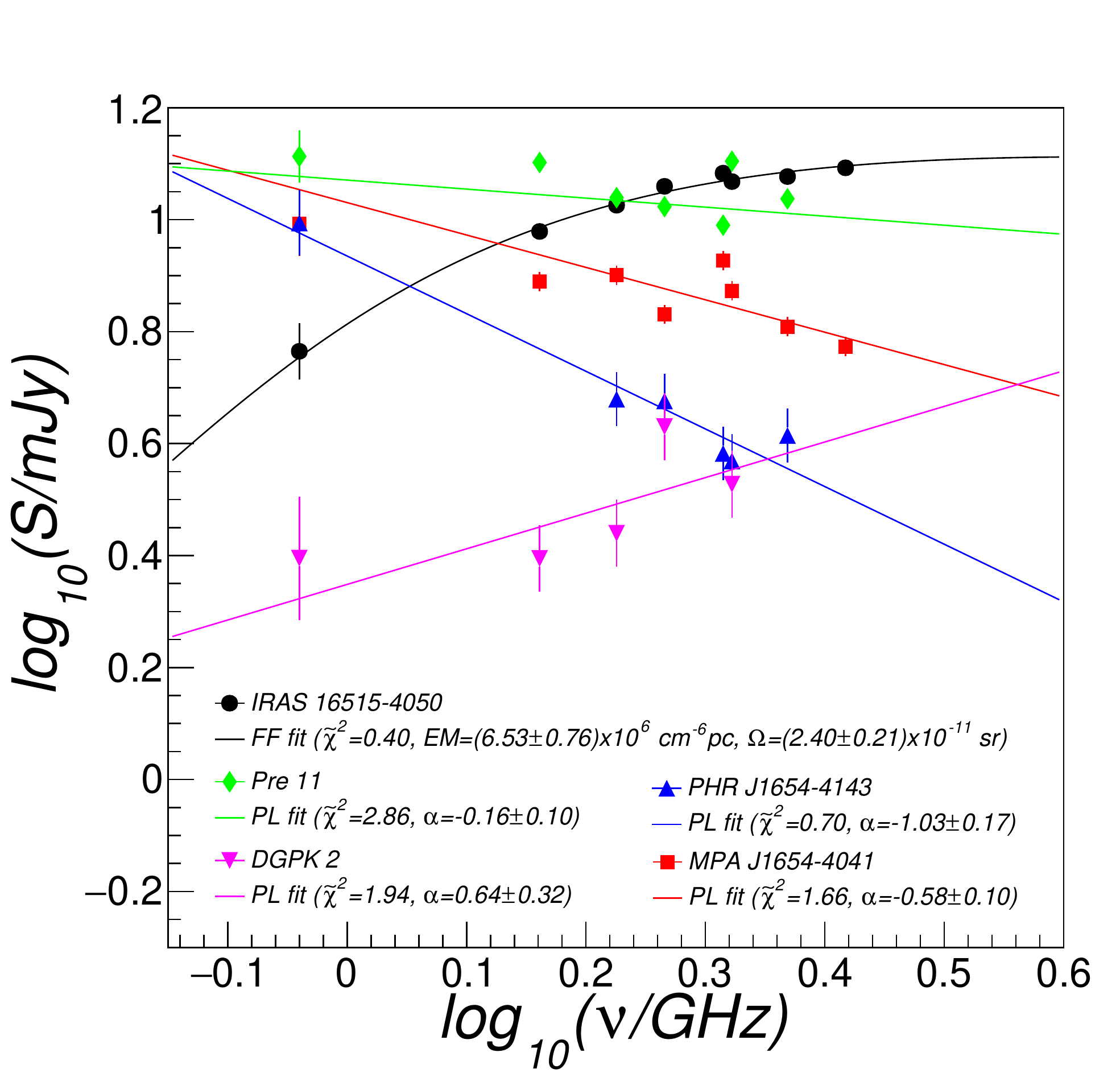}%
\caption{Radio continuum spectra for five PNe obtained using ASKAP and ATCA observations.}
\label{fig:pn_sed}
\end{figure}

Within our surveyed area, HASH has 60 PNe\footnote{This number refers to the number of PNe (32 True PN, 6 Likely PN, 7 Possible PN, 15 New Candidates) listed in the online HASH database version 4.6 (\url{http://202.189.117.101:8999/gpne/dbMainPage.php}) at the time of writing of the paper and found inside the source finding region. If we also consider the mosaic edges the PN count increases to 67 (36 True PN, 7 Likely PN, 8 Possible PN, 16 New Candidates).} that were cross-matched to ASKAP catalogued sources at different matching radii. A number of 38 associations were found assuming a radius of 8". 27 of the 38 matches correspond to confirmed PN objects. The number of chance matches found is compatible with zero. We visually inspected the 5 confirmed PNe (\texttt{MPA J1715-3903}, \texttt{MPA J1717-3945}, \texttt{MPA J1654-3845}, \texttt{MPA J1644-4002}, \texttt{PHR J1709-3931}) that were not cross-matched to any ASKAP source. In the direction of the first four PNe there are no radio sources in the map. A faint radio source is instead found in the direction of \texttt{PHR J1709-3931} but it is detected at an S/N of only 4.1.\\A full list of associated PNe is reported in 
Table~\ref{tab:HASHMatches}. With ASKAP we detected the 9 HASH PNe studied with ATCA 2.1 GHz observation in \cite{Ingallinera2019}. ATCA flux measurement is however reported in this study for only 5 PNe (1 confirmed: \texttt{IRAS 16515-4050}, 4 candidates: \texttt{MPA J1654-4041}, \texttt{PHR J1654-4143}, \texttt{Pre 11}, \texttt{DGPK 2}), shown in Fig.~\ref{fig:pn_sed} and listed as top entries in Table~\ref{tab:HASHMatches}. Four PNe were fitted with simple power-law models (labelled as "PL" in Fig.~\ref{fig:pn_sed}) as their radio spectra do not show evidence of spectral curvature in the studied frequency range. Although the fit models are not highly statistically significant, the results constitute a first measurement of the average spectral index for these objects as no other measurements are available in the literature.\\The radio spectrum of \texttt{IRAS 16515-4050} PN cannot be described by a single power-law model and was instead fitted with a thermal free-free emission model $S_{ff}$:
\begin{align}
S_{ff}(\nu;T_{e},\text{EM},\Omega)= F(\nu;T_{e}) (1-e^{-\tau(\nu;T_{e},\text{EM})}) \Omega
\end{align}
\begin{align}
\tau(\nu;T_{e},EM)=3.014\times10^{-2}T_{e}^{-1.5}\left(\frac{\nu}{\text{GHz}}\right)^{-2}g_{ff}(\nu;T_e)\;\text{EM}
\end{align}
\begin{align}
g_{ff}(\nu;T_e)=\ln\left[4.955\times10^{-2}\left(\frac{\nu}{\text{GHz}}\right)^{-1}\right] + 1.5 \ln{T_e}
\end{align}
where $F(\nu;T_{e})$ is the Planck function, $T_{e}$
is the electron temperature, \text{EM} is the emission measure and $\Omega$ is the source solid angle. A good fit of the data was obtained with $T_{e}$ fixed to 10$^{4}$ K and the other parameters free to vary\footnote{We considered an average value with respect to measured PN electron temperatures found in the range 5000-15000 $K$ (e.g. see \citealt{Zhang2014} for results on a sample of 48 Galactic PNe).}: \text{EM}=6.53$\pm$0.76$\times$10$^{6}$ cm$^{-6}$pc and $\Omega$=2.40$\pm$0.21$\times$10$^{-11}$ sr (angular size $\theta\sim$1").\\%
For the rest of catalogued PNe we were able to estimate the spectral index by combining our measurements with existing ones at 843 MHz \citep{Murphy2007} (MOST), 1.4 GHz \citep{Condon1998,CondonKaplan1998} (VLA), 4.8 GHz \citep{vanDeSteene1993} (ATCA). Some of them, obtained using 843 and 912 MHz measurements only, are clearly unreliable due to the limitations discussed in Section~\ref{subsec:flux-accuracy}.\\The spectral indices obtained with data at higher frequencies are found in general agreement with the expected nature of the emission in PNe, i.e. thermal free-free radiation due to electron-ion interactions in the nebula shell \citep{Kwok2000}. The expected spectral indices, however, vary considerably depending on the thermal and electron density scenario, roughly $\sim-$0.1 in an optically thin regime and positive up to $\sim$2 \citep{Pottasch1984} for optically thick PNe. The first scenario is likely in place in \texttt{MPA J1654-4041}, \texttt{PHR J1654-4143}, \texttt{Pre 11}, \texttt{Vd 1-5} and \texttt{Vd 1-6} for which we observed negative spectral indices. The second scenario seems favoured for some other detected PNe, e.g. \texttt{IRAS 16515-4050}, \texttt{DGPK 2}, \texttt{PM 1-119} and \texttt{PM 1-131}. Due to the limited number of radio observations available, the precision achieved on the spectral indices does not allow us to constrain the PN nature and corresponding emission mechanism. Deeper analysis, possibly in combination with IR data as in \cite{Fragkou2018,Ingallinera2019}, will be therefore performed in the future once new ASKAP observations with different frequency bands become available.

\subsubsection{\hiit{} regions}
\label{sec:hii}
Within the \scorpio{} region used for source finding, we found 356 Galactic \hii{} regions catalogued in \cite{Anderson2014}\footnote{The number of \hii{} regions falling in the full ASKAP mosaic, including borders, is 382, as reported in paper 1 \citep{Umana2020}.}. 256 of these were classified by these authors as "known" (i.e. objects confirmed by \citealt{Anderson2014} or in previous studies, including radio quiet \hii{} regions), with the rest as candidates (including those closely located to a known \hii{} region). Their sizes range from 0.2" to $\sim$23 arcmin with a median of $\sim$1 arcmin. Within a matching radius of 32" we found 67 \hii{} regions associations\footnote{The number of \hii{} regions detected in ASKAP is 261, as reported in paper 1 \citep{Umana2020}, if we also consider the mosaic borders and the extended sources that were excluded from this catalogue.}. The matched \hii{} regions have a reported radius between 12" and 100" (average radius $\sim$52"). From a visual inspection, we were able to confirm $\sim$90\% of the matches found in the automated analysis.\\ 
Thirteen \hii{} regions were detected in both the ASKAP and ATCA maps.
We were able to determine a first measurement of the radio spectral index for 9 of them (6 confirmed objects and 3 candidates). The remaining ATCA sources have either an unreliable spectral fit, due to noisy subband data, or an unclear association with ASKAP data. For instance, ASKAP source associated to \texttt{G344.993-00.265} is located close to another, more compact, \hii{} region (\texttt{G344.989-00.269}). The higher resolution of ATCA observations allows us to distinguish the two sources when estimating their flux densities. This is not the case for ASKAP data in which the two sources are blended. Future ASKAP observations will provide the required sensitivity, resolution and additional spectral data to refine and extend this analysis. Finally, an additional spectral index measurement was obtained for \texttt{G347.921-00.763} using MGPS and NVSS data. More details for each source are reported in Table~\ref{tab:WiseMatches}.

\begin{figure}
\centering%
\subtable{\includegraphics[scale=0.32]{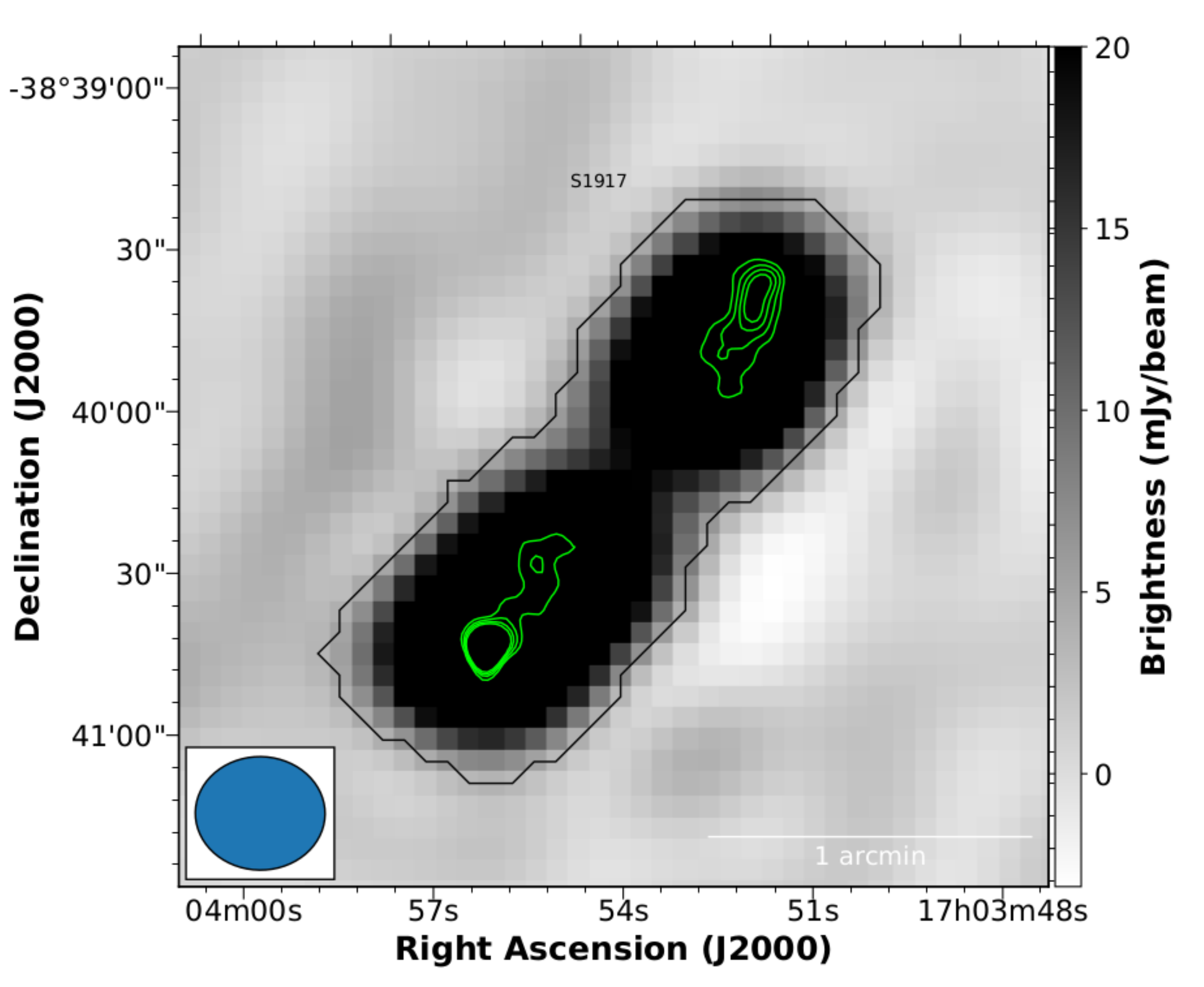}}%
\caption{A sample radio galaxy classified with high-confidence in both the ASKAP \textsc{Scorpio} map (colour scale and solid black contour) and in the VLASS survey (superimposed solid green contours). The ellipse at the bottom left indicates the ASKAP synthesized beam size.}
\label{fig:samplegalaxy}
\end{figure}

\subsection{Extragalactic sources}
\label{sec:extragalsources}
A large number of the extracted sources ($\sim$3800 source islands, $>$95\% of the catalogue) are not classified or associated to a known astrophysical object. 
Following the findings reported in previous radio surveys carried out in the Galactic plane \citep{Wang2018,Cavallaro2018} and the comparison with extragalactic source counts reported in Section~\ref{sec:sourcecounts}, we may reasonably expect that the majority of them are extragalactic objects. A search in the NASA/IPAC Extragalactic Database (NED)\footnote{The NASA/IPAC Extragalactic Database (NED) (\url{http://ned.ipac.caltech.edu}) is operated by the Jet Propulsion Laboratory, California Institute of Technology, under contract with the National Aeronautics and Space Administration.} returned only 20 known objects classified as galaxies (G). Two of them (\texttt{2MASS J17172771-4306573}, \texttt{2MASX J16463421-3903086}) are found associated to \scorpio{} catalogued sources. Another one (\texttt{2MASS J17162433-4225102}) is likely associated to a faint non-catalogued ASKAP radio source (S/N=3.7). For the rest, there are no indications for a co-spatial radio emission at the current map sensitivity.\\
We report here a preliminary analysis to increase the number of classified sources and provide the basis for more advanced studies to be done in the future. Indeed some of the unclassified source islands are found to have a bipolar morphology resembling those of radio galaxy lobes or can be visually associated to neighbor islands connected by a radio jet-like emission. We can therefore visually inspect the ASKAP map to search for this kind of object and label them as candidate radio galaxy. Since a precise identification using morphological considerations only is currently severely limited by the resolution of our ASKAP map, we also considered the ATCA map which provided the needed resolution to address some of the unclear identifications.
\begin{table}
\caption{Number of radio galaxies identified in the Scorpio ASKAP mosaic per number of islands and labelled according to the degree of confidence reached in the identification (see text).}
\begin{tabular}{c|cc|}
\hline%
\hline%
\multirow{2}{*}{\# islands} & \multicolumn{2}{|c|}{Degree of confidence}\\%
\cmidrule(lr){2-3}%
& \textsc{LOW} & \textsc{HIGH} \\%
\hline%
1 & 103 & 131 \\
2 & 10 & 28 \\
3 & 0 & 5 \\
\hline%
All & 113 & 164\\
\hline%
\end{tabular}
\label{tab:galaxycounts}
\end{table}
Results are reported in Table~\ref{tab:galaxycounts}.\\A number of 277 candidate radio galaxies were identified. None of them are associated to known NED extragalactic objects. The majority was found associated to one island detected in the catalogue (e.g. other components are either not visible or below the detection threshold) and only a few associated to two or three islands (which may correspond to a single source). 
The number of source islands associated to the candidate galaxy do not always have a 1-to-1 correspondence to the physical components (e.g. core, lobes). In many cases the latter are in fact visible inside a single island.\\%
For almost half of the cases there were no strong indications whether the radio morphology would suggest a radio galaxy. These have been labelled as candidates with lower degree of confidence in Table~\ref{tab:galaxycounts}.
About 70 radio galaxy candidates, lying north of Dec=$-$40 deg, were visually inspected in the QuickLook data release of the VLASS 2-4 GHz survey \citep{Lacy2020}. Only two of them (one is reported in Fig.~\ref{fig:samplegalaxy}) were firmly confirmed as radio galaxies from their morphology, while the rest are resolved out or only showing the core component.
Given the resolution limitations already discussed, future ASKAP observations at higher resolution will be needed to confirm the nature of these candidates.\\
Spectral information, measured only for 74 source components found in candidate radio galaxy islands, may provide additional hints for these studies, at least for a subset of the sources. Following the classification scheme reported in \cite{Zajacek2019}, 62 of them have a lobe-like steep index ($\alpha<-$0.7), 6 are found with a jet-like index ($-$0.7$\le\alpha\le-$0.4) and 6 with a flatter core-like index ($\alpha>-$0.4).

\begin{figure*}
\subtable{\includegraphics[scale=0.27]{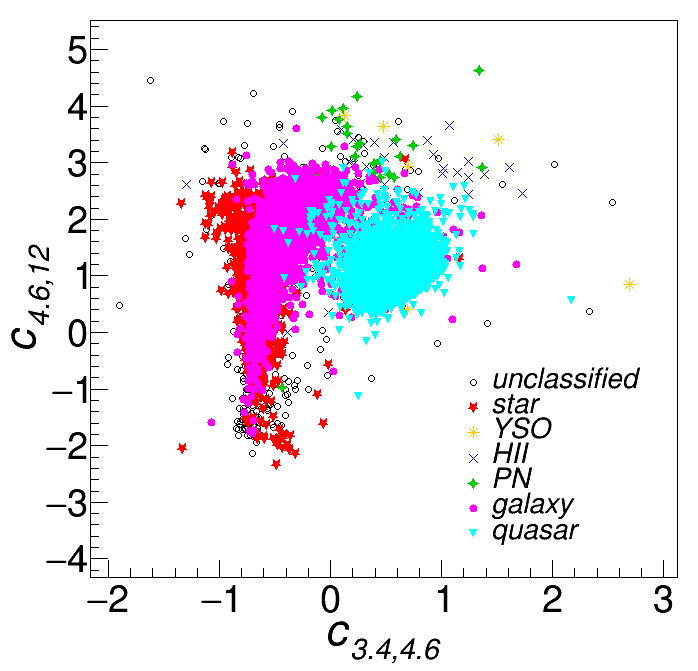}}%
\subtable{\includegraphics[scale=0.27]{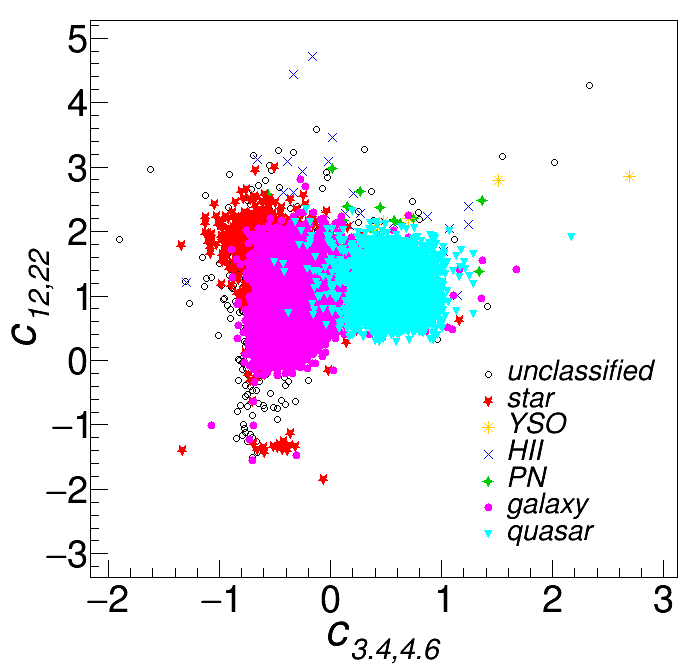}}\\%
\vspace{-0.3cm}%
\subtable{\includegraphics[scale=0.27]{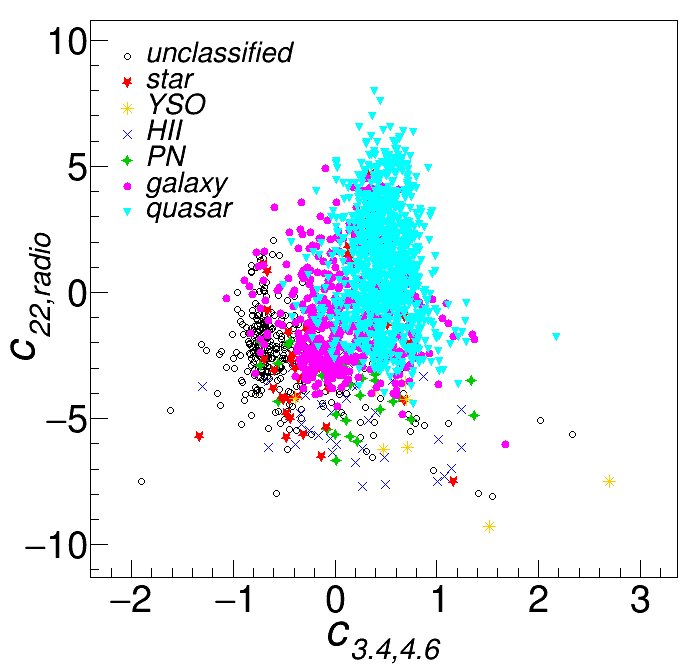}}%
\subtable{\includegraphics[scale=0.27]{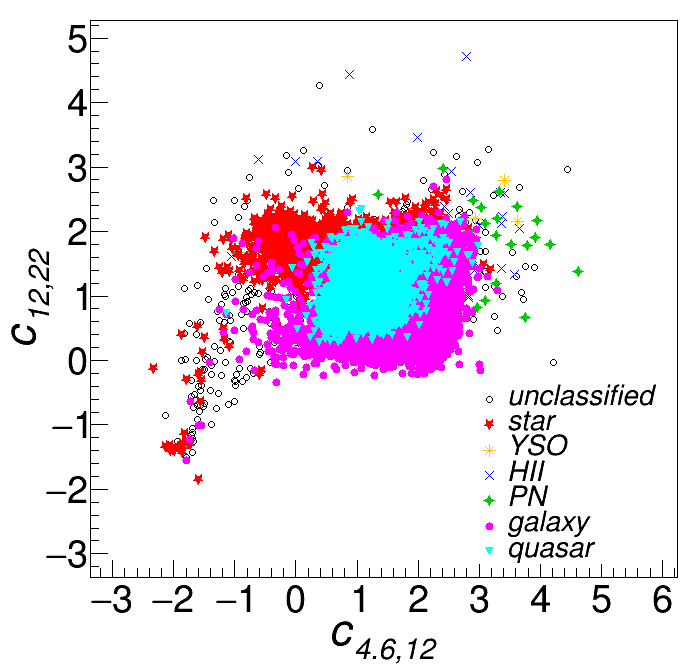}}\\%
\vspace{-0.3cm}%
\subtable{\includegraphics[scale=0.27]{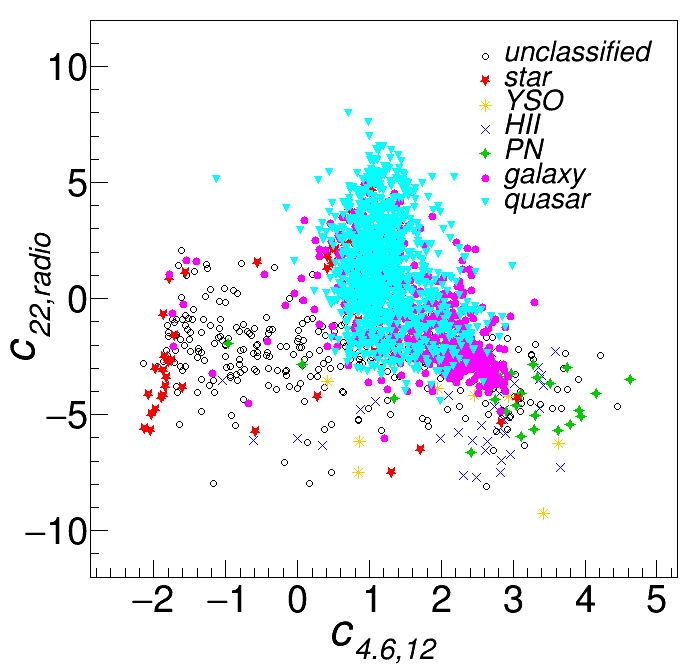}}%
\subtable{\includegraphics[scale=0.27]{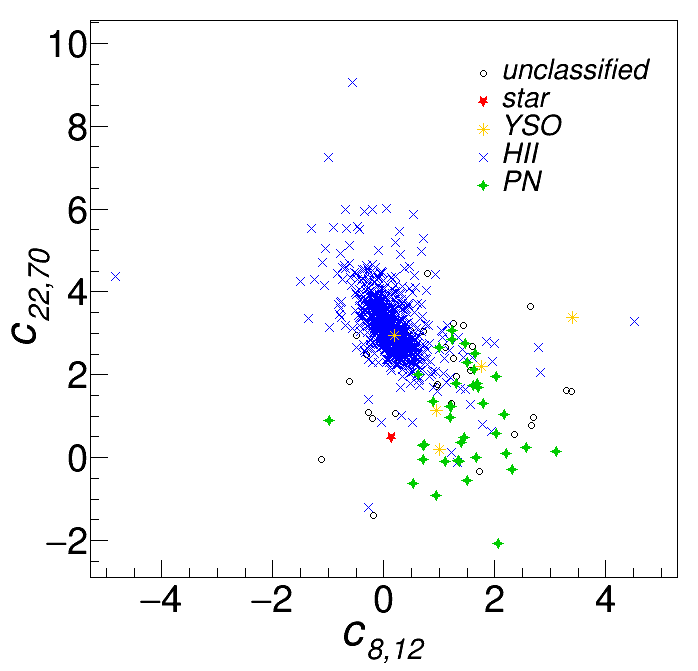}}
\vspace{-0.2cm}%
\caption{Scatter plots of selected infrared-infrared or infrared-radio colors for different infrared bands (3.4$\mic{m}$, 4.6$\mic{m}$, 12$\mic{m}$, 22$\mic{m}$, 70$\mic{m}$). Unclassified objects are shown with gray smaller dots while pre-classified sources with colored markers (\hii{} regions: blue crosses, PNe: green diamonds; stars: red stars; YSOs: orange asterisks; galaxies: purple dots, QSO: cyan triangles).}
\label{fig:radioIRcorr}
\end{figure*}

\subsection{Unclassified sources}
\label{sec:unclassifiedsources}
\begin{figure}
\centering%
\includegraphics[scale=0.35]{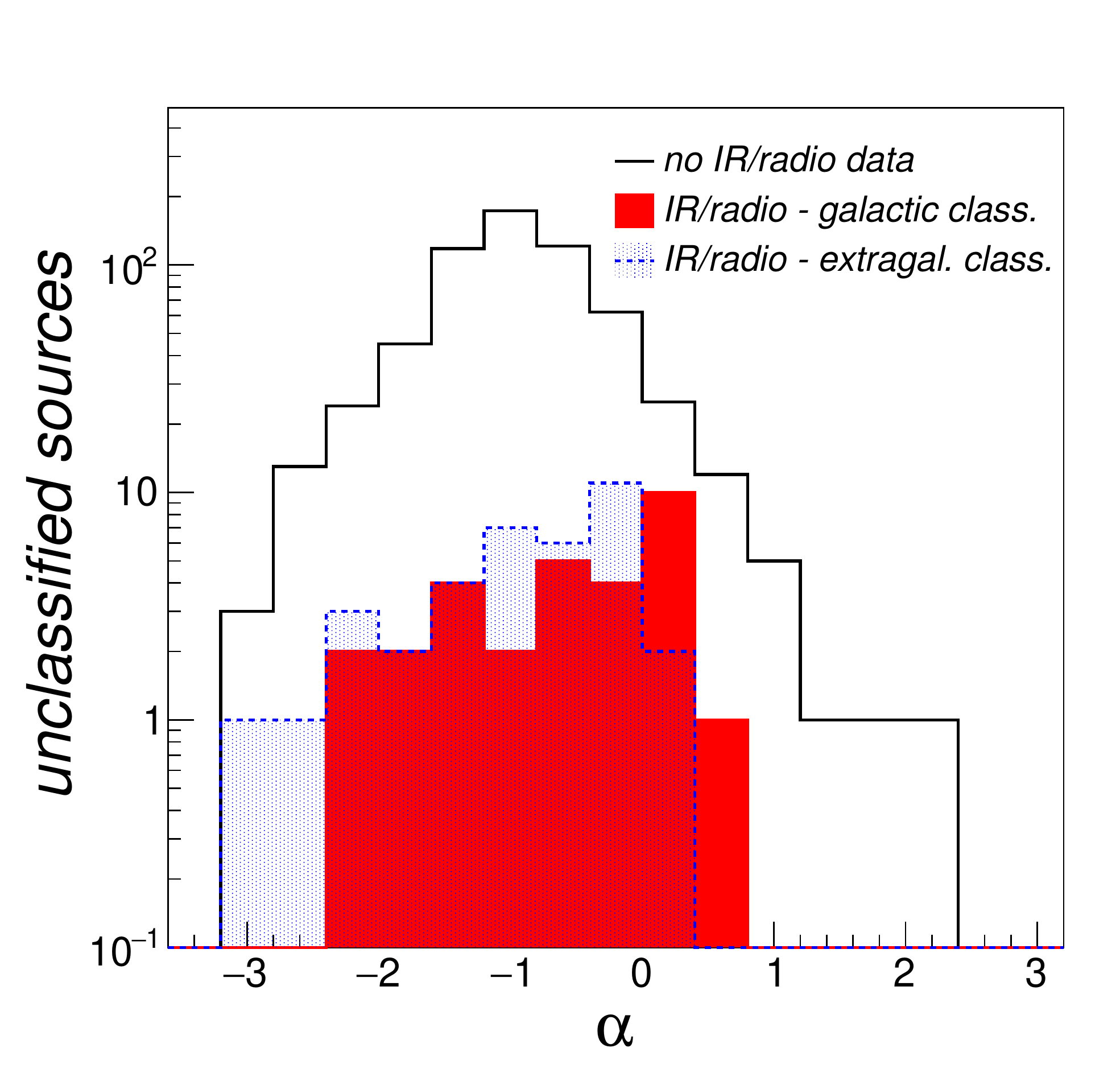}
\caption{Measured spectral index for unclassified sources. The solid black line histogram includes unclassified sources for which no combined infrared-radio parameter data are available. Filled red and shaded blue histograms include unclassified sources that were classified as Galactic and extragalactic sources, respectively, in the combined infrared-radio analysis.}
\label{fig:spectralindex-unclasssources}
\end{figure}

The remaining unclassified sources (3588), which still constitute the major part of the catalogue ($\sim$87\%), cannot be classified on the basis of their morphology given that the majority ($\sim$95\%) does not present any internal structure at the resolution of ASKAP map. For these, the radio spectral index and the correlation with observations at different wavelengths can provide valid hints. The radio-infrared correlation, in particular, was suggested by many authors (e.g. see \citealt{Ingallinera2014,Fragkou2018,Akras2019}) as a powerful tool for the identification of different classes of Galactic objects. Radio continuum emission traces the ionized part of the nebula, whereas infrared traces excited gas and the dust. In this context, the combination of infrared color indices obtained with different wavelength filters (from few $\mic{m}$ to hundreds of $\mic{m}$) has been a widely used technique to probe IR emission from different components \citep{Nikutta2014,Anderson2012}. For example, the correlation of far-infrared to mid-infrared colors allows to compare emission from cold and warmer dust components. The ratio between IR fluxes at 12$\mic{m}$ and 8$\mic{m}$ is instead sensitive to the emission from PAHs (polycyclic aromatic hydrocarbons).\\As discussed in Section~\ref{subsec:ir-crossmatches}, about 2500 unclassified sources ($\sim$70\%) do not have infrared counterparts within a search radius equal to the ASKAP synthesized beam size. These are expected to be mostly extragalactic sources and pulsars. For 607 of these, we obtained a radio spectral index, reported in Fig.~\ref{fig:spectralindex-unclasssources} (solid black histogram), that will be combined with additional observables in future classification studies. Bearing the cross-match limitations in mind (highlighted in Section~\ref{subsec:ir-crossmatches}), we will carry out the following analysis to the other 284 of the unclassified sources found with infrared counterparts within 8", of which 67 have both the radio spectral index information and measured radio-infrared correlation parameters.

\subsubsection{Infrared-radio colors}
We report in Fig.~\ref{fig:radioIRcorr} a series of correlation plots of selected infrared-infrared and infrared-radio color indices for different IR wavelengths (3.4$\mic{m}$, 4.6$\mic{m}$, 8$\mic{m}$, 12$\mic{m}$, 22$\mic{m}$, 70$\mic{m}$). Color $c_{i,j}$ is defined as the magnitude difference between measured fluxes $S_{i}$ and $S_{j}$ in band $i$ and $j$ where $\lambda_j$>$\lambda_i$, e.g. $c_{i,j}=2.5\log_{10}(S_j/S_i)$. As only a subset of the available colors are independent, we selected the most effective ones, in terms of source discrimination power, reported in the literature \citep{Nikutta2014,Anderson2012}: $c_{3.4,4.6}$, $c_{4.6,12}$, $c_{12,22}$, $c_{8,12}$, $c_{22,70}$. An additional color ($c_{22,\text{radio}}$) was considered to include the radio information. Unclassified objects are shown with open black dots, while pre-classified sources found in the \scorpio{} mosaic with colored markers (\hii{}: blue crosses, PNe: green diamonds; stars: red stars; YSOs: orange asterisks; galaxies: purple dots).\\
We superimposed data for different classes of objects taken from previous radio surveys crossmatched to the aforementioned infrared surveys. Radio star data (shown as red stars) at 843 MHz and 1.4 GHz are taken from \cite{Kimball2008}\footnote{Catalogue data available at \url{http://www.aoc.nrao.edu/~akimball/radiocat_2.0.shtml}} and from \cite{Wendker1995}. Data for galaxy and QSO with point-like morphology at 1.4 GHz (shown respectively with purple dots and cyan triangles) are taken from \cite{Kimball2008}. Additionally, we included infrared data for stars, galaxies and quasars from \cite{Clarke2020}, taking only sources reported with classification probability equal to 1 and requiring infrared fluxes above the AllWISE 5$\sigma$ limits for all bands \citep{Cutri2013}.
\hii{} regions (blue crosses) and PNe (green diamonds) at 843 MHz and 1.4 GHz data are taken from the \cite{Anderson2014} and HASH \citep{Parker2016} databases, respectively. Radio data at 5 GHz for YSO (orange asterisks) are taken from \cite{Urquhart2009}. Pulsar data from the ATNF database \citep{Manchester2005} were also considered but since no match was found with IR data they were not reported in Fig.~\ref{fig:radioIRcorr}.\\
From Fig.~\ref{fig:radioIRcorr} we observe that most color correlations provide valuable information to separate Galactic and extragalactic sources. Longer wavelength bands are found more sensitive to object identity information for classification purposes. For example known Galactic objects tend to have larger $S_{12,22}$/$S_{3.4,4.6,radio}$ flux ratios on average compared to galaxies. From the data shown in the bottom right panel, one may be tempted to conclude that the sole presence of an associated 70$\mic{m}$ emission is a clear indication of the source's Galactic origin. This is rather due to a lack of reference data for extragalactic background objects in this portion of the parameter space. Indeed, far infrared emission (also at 70$\mic{m}$) from several nearby galaxies was reported in different works (e.g. see \citealt{Dale2012}) and expected from our calculations\footnote{Following \cite{Silva1998,Michalowski2008,Michalowski2010,Iglesias2007} we computed the expected 70$\mic{m}$ emission as a function of redshift for different SED models and compared it with the Hi-GAL detection threshold of $\sim$0.5 mJy \citep{Molinari2016}. For radio fluxes close to the catalogue detection threshold ($\sim$5 mJy) only galaxies ($z<$0.6) with specific SED models can be detected. Above the ASKAP detection threshold we expect galaxies to be detectable at all redshifts and, thus, to populate the corresponding parameter space in Fig.~\ref{fig:radioIRcorr}}.\\
Reference data have some limitations to be considered for further analysis. First, the currently available radio data are obtained at slightly different frequencies (843 MHz, 1.4 GHz, 5 GHz) with respect to the ASKAP \scorpio{} data (912 MHz). The introduced spread in the IR/radio colors is expected to be negligible for Galactic sources with thermal emission or flat spectra. A correction for sources with steeper spectra (expected of the order of 0.5 and 1.8 at 1.4 and 5 GHz, respectively, with $\alpha$=$-$1) is however not possible as the spectral index information is only available for very few sources. A second limitation is that the source parameter data have many missing values for multiple reasons that can not easily be distinguished from one another without an in-depth revision of the external catalogue data. This prevented us from encoding and imputing missing information in an effective way. For example, some catalogues do not always provide match information for all IR bands, e.g. WISE W1 and W2 information is missing in the \hii{} region and PN catalogues, but we may expect for a subset of these sources an emission at these wavelengths\footnote{We visually inspected some of the catalogued \hii{} regions in all WISE bands. The infrared emission in bands W1 and W2 was effectively found absent or not correlated with the W3 and W4 emission but we cannot exclude such a correlation for the rest of catalogued objects that were not inspected. In fact in \cite{Assef2018} the authors conservatively removed the areas of the sky around WISE \hii{} regions to eliminate the Galactic source background when searching for potential AGN in W1 and W2 bands. An emission at these shorter wavelengths, albeit smaller compared to W3 and W4 bands, can be therefore present and was indeed observed in some of the \hii{} regions catalogued by CORNISH \citep{Purcell2013} (for example in \texttt{G019.6062-00.9018})}. Flux information may also be missing for some sources either because the reference survey covers a limited portion of the sky (e.g. radio galaxy data obtained in extragalactic surveys do not have 8 and 70$\mic{m}$ information) or because the source is not emitting or below the detection threshold. We eventually removed sources with missing data (e.g. using listwise deletion), reducing the sample size usable for analysis.

\subsubsection{Classification analysis}
We attempted to perform a simple source classification analysis to identify Galactic source candidates using four color parameters ($c_{3.4,4.6}$, $c_{4.6,12}$, $c_{12,22}$, $c_{22,\text{radio}}$) for which we have the largest availability of data for all bands. The resulting training sample ($\sim$1650 sources) is unbalanced by a factor 10 towards extragalactic objects. A decision tree classifier was then trained on this dataset and tested on the unclassified sample (284 sources). The optimal tree maximum depth size (3) was determined by comparing classification performances on multiple random validation samples. On the full training sample we obtained these classification metrics (recall/precision/F1-score): 0.99/0.97/0.98 for extragalactic sources and 0.69/0.89/0.78 for Galactic sources. As expected from Fig.~\ref{fig:radioIRcorr}, stars are the most misclassified type of object ($\sim$70\% of the wrong identifications). From the decision rules it can be seen that one of the color parameter ($c_{12,22}$) is not effectively used for the classification. Decision rules can be expressed in pseudocode as follows:
\removelatexerror
\begin{algorithm}[H]
\scriptsize%
\uIf{$c_{22,\text{radio}}\le-$4.05}{
 Galactic
}
\Else{
  \uIf{$c_{4.6,12}\le-$0.51}{
    \eIf{$c_{3.4,4.6}\le-$0.68}{
      extragalactic
    }{
      Galactic
    }
  }
  \uElseIf{$c_{4.6,12}>-$0.51 \& $c_{4.6,12}\le$2.97}{
     extragalactic
  }
  \Else{
    Galactic
  }
}
\end{algorithm}
The trained tree classifies 99 sources as Galactic objects, 75 of them with classification probability larger than 90\%. In Fig.~\ref{fig:spectralindex-unclasssources} we report the measured spectral indices for a subset (30) of them (filled red histogram) in comparison with sources classified as extragalactic (shaded blue histogram).\\Unfortunately, we do not have enough training data to classify different types of Galactic objects. Nevertheless, it is interesting to inspect the predictions of other classifiers (employing additional IR colors) on our data. \cite{Akras2019}, for example, obtained a PN classification efficiency of $\sim$50\% and a contamination from mimics (\hii{}, stars, YSO) ranging from 30\% to 40\% using these color cuts:
\begin{equation}
\left.
\begin{aligned}
&J-H<1.10,\;W1-W4\ge7.87\;\;\;\;\text{or}\\%
&J-H<1.31,\;K-W3\ge6.42%
\end{aligned}    
\right.
\end{equation}
No unclassified source was found to satisfy both cuts. 
In particular the second condition (involving WISE colors) of both cuts was not fulfilled by any source passing the first condition.\\\cite{Anderson2012} reported an \hii{} region classification efficiency better than 95\% with a PN contamination around 20\% requiring the following criteria (at least one of them):
\begin{equation}
\left.
\begin{aligned}
&\log_{10}(S_{12}/S_{8})<0.3\\%
&\log_{10}(S_{22}/S_{8})<1\\%
&\log_{10}(S_{70}/S_{12})>1.3\\%
&\log_{10}(S_{70}/S_{22})>0.8\\%
\end{aligned}    
\right.
\end{equation}
We found that 47 sources, previously classified as Galactic objects, satisfy the first two criteria, while only 2 are passing all criteria (requiring emission at 70$\mic{m}$). These are likely new \hii{} region candidates to be explored and confirmed in follow-up studies.\\To make further advances we will need to individually assess the reliability of each radio-IR association (currently quoted in Section~\ref{subsec:ir-crossmatches} as $\sim$60\% with a match radius of 8") and possibly increase the sample of pre-classified objects to constrain the parameter space with a higher degree of confidence. Data validation efforts have already started and will be reported in a future paper.

\section{Summary}
\label{sec:summary}
We have analyzed ASKAP Early Science observations of the \scorpio{} field ($\sim$40 deg$^{2}$, $\sigma_{rms}$>200 $\mu$Jy/beam, 24"$\times$21" angular resolution) at 912 MHz and produced a first catalogue of compact radio sources and their components using the \caesar{} source finder. We summarize below the main results:
\begin{enumerate}
\item The catalogue contains 4144 source components with flux densities ranging from 3.9 Jy down to 0.3 mJy. About $\sim$87\% were detected at a significance higher than 5$\sigma$. From simulations, we estimated a catalogue completeness of at least 90\% above 5 mJy. The differential source counts are compatible with literature data at 1.4 GHz obtained outside the Galactic plane, after taking into account the different observation frequency and analysis systematics. This suggests a majority of extragalactic sources in the catalogue.
\item Through comparison with different radio surveys, including our ATCA observations at 1.4 - 2.6 GHz, we estimated the spectral indices for a subset of the catalogued sources (780) finding a median value of $\langle\alpha\rangle$=$-$0.92 (IQR=0.68) in agreement with previous works within the discussed uncertainties. The higher resolution ATCA data also allowed us to derive a lower limit ($\sim$8\%) on the fraction of the extended sources present in the catalogue. 
\item We cross-matched the source catalogue with different astronomical databases (SIMBAD, HASH, WISE \hii{} regions, Wolf-Rayet catalogue) to search for possible associations to known Galactic objects. Including candidate objects, we have found 20 stars (including 7 YSO), 21 pulsars, 38 PNe, 67 \hii{} regions, providing new spectral index measurements for a subsample of these, detected in both ASKAP and ATCA observations.
\item The vast majority ($>$96\%) of the catalogued sources are unclassified, likely dominated by extragalactic background sources. About 300 sources have been identified as radio galaxy candidates from morphological considerations. The remaining sources were analyzed in correlation with existing mid-infrared data, in particular with the AllWISE survey. The majority of the unclassified sources do not have IR counterparts at 12 and 22$\mic{m}$ above IR flux sensitivity levels ($\sim$6 mJy) within the ASKAP beam radius. About one hundred sources found with IR counterparts were preliminarily classified as Galactic objects on the basis of their radio-infrared colors. A summary of the obtained results is reported in Table~\ref{tab:source-summary}.
\end{enumerate}
The source catalogue will be updated before the beginning of the ASKAP Galactic plane survey once that the scheduled Early Science data of the ASKAP \textsc{Scorpio} field with 36 antennas at three different frequencies (920 MHz, 1296 MHz, 1630 MHz) are available and fully reduced. 
The multi-frequency data and the expected boost in sensitivity and spatial resolution will enable us to measure the spectral index for all \scorpio{} sources, including those located outside the ATCA survey and near the current ASKAP detection threshold, enabling further advances in our classification studies.

\section*{Acknowledgements}
The Australian SKA Pathfinder is part of the Australia Telescope National
Facility which is managed by CSIRO. Operation of ASKAP is funded by the
Australian Government with support from the National Collaborative Research
Infrastructure Strategy. Establishment of the Murchison Radio-astronomy
Observatory was funded by the Australian Government and the Government of
Western Australia. This work was supported by resources provided by the Pawsey Supercomputing Centre with funding from the Australian Government and the Government of Western Australia. We acknowledge the Wajarri Yamatji people as the traditional owners of the Observatory site.\\
We thank the authors of the following software tools and libraries that have been extensively used for data reduction, analysis and visualization: ASKAPsoft \citep{Whiting2019}, \caesar{} \citep{Riggi2016,Riggi2019}, astropy \citep{astropy2013,astropy2018}, \textsc{Root} \citep{Brun1996}, \textsc{Topcat} \citep{Taylor2005, Taylor2011}, ds9 \citep{Joye2003}, APLpy \citep{Robitaille2012}.\\
This publication makes use of data products from the Wide-field Infrared Survey Explorer, which is a joint project of the University of California, Los Angeles, and the Jet Propulsion Laboratory/California Institute of Technology, funded by the National Aeronautics and Space Administration. Additionally, this research has made use of the SIMBAD database, operated at CDS, Strasbourg, France \citep{Wenger2000} and the VizieR catalogue access tool, CDS, Strasbourg, France (DOI: 10.26093/cds/vizier). The original description of the VizieR service was published in \cite{Ochsenbein2000}.\\%
Some of the authors benefited from grant No. 863448 (NEANIAS) of the Horizon 2020 European Commission programme.
H.A. benefited from grant CIIC 90/2020 of Universidad de Guanajuato, Mexico. M.J.M.~acknowledges the support of the National Science Centre,
Poland through the SONATA BIS grant 2018/30/E/ST9/00208.

\section*{Data Availability}
Source tables are provided as online supplementary material.\\
The following catalogue data products are available in the Zenodo repository\footnote{\url{https://zenodo.org/}} at \url{http://doi.org/10.5281/zenodo.4386692}:
\begin{enumerate}
\item \textbf{Source catalogue in tabular format}: Two ascii/FITS table files with a series of summary parameters for each catalogued source islands and fitted components, respectively. Table format (number of data columns and column description) is detailed in the \caesar{} documentation at \url{https://caesar-doc.readthedocs.io}. Additionally, we provide an added-value source component catalogue table (ascii and FITS formats) with extra-information (corrected fluxes, radio/infrared cross-match info, spectral indices, etc.). Its format is described in 
Table B1 (available in the online version of this paper)
\item \textbf{Source catalogue in \textsc{Root} format}: A \textsc{Root}\footnote{\url{https://root.cern.ch/}} file storing the list of catalogued sources and relative components as a \caesar{} \texttt{Source} C++ object. For each source the summary parameters plus detailed information at pixel level are available. The detailed format is described in the \caesar{} API documentation at \url{https://caesar-doc.readthedocs.io}.
\item \textbf{Source list in region format}: Two \textsc{ds9}\footnote{\url{http://ds9.si.edu/}} region files with the list of catalogued source islands and fitted components, respectively reported as labelled polygons or ellipses.
\item \textbf{Background maps in \textsc{FITS} format}: Two \textsc{FITS} files with background and noise maps obtained in the source finding process. 
\end{enumerate}



\bsp	
\label{lastpage}


\appendix%
\thispagestyle{empty}

\onecolumn%

\section{Source finder parameters}
\label{appendix:finderpars}

\begin{center}
\footnotesize%
\captionsetup{labelfont=bf}
\captionof{table}{
Main parameters used in the \caesar{} source finder to build the source catalogue. More details are available in \citealt{Riggi2019} and in the \caesar{} online documentation at \url{https://caesar-doc.readthedocs.io/en/latest}.}%
\begin{threeparttable}
\begin{tabular}{llll}
\hline%
\hline%
Stage & Par. Name & Par. Value & Description\\
\hline%
\emph{Background} & \texttt{method} & local & Compute local background and noise maps\\%
\emph{finder} & \texttt{bkgEstimator} & median & Estimator used to compute background level\\%
& \texttt{rmsEstimator} & MAD & Estimator used to compute background noise rms\\%
& \texttt{bkgBoxSize} & 10$\times$beam & Size of sampling box for local bkg
calculation\\%
& \texttt{bkgGridSize}& 0.2$\times$box size & Size of grid for local bkg
interpolation\\%
\hline%
\emph{Source} & $\sigma_{seed}$ & 5 & Seed threshold in blob finding given as number of sigmas above background\\%
\emph{detection} & $\sigma_{merge}$ & 2.5 & Merge threshold in blob finding given as
number of sigmas above background.\\%
& $n_{pix}$ & 5 & Minimum number of pixels to consider a blob as source candidate\\
& $\Delta\sigma$ & 0.5 & Decrease step applied to the seed threshold at each iteration.
Effective only when $n_{\text{iters}}$>1\\%
& $n_{\text{iters}}$ & 2 & Number of iterations to be performed in compact source
search\\%
& $\sigma_{seed}^{nested}$ & 5 & Seed threshold for nested blob search\\%
& $\sigma_{merge}^{nested}$ & 2.5 & Merge threshold for nested blob search\\%
& \texttt{nestedBlobMinScale} & 1$\times$beam & Nested blob search min scale\\%
& \texttt{nestedBlobMaxScale} & 2$\times$beam & Nested blob search max scale\\%
& \texttt{nestedParentSourceMinDist} & 2 & Minimum distance in pixels between nested and parent blob centroids below which \\%
& & & nested source is skipped\\%
& \texttt{maxMatchingPixFraction} & 0.5 & Max fraction of matching pixels between nested and parent blob above which  \\%
& & & nested source is skipped\\%
\hline%
\emph{Source} & \texttt{method} & peak search+& Search blended components around detected peaks using a multiscale blob detection \\%
\emph{deblending} & & blob detection & algorithm based on a Watershed segmentation of the Laplacian of Gaussian (LoG)\\%
& & & filtered map\\%
& \texttt{peakMinKernSize} & 3 & Min dilation kernel size (in pixels) used to detect peaks \\%
& \texttt{peakMaxKernSize} &7 & Max dilation kernel size (in pixels) used to detect peaks\\%
& \texttt{peakKernMultiplicityThr} & 1  & Requested peak multiplicity across different dilation kernels\\%
& \texttt{peakShiftTolerance} & 2 & Shift tolerance (in pixels) used to compare peaks in different dilation kernels\\%
& \texttt{peakZThrMin} & 1 & Minimum peak flux significance below which peak is skipped\\%
\hline%
\emph{Source} & \texttt{method} & $N$-gaus fit & Fit a mixture of 2D gaussians to source islands \\%
\emph{fitting} & \texttt{nBeamsMaxToFit} & 100 & Threshold above which source fitting is not performed\\%
& \texttt{nMaxComponents} & 5 & Max number of fit components\\%
\hline%
\end{tabular}
\end{threeparttable}
\label{tab:caesar-pars}
\end{center}

\pagebreak[4]
\thispagestyle{empty}
\pagestyle{empty}


\section{Catalogue format}
\label{appendix:catalogformat}

\begin{flushleft}
\footnotesize%
\captionsetup{labelfont=bf}
\captionof{table}{
Format of the added-value catalogue for ASKAP \scorpio{} source components (one component per table row, 135 data columns per component). More details of each data column are available in the \caesar{} online documentation at \url{https://caesar-doc.readthedocs.io/en/latest/}.
}%
\label{tab:catalogue-format}
\addtocounter{table}{-1}%
\begin{longtable}{c|l|c|l}
\hline%
\hline%
Col. Num. & Name & Unit & Description\\%
\hline%
\endfirsthead%

\multicolumn{4}{c}%
{{\textbf{\tablename\ \thetable{}}  -- \emph{continued from previous page}}}\\%
\hline%
\hline%
Col. Num. & Name & Unit & Description\\%
\hline%
\endhead%

\hline%
\multicolumn{4}{|r|}{{\emph{Continued on next page}}} \\%
\hline%
\endfoot%

\endlastfoot%

1 & \emph{sname} & - & Source island name\\%
2 & \emph{npix} & - & Number of pixels in island\\%
3 & \emph{compId} & - & Component id\\%
4 & \emph{sname\_iau} & - & Component name in IAU format\\%
5-6 & ($x$,$y$) & - & Component x, y position\\%
7-8 & ($\Delta x$,$\Delta y$) & - & Component x, y position error\\%
9-10 & (\emph{ra},\emph{dec}) & deg & Component RA and Dec\\%
11-12 & ($\Delta$\emph{ra},$\Delta$\emph{dec}) & deg & Component RA and Dec errors\\%
13 & $\nu$ & GHz & Data frequency from map header\\
14-15 & ($S_{peak}$, $\Delta S_{peak}$) & Jy/beam & Component peak brightness and its error\\
16-17 & ($S$,$\Delta S$) & Jy & Component (corrected) flux density and its total error\\ 
18-19 & ($S_{island}$,$\Delta S_{island}$) & Jy/beam & Island flux brightness and its error\\ 
20-22 & ($a$,$b$,$\theta$) & -,-,deg & Component ellipse pars\footnote{Position angle is measured counterclock-wise from North}\\
23-25 & ($\Delta a$,$\Delta b$,$\Delta\theta$) & - & Component ellipse par errors\\
26-28 & $(a$,$b$,$\theta$)$^{wcs}$ & ",",deg & Component ellipse pars in J2000 coords\\
29-31 & ($\Delta a$,$\Delta b$,$\Delta\theta$)$^{wcs}$ & ",",deg & Component ellipse par errors in J2000 coords\\
32-34 & $(a$,$b$,$\theta$)$_{deconv}^{wcs}$ & ",",deg & Component ellipse pars in J2000 coords, deconvolved by beam\\
35 & $E/E_{beam}$ & - & Ratio of fitted and beam ellipse eccentricities\\
36 & $A/A_{beam}$ & - & Ratio of fitted and beam ellipse areas\\
37 & $\psi$ & deg & Angle between fitted major axis and beam major axis\\
38 & $\mu_{bkg}$ & Jy/beam & Background averaged over island pixels\\
39 & $\sigma_{rms}$ & Jy/beam & Background noise averaged over island pixels\\
40-41 & $\chi^{2}$,\emph{ndf} & - & Source island fit chisquare and degrees of freedom\\
42 & $F_{1}$ & - & Source fit quality flag \{0=\texttt{BAD}, 1=\texttt{LOW}, 2=\texttt{MEDIUM}, 3=\texttt{HIGH}\}\\
43 & $F_{2}$ & - & Component sourceness flag \{1=\texttt{REAL}, 2=\texttt{CANDIDATE}, 3=\texttt{FAKE}\}\\
44 & $F_{3}$ & - & Component morphology flag \{1=\texttt{COMPACT}, 2=\texttt{POINT-LIKE}, 3=\texttt{EXTENDED}, 4=\texttt{COMPACT-EXTENDED}\}\\
45 & \emph{sname\_atca} & - & Name of matched source island in ATCA full band map\\
46 & \emph{compId\_atca} & - & ATCA component id\\%
47 & \emph{npix\_atca} & - & Number of pixels in ATCA island\\%
48-49 & (\emph{ra\_atca},\emph{dec\_atca}) & deg & ATCA component RA and Dec\\%
50-51 & ($\Delta$\emph{ra\_atca},$\Delta$\emph{dec\_atca}) & deg & ATCA component RA and Dec errors\\%
52-53 & ($S_{peak}^{atca}$, $\Delta S_{peak}^{atca}$) & Jy/beam & ATCA component peak brightness and its error\\
54 & $\mu_{bkg}^{atca}$ & Jy/beam & Background averaged over ATCA island pixels\\
55 & $\sigma_{rms}^{atca}$ & Jy/beam & Background noise averaged over ATCA island pixels\\
56-57 & ($S^{atca}$,$\Delta S^{atca}$) & Jy & ATCA component (corrected) flux density and its total error \\
58-71 & ($S_{k}^{atca}$,$\Delta S_{k}^{atca}$) & Jy & Component (corrected) flux densities and their total errors for ATCA subbands ($k$=1,\dots,7)\\ 
72 & \emph{matchFlag\_atca} & - & ASKAP-ATCA match flag \{-1=\texttt{BAD/NO-MATCH}, 0=\texttt{UNCERTAIN}, 1=\texttt{SINGLE}, 2=\texttt{DOUBLE}\}\\
73 & \emph{sname\_mgps} & - & Matched source name in MGPS\\
74-75 & (\emph{ra\_mgps},\emph{dec\_mgps}) & deg & MGPS source RA and Dec\\%
76-77 & ($S^{mgps}$,$\Delta S^{mgps}$) & Jy & MGPS flux and its error \\
78 & \emph{matchFlag\_mgps} & - & ASKAP-MGPS match flag \\
79 & \emph{sname\_nvss} & - & Matched source name in NVSS\\
80-81 & (\emph{ra\_nvss},\emph{dec\_nvss}) & deg & NVSS source RA and Dec\\%
82-83 & ($S^{nvss}$,$\Delta S^{nvss}$) & Jy & NVSS flux and its error \\
84 & \emph{matchFlag\_nvss} & - & ASKAP-NVSS match flag \\
85 & \emph{sname\_tgss} & - & Matched source name in TGSS\\
86-87 & (\emph{ra\_tgss},\emph{dec\_tgss}) & deg & TGSS source RA and Dec\\%
88-89 & ($S^{tgss}$,$\Delta S^{tgss}$) & Jy & TGSS flux and its error \\
90 & \emph{matchFlag\_tgss} & - & ASKAP-TGSS match flag \\
91 & \emph{sname\_gleam} & - & Matched source name in GLEAM\\
92-93 & (\emph{ra\_gleam},\emph{dec\_tgss}) & deg & GLEAM source RA and Dec\\%
94-95 & ($S^{gleam}$,$\Delta S^{gleam}$) & Jy & GLEAM flux and its error \\
96 & \emph{matchFlag\_gleam} & - & ASKAP-GLEAM match flag \\
97 & $d_{wise}$ & " & Distance ($r$<24") between ASKAP and matched AllWISE source\\
98 & \emph{sname\_wise} & - & Matched source name in AllWISE\\
99-100 & (\emph{ra\_wise},\emph{dec\_wise}) & deg & AllWISE source RA and Dec\\%
101-108 & ($S_{k}^{wise}$,$\Delta S_{k}^{wise}$) & Jy & AllWISE source flux and errors in all bands ($k$=W1,W2,W3,W4)\\ 
109-114 & ($S_{k}^{2MASS}$,$\Delta S_{k}^{2MASS}$) & Jy & 2MASS source flux densities and their errors in all bands ($k$=J,H,K)\\
115 & \emph{sname\_glimpse} & - & Matched source name in GLIMPSE\\
116-117 & (\emph{ra\_glimpse},\emph{dec\_glimpse}) & deg & GLIMPSE source RA and Dec\\%
118-119 & ($S^{glimpse}$,$\Delta S^{glimpse}$) & Jy & GLIMPSE 8$\mic{m}$ flux and its error \\
120 & \emph{sname\_higal} & - & Matched source name in Hi-GAL\\
121-122 & (\emph{ra\_higal},\emph{dec\_higal}) & deg & Hi-GAL source RA and Dec\\%
123-124 & ($S^{higal}$,$\Delta S^{higal}$) & Jy & Hi-GAL 70$\mic{m}$ flux and its error\\
125 & \emph{resolvedFlag} & - & Resolved flag \{0=\texttt{POINT-SOURCE}, 1=\texttt{RESOLVED}, 2=\texttt{ATCA-RESOLVED}\}\\
126-127 & ($\alpha$,$\Delta\alpha$) & - & Spectral index and its error\\
128-129 & ($\chi^{2}$,\emph{ndf})$_{sed}$ & - & Radio spectrum fit $\chi^{2}$ and degrees of freedom\\
130 & \emph{sedRobustFitFlag} & - & Spectral data fit flag \{0=\texttt{STANDARD}, 1=\texttt{ROBUST}\}\\
131 & \emph{objClassStrId} & - & Matched astro object class string id\\
132 & \emph{objClassId} & - & Matched astro object class integer id\\
&  &  & \{0=\texttt{UNKNOWN}, -1=\texttt{MULTI-ID}, 1=\texttt{STAR}, 2=\texttt{GALAXY}, 3=\texttt{PN}, 6=\texttt{HII}, 23=\texttt{PULSAR}, 24=\texttt{YSO}\}\\
133 & \emph{objClassSubId} & - & Matched astro object class integer sub-category id\\
134 & \emph{objConfirmedFlag} & - & Matched astro object flag \{0=\texttt{CANDIDATE}, 1=\texttt{CONFIRMED}\}\\
135 & \emph{objName} & - & Matched astro object name(s) (separated by semicolons for multiple matches)\\
\hline%
\end{longtable}
\end{flushleft}

\pagebreak[4]
\thispagestyle{empty}
\onecolumn

\section{\textsc{Scorpio} ATCA map}
\label{appendix:atcamap}

\begin{center}
\centering%
\includegraphics[scale=0.9]{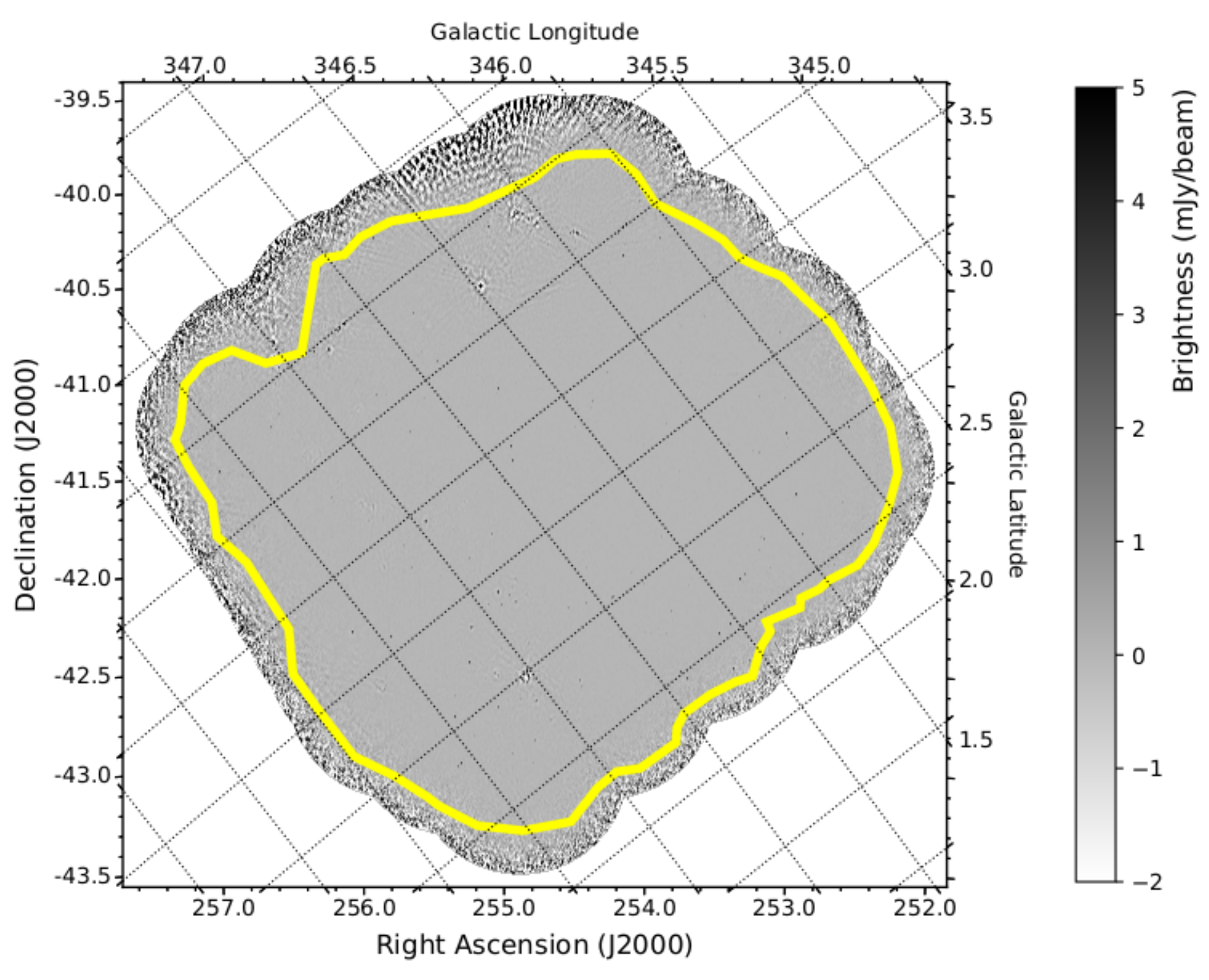}
\captionof{figure}{%
2.1 GHz mosaic of the \scorpio{} region observed with ATCA. The yellow contour delimits the mosaic area considered for source catalogue extraction with \caesar{}.
}%
\label{fig:atcamosaic}
\end{center}

\pagebreak[4]
\thispagestyle{empty}

\section{Source catalogue map}
\label{appendix:catalogmap}

\begin{center}
\centering%
\includegraphics[scale=0.8]{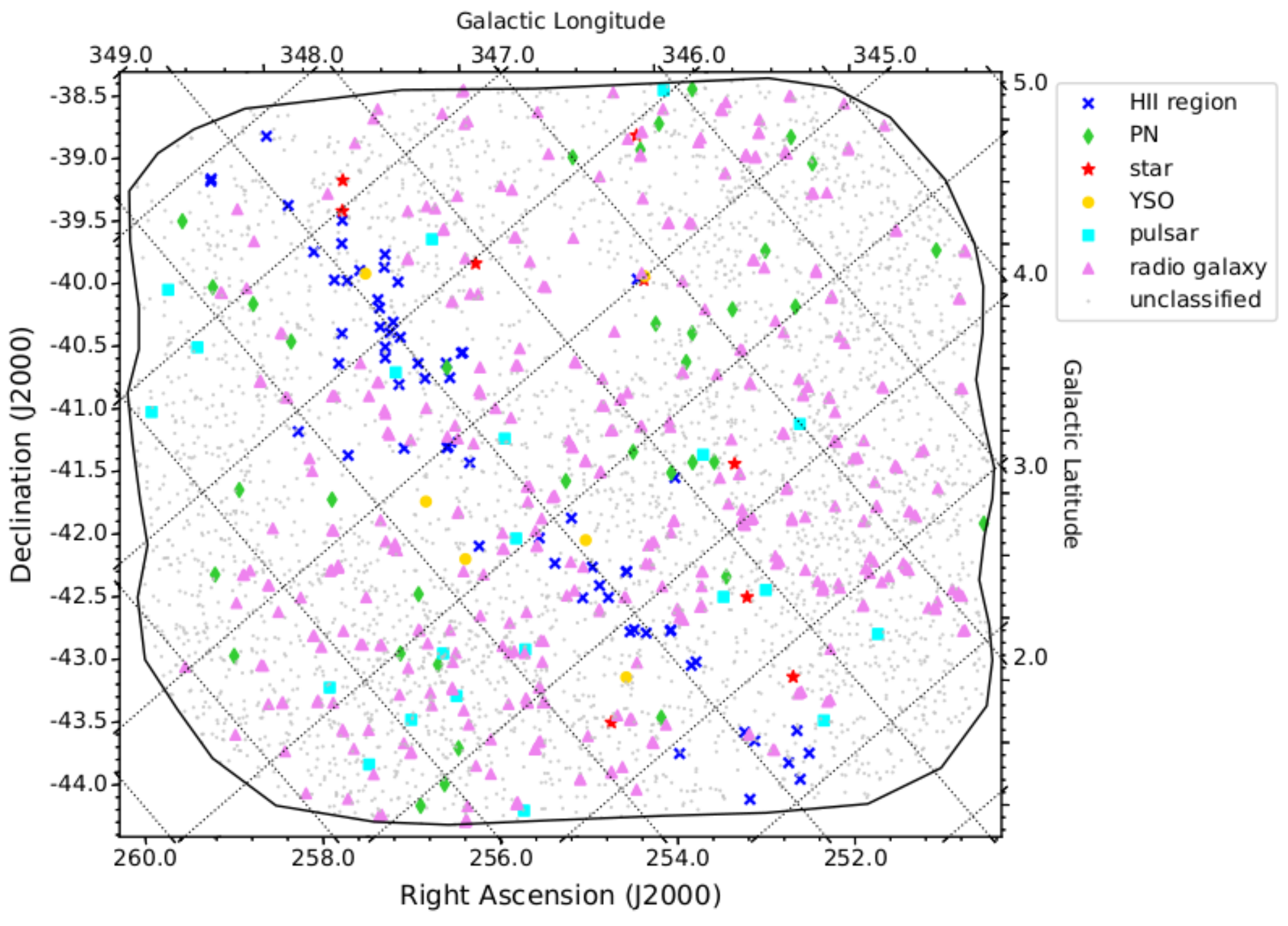}%
\captionof{figure}{%
\textsc{Scorpio} ASKAP region in J2000 sky coordinates (solid black line contour). Colored markers represent the ASKAP source components that were classified on the basis of their morphology or match association to objects reported in astronomical databases (see text). The sample includes both confirmed and candidate objects. Small gray dots represent the unclassified source components.}%
\label{fig:catcrossmatches}
\end{center}

\pagebreak[4]
\thispagestyle{empty}

\section{Source tables}
\label{appendix:catalogtables}

\begin{center}
\footnotesize%
\captionsetup{labelfont=bf}
\captionof{table}{
Summary of source component classification information. Column 3 reports the total number of source components ($n_{tot}$) classified in a given category, including number of confirmed ($n_{confirmed}$) and candidate ($n_{candidate}$) objects, which are reported in columns 4 and 5, respectively. The number of source components ($n_{\alpha}$) found with spectral index information is reported in column 6. Column 7 reports the number of source components ($n_{IR}$) found with infrared counterparts in all four AllWISE bands. Column 8 reports the number of source components ($n_{\alpha+IR}$) found with both measured spectral indices and IR data.
}%
\label{tab:source-summary}%
\begin{threeparttable}
\begin{tabular}{l|l|c|c|c|c|c|c|l|}
\hline%
\hline%
Type (1) & Class. (2) & $n_{tot}$ (3) & $n_{confirmed}$ (4) & $n_{candidates}$ (5) & $n_{\alpha}$ (6) & $n_{IR}$ (7) & $n_{\alpha+IR}$ (8)\\%
\hline%
\multirow{6}{*}{GALACTIC} & STAR & 9 & 8 & 1 & 5 & 7 & 5\\%
& WR STAR & 0 & $-$ & $-$ & $-$ & $-$ & $-$\\%
& YSO & 6 & 2 & 4 & 4 & 5 & 3 \\%
& PULSAR & 21 & 21 & 0 & 3 & 0 & 0 \\%
& PN & 36 & 26 & 10 & 11 & 29 & 10 \\%
& \hii{} REGIONS & 64 & 32 & 32 & 7 & 34 & 3\\%
& UNKNOWN & 104 & 5 & 99 & 32 & 101 & 31\\%
\hline%
\multirow{2}{*}{EXTRAGALACTIC} & RG & 415 & 262 & 153 & 74 & 23 & 5\\%
& UNKNOWN & 185 & 0 & 185 & 37 & 185 & 37 \\%
\hline%
\multirow{1}{*}{UNCLASSIFIED} & $-$ & 3304 & $-$ & $-$ & 607 & 0 & 0\\%
\hline%
ALL & $-$ & 4144 & 356 & 484 & 780 & 384 & 94\\%
\hline%
\end{tabular}
\end{threeparttable}
\end{center}

\pagebreak[4]
\thispagestyle{empty}
\onecolumn

\begin{landscape}

\begin{center}
\footnotesize%
\captionsetup{labelfont=bf}
\captionof{table}{
List of SIMBAD stellar objects found associated to ASKAP \scorpio{} sources. Column 1 indicates the source name following EMU IAU-approved convention (\texttt{EMU ES JHHMMSS.s+DDMMSS}) where \texttt{'E'} stands for Early Science survey and 'S' for source. Column 2 and 3 indicate the island name and component id assigned by the \caesar{} finder. Column 4 and 5 represent the spectral type (or object classification) and object name reported in the SIMBAD catalogue. Missing entries in column 4 are generically classified as 'star' in SIMBAD. Column 6 indicates if the object is confirmed (1) or a candidate (0). The object position in SIMBAD is reported in columns 7-8 in J2000 coordinates. Column 9 represents the distance between the object and ASKAP measured source positions in arcsec. Column 10-13 are the source flux density and total error in mJy at 912 MHz and 2.1 GHz respectively. Column 14-15 represents the measured spectral index and its error obtained from ASKAP, ATCA, NVSS and MGPS data (when available).}%
\scriptsize%
\label{tab:StarMatches}%
\begin{threeparttable}
\begin{tabular}{l|l|c|c|l|c|c|c|c|c|c|c|c|c|c}
\hline%
\hline%
Source name (1) & Island (2) & Comp. (3) & Type (4) & Obj. name (5) & C (6) & RA (7) & Dec (8) & $d$ (9) & S$_{912}$ (10) & $\Delta$S$_{912}$ (11) & S$_{2100}$ (12) & $\Delta$S$_{2100}$ (13) & $\alpha$ (14) & $\Delta\alpha$ (15)\\%
(prefix=EMU ES) & name & id & &  &  & (deg) & (deg) & (arcsec) & (mJy) & (mJy) & (mJy) & (mJy) & &\\%
\hline%
\texttt{J165935.4-421931} & S2609 & 1 & YSO & \texttt{SSTGLMC G343.7018+00.0861} & 0 & 254.8971 & $-$42.3250 & 1.2 & 9.83 & 1.27 & 12.91 & 0.42 & 0.21 & 0.06\\%
\texttt{J165302.9-414508} & S3496 & 1 & - & \texttt{IRAS 16495-4140} & 1 & 253.2622 & $-$41.7524 & 0.6 & 4.39 & 0.62 & 7.66 & 0.25 & 0.58 & 0.08\\%
\texttt{J170624.8-415654} & S1694 & 1 & YSO & \texttt{2MASS J17062471-4156536} & 0 & 256.6030 & $-$41.9482 & 1.7 & 3.41 & 0.64 & 2.95 & 0.20 & 0.02 & 0.15\\%
\texttt{J170451.3-422541} & S1951 & 1 & YSO & \texttt{SSTGLMC G344.2155-00.7460} & 0 & 256.2133 & $-$42.4281 & 1.5 & 6.86 & 0.82 & 7.57 & 0.29 & 0.17 & 0.09\\%
\texttt{J165654.7-412824} & S2931 & 1 & -\tnote{a} & \texttt{IRAS 16534-4123} & 1 & 254.2278 & $-$41.4737 & 1.1 & 12.57 & 1.31 & 11.85 & 0.38 & $-$0.06 & 0.07\\%
\texttt{J165640.0-401334} & S2933 & 1 & YSO & \texttt{[MHL2007] G345.0052+01.8209 1} & 0 & 254.1658 & $-$40.2261 & 3.0 & 125.32 & 12.82 & 204.42 & 6.33 & 0.43 & 0.08\\%
\texttt{J165847.4-434741} & S2740 & 1 & - & \texttt{HD 326586} & 1 & 254.6974 & $-$43.7943 & 2.1 & 23.20 & 2.39 & - & - & - & -\\%
\texttt{J165803.1-432615} & S2820 & 1 & YSO & \texttt{SSTGLMC G342.6544-00.3827} & 1 & 254.5129 & $-$43.4375 & 0.5 & 3.79 & 0.68 & - & - & - & -\\%
\texttt{J165054.3-440704} & S3841 & 1 & -\tnote{c} & \texttt{IRAS 16472-4401} & 1 & 252.7268 & $-$44.1173 & 2.1 & 22.96 & 2.44 & - & - & - & -\\%
\texttt{J165040.6-432812} & S3867 & 1 & AB & \texttt{2MASS J16504054-4328122} & 0 & 252.6690 & $-$43.4701 & 1.1 & 2.72 & 0.67 & - & - & - & -\\%
\texttt{J165238.2-424925} & S3587 & 1 & B8 &\texttt{HD 326392} & 1 & 253.1596 & $-$42.8243 & 2.3 & 0.75 & 0.32 & - & - & - & -\\%
\texttt{J171222.2-423041} & S968 & 1 & YSO\tnote{b} & \texttt{2MASS J17122205-4230414} & 0 & 258.0919 & $-$42.5115 & 2.8 & 2.98 & 0.38 & - & - & - & -\\%
\texttt{J165641.9-401520} & S2934 & 1 & - & \texttt{TYC 7872-1355-1} & 1 & 254.1740 & $-$40.2552 & 2.3 & 7.81 & 2.55 & 1.44 & 0.57 & $-$2.0 & 0.9\\%
\texttt{J170822.6-400527} & S1348 & 1 & YSO & \texttt{MSX6C G346.4809+00.1320} & 1 & 257.0948 & $-$40.0906 & 2.0 & 6.69 & 2.33 & - & - & - & -\\%
\texttt{J170741.7-403125} & S1455\_N1 & 1 & -\tnote{d} & \texttt{2MASS J17074166-4031240} & 1 & 256.9236 & $-$40.5233 & 0.8 & 365.30 & 38.39 & - & - & - & -\\%
\texttt{J170343.7-400350} & S2012 & 1 & - & \texttt{TYC 7873-953-1} & 1 & 255.9318 & $-$40.0648 & 3.5 & 35.81 & 3.66 & - & - & - & -\\%
\texttt{J165648.7-390541} & S2886 & 1 & er* (M3Ve+M4Ve) & \texttt{CD-38 11343} & 1 & 254.2024 & $-$39.0939 & 3.4 & 3.56 & 0.73 & - & - & - & -\\%
\texttt{J170910.1-393433} & S1212 & 1 & - & \texttt{IRAS 17056-3930} & 1 & 257.2918 & $-$39.5753 & 1.8 & 27.86 & 2.96 & - & - & - & -\\%
\texttt{J170904.0-391949} & S1216 & 1 & - & \texttt{IRAS 17056-3916} & 1 & 257.2664 & $-$39.3301 & 1.1 & 7.10 & 0.87 & - & - & $-$1.65 & 0.82 \\%
\texttt{J171632.4-392718} & S324 & 1 & Cl*\tnote{e} & \texttt{Cl* NGC 6318 PCA 7229} & 1 & 259.1343 & $-$39.4549 & 2.9 & 148.69 & 15.27 & - & - & $-$0.74 & 0.10\\%
\hline%
\end{tabular}
\begin{tablenotes}
\scriptsize%
\item[a] This source matches also with a candidate PN (\texttt{Pre 38})
\item[b] This source matches also with a confirmed PN (\texttt{PM 1-131})
\item[c] This source matches also with a confirmed \hii{} region (\texttt{G341.314+00.190})
\item[d] This source matches also with a confirmed \hii{} region (\texttt{G346.056-00.020})
\item[e] This source matches also with a confirmed \hii{} region (\texttt{G347.921-00.763})
\end{tablenotes}
\end{threeparttable}
\end{center}
\end{landscape}

\pagebreak[4]
\thispagestyle{empty}
\onecolumn

\begin{landscape}
\begin{center}
\footnotesize%
\captionsetup{labelfont=bf}
\captionof{table}{
List of pulsars in the ATNF database associated to ASKAP \scorpio{} sources. Column 1 indicates the source name following EMU IAU-approved convention (\texttt{EMU ES JHHMMSS.s+DDMMSS}) where \texttt{'E'} stands for Early Science survey and 'S' for source. Column 2 and 3 indicate the island name and component id assigned by the \caesar{} finder. Column 4 indicates the object name as reported in the database. The object position from the pulsar database is reported in columns 5-6 in J2000 coordinates. Column 7 represents the distance between the ATNF pulsar database position and the ASKAP source centroid in arcsec. Columns 8-11 are the source flux density and total error in mJy at 912 MHz and 2.1 GHz respectively. Columns 12-13 represents the spectral index and its error reported in the ATNF database (when available) and taken from measurements at 5 GHz \citep{Zhao2019}
728, 1382 and 3100 MHz \citep{Jankowski2018} and 
1.4 GHz \citep{Han2016}. Columns 14-15 represents the spectral index and its error measured using ASKAP, ATCA and literature data at 0.8, 1.4 and 3.0 GHz reported in the ATNF catalogue \citep{Jankowski2018,Kramer2003,Johnston2018,Johnston1992}.}%
\scriptsize%
\label{tab:ATNFMatches}%
\begin{threeparttable}
\begin{tabular}{l|l|c|l|c|c|c|c|c|c|c|c|c|c|c}
\hline%
\hline%
Source name (1) & Island (2) & Comp. (3) & Obj. name (4) & RA (5) & Dec (6) & $d$ (7) & S$_{912}$ (8) & $\Delta$S$_{912}$ (9) & S$_{2100}$ (10) & $\Delta$S$_{2100}$ (11) & $\alpha_{atnf}$ (12) & $\Delta\alpha_{atnf}$ (13) & $\alpha$ (14) & $\Delta\alpha$ (15)\\%
(prefix=EMU ES) & name & id & & (deg) & (deg) & (arcsec) & (mJy) & (mJy) & (mJy) & (mJy) & & & &\\%
\hline%
\texttt{J165423.8-414026} & S3298 & 1 & \texttt{J1654-4140} & 253.5979 & $-$41.6733 & 5.1 & 2.33 & 0.33 & 0.36 & 0.03 & - & - & $-$2.00 & 0.16 \\%
\texttt{J170721.8-405355} & S1521 & 1 & \texttt{B1703-40}\tnote{a} & 256.8405 & $-$40.8989 & 1.8 & 29.02 & 3.31 & 3.33 & 0.42 & $-$2.00 & 0.10 & $-$2.82 & 0.11\\%
\texttt{J170252.6-412848} & S2180 & 1 & \texttt{J1702-4128} & 255.7188 & $-$41.4801 & 1.8 & 1.91 & 0.44 & 1.29 & 0.08 & $-$0.20 & 0.20 & $-$0.41 & 0.10\\%
\texttt{J170236.3-421708} & S2246 & 1 & \texttt{J1702-4217} & 255.6518 & $-$42.2837 & 6.8 & 7.13 & 0.85 & - & - & $-$1.20 & 0.19 & -\tnote{b} & -\\%
\texttt{J164920.7-434923} & S4085 & 1 & \texttt{J1649-4349} & 252.3351 & $-$43.8228 & 4.3 & 2.59 & 0.52 & - & - & - & - & - & -\\%
\texttt{J171110.5-432256} & S1142 & 1 & \texttt{J1711-4322} & 257.7940 & $-$43.3814 & 2.6 & 0.83 & 0.41 & - & - & - & - & - & -\\%
\texttt{J170941.3-440115} & S1371 & 1 & \texttt{J1709-4401} & 257.4225 & $-$44.0198 & 4.0 & 0.72 & 0.25 & - & - & - & - & - & -\\%
\texttt{J170740.3-434112} & S1608 & 2 & \texttt{J1707-4341} & 256.9171 & $-$43.6867 & 2.2 & 1.41 & 0.27 & - & - & $-$1.85 & 0.50 & - & -\\%
\texttt{J170535.7-433112} & S1887 & 1 & \texttt{J1705-4331} & 256.3996 & $-$43.5204 & 3.1 & 1.14 & 0.24 & - & - & - & - & - & -\\%
\texttt{J170252.9-442805} & S2279 & 1 & \texttt{J1702-4428} & 255.7192 & $-$44.4675 & 4.8 & 1.54 & 0.47 & - & - & - & - & - & -\\%
\texttt{J164655.4-430802} & S4499 & 1 & \texttt{J1646-4308} & 251.7304 & $-$43.1353 & 5.6 & 1.27 & 0.29 & - & - & - & - & - & -\\%
\texttt{J165340.2-424904} & S3421 & 1 & \texttt{J1653-4249} & 253.4176 & $-$42.8176 & 0.3 & 2.93 & 0.51 & - & - & 1.00 & 0.60 & - & -\\%
\texttt{J165148.8-424611} & S3699 & 1 & \texttt{B1648-42}\tnote{c} & 252.9533 & $-$42.7697 & 0.6 & 57.22 & 5.82 & - & - & $-$2.03 & 0.10 & - & -\\%
\texttt{J170604.4-431020} & S1804 & 1 & \texttt{J1706-4310} & 256.5188 & $-$43.1725 & 1.4 & 0.43 & 0.19 & - & - & - & - & - & -\\%
\texttt{J170226.9-431042} & S2288 & 1 & \texttt{J1702-4310} & 255.6123 & $-$43.1778 & 2.3 & 1.31 & 0.32 & - & - & $-$1.10 & 0.30 & - & -\\%
\texttt{J165013.3-412636} & S3915 & 1 & \texttt{J1650-4126} & 252.5549 & $-$41.4427 & 2.9 & 0.71 & 0.28 & - & - & - & - & - & -\\%
\texttt{J170529.9-395057} & S1736 & 1 & \texttt{J1705-3950} & 256.3743 & $-$39.8497 & 2.3 & 2.38 & 0.44 & - & - & $-$0.84 & 0.30 & - & -\\%
\texttt{J171752.4-410315} & S284 & 1 & \texttt{B1713-40}\tnote{d} & 259.4676 & -41.0547 & 3.0 & 2.85 & 0.55 & - & - & $-$1.40 & 0.50 & - & -\\%
\texttt{J171642.3-400525} & S339 & 1 & \texttt{J1716-4005} & 259.1753 & $-$40.0908 & 3.8 & 3.78 & 1.55 & - & - & - & - & - & -\\%
\texttt{J171541.1-403421} & S459 & 1 & \texttt{J1715-4034} & 258.9208 & $-$40.5728 & 1.8 & 4.23 & 0.70  & - & - & $-$2.20 & 0.20 & - & -\\%
\texttt{J165538.3-384414} & S3051 & 1 & \texttt{J1655-3844} & 253.9111 & $-$38.7358 & 7.5 & 0.97 & 0.41 & - & - & - & - & - & -\\%
\hline%
\end{tabular}
\begin{tablenotes}
\scriptsize%
\item[a] Alternative name: \texttt{J1707-4053}
\item[b] Spectral index not estimated as the ASKAP flux is including flux from a neighbor background source that cannot be properly accounted as in ATCA map.).
\item[c] Alternative name: \texttt{J1651-4246}
\item[d] Alternative name: \texttt{J1717-4054}
\end{tablenotes}
\end{threeparttable}
\end{center}
\end{landscape}

\pagebreak[4]
\thispagestyle{empty}
\onecolumn

\begin{landscape}
\begin{center}
\footnotesize%
\captionsetup{labelfont=bf}
\captionof{table}{
List of HASH PNe associated to ASKAP \scorpio{} sources. Column 1 indicates the source name following EMU IAU-approved convention (\texttt{EMU ES JHHMMSS.s+DDMMSS}) where \texttt{'E'} stands for Early Science survey and 'S' for source. Column 2 and 3 indicate the island name and component id assigned by the \caesar{} finder. Column 4 and 5 indicate the object name as reported in the HASH catalogue and if the object is confirmed (1) or a candidate (0). The object position from HASH is reported in columns 6-7 in J2000 coordinates. Column 8 represents the position offset between the HASH PN and the ASKAP source in arcsec. Column 9 is the PN major diameter in arcsec as reported in the database (when available). Columns 10-11 and 16-17 are the measured source flux density and total error at 912 MHz (ASKAP) and 2.1 GHz (ATCA) respectively. Columns 12-13 are the measured source flux density and error at 843 MHz reported in the HASH catalogue and taken with the MOST \citep{Murphy2007}. Columns 14-15 are the measured source flux density and error at 1.4 GHz reported in the HASH catalogue and obtained with the VLA \citep{Condon1998,CondonKaplan1998}. Columns 18-19 are the measured source flux density and error at 4.8 GHz reported in the HASH catalogue and obtained with the ATCA \citep{vanDeSteene1993}. Column 20 and 21 are the estimated spectral index and its error from ASKAP, ATCA and literature data (when available).
}%
\scriptsize%
\vspace{-0.06cm}%
\label{tab:HASHMatches}%
\setlength\tabcolsep{1.6pt}
\begin{threeparttable}
\begin{tabular}{l|l|c|l|c|c|c|c|c|c|c|c|c|c|c|c|c|c|c|c|c}
\hline%
\hline%
Source name (1) & Island (2) & Comp. (3) & Obj. name (4) & C (5) & RA (6) & Dec (7) & $d$ (8) & $d_{maj}$ (9) & S$_{912}$ (10) & $\Delta$S$_{912}$ (11) & S$_{843}$ (12) & $\Delta$S$_{843}$ (13) & S$_{1400}$ (14) & $\Delta$S$_{1400}$ (15) & S$_{2100}$ (16) & $\Delta$S$_{2100}$ (17) & S$_{4800}$ (18) & $\Delta$S$_{4800}$ (19) & $\alpha$ (20) & $\Delta\alpha$ (21)\\%
(prefix=EMU ES) & name & id & & & (deg) & (deg) & (arcsec) & (arcsec) & (mJy) & (mJy) & (mJy) & (mJy) & (mJy) & (mJy) & (mJy) & (mJy) & (mJy) & (mJy) & &\\%
\hline%
\texttt{J165500.6-405534} & S3192 & 1 & \texttt{IRAS 16515-4050} & 1 & 253.7518 & $-$40.9263 & 2.3 & - & 5.82 & 0.67 & - & - & - & - & 11.69 & 0.37 & - & - & -\tnote{a} & -\\%
\texttt{J165443.4-404146} & S3227 & 1 & \texttt{MPA J1654-4041} & 0 & 253.6799 & $-$40.6963 & 2.9 & 15.0 & 9.83 & 1.30 & - & - & - & - & 7.46 & 0.30  & - & - & $-$0.58 & 0.10\\%
\texttt{J165451.2-414356} & S3226 & 1 & \texttt{PHR J1654-4143}\tnote{b} & 0 & 253.7125 & $-$41.7305 & 6.8 & 65.0 & 9.86 & 1.33 & - & - & - & - & 3.85 & 0.28 & - & - & $-$1.03 & 0.17\\%
\texttt{J165615.2-403641} & S2998 & 1 & \texttt{Pre 11} & 0 & 254.0629 & $-$40.6120 & 1.8 & 6.0 & 12.97 & 1.40 & - & - & - & - & 12.72 & 0.42 & - & - & $-$0.16 & 0.10\\%
\texttt{J165300.0-403047} & S3479 & 1 & \texttt{DGPK 2} & 0 & 253.2483 & $-$40.5129 & 5.7 & 4.0 & 2.48 & 0.63 & - & - & - & - &  3.37 & 0.46 & - & - & 0.64 & 0.32\\%
\texttt{J165547.2-414840} & S3101 & 1 & \texttt{MGE 343.6641+00.9584} & 0 & 253.9454 & $-$41.8117 & 4.4 & 21.0 & 6.18 & 0.95 & - & - & - & - & -\tnote{c} & - & - & - & - & -\\%
\texttt{J170020.1-415043} & S2499 & 1 & \texttt{MGE 344.1648+00.2733} & 0 & 255.0833 & $-$41.8453 & 2.3 & 24.0 & 37.85 & 3.94 & 27.90 & 3.20 & - & - & -\tnote{c} & - & - & - & - & -\\%
\texttt{J165634.1-434615} & S3028 & 1 & \texttt{PM 1-119} & 1 & 254.1416 & $-$43.7708 & 1.2 & 4.7 & 113.79 & 11.47 & 85.30 & 3.50 & - & - & - & - & 251.00 & 4.00 & 0.60 & 0.02\\%
\texttt{J170730.6-442250} & S1651 & 1 & \texttt{WRAY 16-251} & 1 & 256.8774 & $-$44.3806 & 0.9 & 15.0 & 33.66 & 3.45 & 32.80 & 2.40 & - & - & - & - & - & - & - & -\\%
\texttt{J170622.6-441311} & S1811 & 1 & \texttt{Pe 1-8} & 1 & 256.5940 & $-$44.2194 & 1.1 & 23.0 & 77.39 & 7.86 & 80.30 & 3.20 & - & - & - & - & - & - & - & - \\%
\texttt{J170538.9-435620} & S1901 & 1 & \texttt{Vd 1-9} & 1 & 256.4121 & $-$43.9390 & 0.5 & - &  9.75 & 1.02 & - & - & - & - & - & - & - & - & - & -\\%
\texttt{J164218.7-421445} & S5124 & 1 & \texttt{PHR J1642-4214} & 1 & 250.5775 & $-$42.2459 & 1.1 & 10.0 & 6.00 & 0.76 & - & - & - & - & - & - & - & - & - & -\\%
\texttt{J165723.9-413758} & S2867 & 1 & \texttt{H 1-5} & 1 & 254.3489 & $-$41.6327 & 1.8 & 5.2 & 147.89 & 14.91 & 114.80 & 4.10 & - & - & - & - & - & - & - & -\\%
\texttt{J165355.5-414400} & S3365 & 1 & \texttt{PHR J1653-4143} & 1 & 253.4804 & $-$41.7333 & 3.1 & 15.0 & 13.86 & 1.46 & 18.70 & 1.90 & - & - & - & - & - & - & - & -\\%
\texttt{J165331.4-423923} & S3438 & 1 & \texttt{H 1-3} & 1 & 253.3806 & $-$42.6562 & 1.2 & 19.0 & 33.86 & 3.56 & 32.40 & 2.50 & - & - & - & - & - & - & - & -\\%
\texttt{J171027.4-415249} & S1151 & 1 & \texttt{H 1-7} & 1 & 257.6141 & $-$41.8804 & 0.9 & 10.6 & 187.77 & 18.99 & 159.00 & 5.60 & - & - & - & - & - & - & - & - \\%
\texttt{J170757.2-430905} & S1552 & 1 & \texttt{PHR J1707-4309} & 1 & 256.9874 & $-$43.1514 & 3.4 & 16.0 & 7.74 & 0.83 & - & - & - & - & - & - & - & - & - & -\\%
\texttt{J170659.2-424109} & S1637 & 1 & \texttt{H 1-6} & 1 & 256.7464 & $-$42.6859 & 1.5 & 14.0 & 34.62 & 3.52 & 30.70 & 1.80 & - & - & - & - & - & - & - & -\\%
\texttt{J170619.7-431533} & S1770 & 1 & \texttt{Kn 98} & 0 & 256.5817 & $-$43.2592 & 1.4 & 4.0 & 7.11 & 0.76 & - & - & - & - & - & - & - & - & - & -\\%
\texttt{J171547.0-422406} & S560 & 1 & \texttt{CBF 3} & 1 & 258.9450 & $-$42.4017 & 2.5 & 7.6 & 3.61 & 0.46 & - & - & - & - & - & - & - & - & - & -\\%
\texttt{J171516.0-430354} & S663 & 1 & \texttt{MPA J1715-4303} & 1 & 258.8163 & $-$43.0649 & 1.8 & 7.0 & 10.41 & 1.11 & 17.60 & 2.20 & - & - & - & - & - & - & - & -\\%
\texttt{J171425.2-414433} & S697 & 1 & \texttt{Kn 102} & 0 & 258.6042 & $-$41.7428 & 2.4 & - & 2.48 & 0.42 & - & - & - & - & - & - & - & - & - & -\\%
\texttt{J171222.2-423041} & S968 & 1 & \texttt{PM 1-131}\tnote{e} & 1 & 258.0918 & $-$42.5115 & 3.1 & 2.2 & 2.98 & 0.38 & - & - & - & - & - & - & 13.5 & 0.1 & 0.91 & 0.08 \\%
\texttt{J164424.1-400321} & S4880 & 1 & \texttt{SuWt 3} & 1 & 251.1006 & $-$40.0557 & 0.6 & 31.9 & 2.35 & 0.52 & - & - & 4.50 & 0.70 & - & - & - & - & 1.52 & 0.88\\%
\texttt{J165654.7-412824} & S2931 & 1 & \texttt{Pre 38}\tnote{f} & 0 & 254.2275 & $-$41.4733 & 2.0 & 6.0 & 12.57 & 1.31 & - & - & - & - & - & - & - & - & - & - \\%
\texttt{J171142.2-403527} & S957 & 1 & \texttt{Mo 16} & 0 & 257.9251 & $-$40.5910 & 2.4 & 18.0 & 21.35 & 2.25 & - & - & - & - & - & - & - & - & - & -\\%
\texttt{J165929.0-391500} & S2535 & 1 & \texttt{PreRo 2} & 0 & 254.8708 & $-$39.2503 & 0.9 & 6.0 & 5.57 & 1.03 & - & - & - & - & - & - & - & - & - & -\\%
\texttt{J165133.9-400256} & S3701 & 1 & \texttt{Vd 1-5} & 1 & 252.8902 & $-$40.0488 & 3.2 & 11.6 & 5.62 & 0.62 & - & - & 3.70 & 0.60 & - & - & - & - & $-$0.97 & 0.64\\%
\texttt{J165020.3-403004} & S3878 & 1 & \texttt{MPA J1650-4030} & 1 & 252.5840 & $-$40.5009 & 2.0 & 9.0 & 5.09 & 0.58 & - & - & - & - & - & - & - & - & - & -\\%
\texttt{J170510.7-405310} & S1829 & 1 & \texttt{IC 4637} & 1 & 256.2938 & $-$40.8857 & 2.8 & 18.9 & 237.76 & 24.12 & 215.80 & 7.30 & - & - & - & - & 401\tnote{d} & - & 0.33 & 0.04 \\%
\texttt{J171449.4-400608} & S552 & 1 & \texttt{PHR J1714-4006} & 1 & 258.7055 & $-$40.1025 & 1.1 & 20.0 & 5.79 & 0.95 & 25.60 & 6.80 & - & - & - & - & - & - & - & -\\%
\texttt{J171310.9-401556} & S766 & 1 & \texttt{MPA J1713-4015} & 1 & 258.2950 & $-$40.2656 & 2.0 & 7.0 & 10.32 & 1.18 & - & - & - & - & - & - & - & - & - & -\\%
\texttt{J165640.2-391237} & S2909 & 1 & \texttt{MPA J1656-3912} & 1 & 254.1671 & $-$39.2102 & 1.0 & 4.0 & 4.16 & 0.54 & - & - & 4.90 & 0.80 & - & - & - & - & 0.38 & 0.69\\%
\texttt{J165552.0-390023} & S3020 & 1 & \texttt{PHR J1655-3900} & 1 & 253.9663 & $-$39.0058 & 3.3 & 12.0 & 1.82 & 0.48 & - & - & - & - & - & - & - & - & - & -\\%
\texttt{J165427.5-384412} & S3228 & 1 & \texttt{Vd 1-6} & 1 & 253.6138 & $-$38.7362 & 3.0 & 16.0 & 13.44 & 1.59 & 13.00 & 2.10 & 11.20 & 0.60 & - & - & - & - & $-$0.37 & 0.24\\%
\texttt{J165025.6-390819} & S3853 & 1 & \texttt{Vd 1-4} & 1 & 252.6055 & $-$39.1386 & 4.0 & 5.0 & 3.31 & 0.40 & - & - & 4.20 & 0.60 & - & - & - & - & 0.55 & 0.62\\%
\texttt{J164933.0-392109} & S3994 & 1 & \texttt{Vd 1-3} & 1 & 252.3870 & $-$39.3525 & 2.7 & - & 3.12 & 0.38 & - & - & 5.00 & 0.50 & - & - & - & - & 1.10 & 0.52\\%
\texttt{J171551.7-393309} & S388 & 1 & \texttt{RPZM 8} & 1 & 258.9651 & $-$39.5523 & 1.0 & - & 10.33 & 1.32 & - & - & 12.30 & 1.80 & - & - & - & - & 0.41 & 0.64\\%
\hline%
\end{tabular}

\begin{tablenotes}
\scriptsize%
\item[a] \texttt{IRAS 16515-4050} radio spectrum cannot be well described by a single power-law model and was fitted with a thermal free-free emission model (see text)
\item[b] \texttt{PHR J1654-4143} associated to a two-component island in both ASKAP and ATCA maps. The flux considered for the spectral index is the sum of both components.
\item[c] \texttt{MGE 343.6641+00.9584} and \texttt{MGE 344.1648+00.2733} PN candidates are detected and resolved in ATCA observations at 2.1 GHz. We did not report the flux density measurements in the Table.
\item[d] This flux measurement was obtained with Parkes at a reference frequency of 5000 MHz \citep{Milne1975}. No flux error is provided in the reference paper. The spectral index fit was thus performed assuming no flux errors in all measurements. An additional flux measurement ($S$=117 mJy) obtained with Parkes at 14.7 GHz \citep{Milne1982} is reported in the HASH catalogue but not shown in the Table and not included in the fit. 
\item[e] This source matches also with a candidate YSO (\texttt{2MASS J17122205-4230414})
\item[f] This source matches also with a confirmed star (\texttt{IRAS 16534-4123})
\end{tablenotes}
\end{threeparttable}

\end{center}

\end{landscape}

\pagebreak[4]
\thispagestyle{empty}
\onecolumn

\begin{landscape}
\begin{center}
\footnotesize%
\captionsetup{labelfont=bf}
\captionof{table}{
List of WISE \hii{} regions found associated to ASKAP \scorpio{} sources. Column 1 indicates the source name following EMU IAU-approved convention (\texttt{EMU ES JHHMMSS.s+DDMMSS}) where \texttt{'E'} stands for Early Science survey and 'S' for source. Columns 2 and 3 indicate the island name and component id assigned by the \caesar{} finder. Columns 4 and 5 indicate the object name as reported in the WISE catalogue and if the object is confirmed (1) or a candidate (0). The object position from \protect{\citealt{Anderson2014}} 
is reported in columns 6-7 in J2000 coordinates. Column 8 represents the distance between source centroids. Column 9 is the radius of the \hii{} region as reported in the WISE catalogue. Columns 10-13 are the measured source flux density and total error at 912 MHz and 2.1 GHz respectively. Columns 14 and 15 are the estimated spectral index and its error from ASKAP, ATCA and literature data.}%
\scriptsize%
\addtocounter{table}{-1}%
\label{tab:WiseMatches}%
\vspace{-0.2cm}%
\begin{longtable}{llllccccccccccc}

\hline%
\hline%
Source name (1) & Island (2) & Comp. (3) & Obj. name (4) & C (5) & RA (6) & Dec (7) & $d$ (8) & Radius (9) & S$_{912}$ (10) & $\Delta$S$_{912}$ (11) & S$_{2100}$ (12) & $\Delta$S$_{2100}$ (13) & $\alpha$ (14) & $\Delta\alpha$ (15)\\%
(prefix=EMU ES) & name & id & &  & (deg) & (deg) & (arcsec) & (arcsec) & (mJy) & (mJy) & (mJy) & (mJy) & &\\%
\hline%
\endfirsthead%

\multicolumn{15}{c}%
{{\textbf{\tablename\ \thetable{}}  -- \emph{continued from previous page}}}\\%
\hline%
\hline%
Source name & Island & Comp. & Obj. name & C & RA  & Dec & $d$ & Radius & S$_{912}$ & $\Delta$S$_{912}$ & S$_{2100}$ & $\Delta$S$_{2100}$ & $\alpha$ & $\Delta\alpha$\\%
(prefix=EMU ES) & name & id & &  & (deg) & (deg) & (arcsec) & (arcsec) & (mJy) & (mJy) & (mJy) & (mJy) & &\\%
\hline%
\endhead%

\hline%
\multicolumn{15}{|r|}{{\emph{Continued on next page}}} \\%
\hline%
\endfoot%

\endlastfoot%

\texttt{J170426.2-413932} & S1977 & 1 & \texttt{G344.781-00.217} & 1 & 256.1087 & $-$41.6577 & 4.5 & 100 & 2.37 & 0.86 & 1.78 & 0.18 & $-$0.56 & 0.40\\%
\texttt{J170135.9-421727} & S2365 & 1 & \texttt{G343.961-00.183} & 1 & 255.3994 & $-$42.2866 & 15.0 & 54 & 1.76 & 0.62 & 1.15$^{a}$ & 0.70 & $-$0.51 & 1.14\\%
\texttt{J170058.2-423000} & S2454 & 1 & \texttt{G343.721-00.223} & 0 & 255.2426 & $-$42.5005 & 2.3 & 28 &  26.97 & 3.20 & 20.67 & 0.69 & $-$0.14 & 0.08\\%
\texttt{J165658.2-401456} & S2892 & 1 & \texttt{G345.029+01.764} & 1 & 254.2423 & $-$40.2424 & 23.6 & 100 & 5.03 & 2.05 & 2.99$^{a}$ & 0.53 & $-$0.62 & 0.70 \\%
\texttt{J170715.4-405952} & S1547 & 1 & \texttt{G345.631-00.234} & 1 & 256.8101 & $-$40.9903 & 30.6 & 57 & 4.31 & 0.99 & 4.87 & 0.27 & $-$0.08$^{b}$ & 0.16\\%
\texttt{J170712.6-413038} & S1566 & 1 & \texttt{G345.211-00.538} & 0 & 256.7978 & $-$41.5087 & 17.2 & 50 & 34.35 & 3.60 & 24.18$^{a}$ & 1.02 & -$^{c}$ & - \\%
\texttt{J170504.8-405806} & S1843 & 1 & \texttt{G345.402+00.103} & 0 & 256.2710 & $-$40.9698 & 6.4 & 23 & 9.50 & 1.82 & 2.89 & 0.29 & $-$1.54 & 0.25\\%
\texttt{J170511.3-412905} & S1846 & 1 & \texttt{G345.004-00.224} & 0 & 256.2968 & $-$41.4845 & 1.7 & 65 & 26.44 & 3.08 & 61.39 & 1.93 & 1.04 & 0.07 \\%
\texttt{J165842.1-424732} & S2721\_N1 & 1 & \texttt{G343.234-00.078} & 1 & 254.6776 & $-$42.7934 & 8.6 & 21 & 89.65 & 9.74 & 40.23 & 1.56 & $-$0.12$^{d}$ & 0.07\\%
\texttt{J170522.0-413142} & S1831 & 2 & \texttt{G344.989-00.277} & 1 & 256.3410 & $-$41.5286 & 2.8 & 27 & 6.58 & 1.25 & 4.58 & 0.21 & $-$0.38 & 0.11\\%
\texttt{J170522.0-413142} & S1831 & 1 & \texttt{G344.993-00.265}$^{e}$ & 1 & 256.3304 & $-$41.5175 & 7.9 & 41 & 13.73 & 1.81 & -$^{e}$ & - & - & -\\%
\texttt{J165737.2-430328} & S2874 & 1 & \texttt{G342.903-00.083} & 1 & 254.4031 & $-$43.0565 & 8.7 & 50 & 68.32 & 6.95 & -$^{f}$ & - & - & -\\%
\texttt{J165538.2-415104} & S3114 & 1 & \texttt{G343.615+00.955} & 1 & 253.9079 & $-$41.8520 & 5.7 & 88 & 28.40 & 3.51 & -$^{g}$ & - & - & -\\%
\texttt{J165054.3-440704} & S3841 & 1 & \texttt{G341.314+00.190}$^{h}$ & 1 & 252.7273 & $-$44.1177 & 3.2 & 22 & 22.96 & 2.44 & - & - & - & -\\%
\texttt{J165457.1-431956} & S3250 & 1 & \texttt{G342.384+00.122} & 0 & 253.7382 & $-$43.3330 & 3.1 & 20 & 19.19 & 2.23 & - & - & - & -\\%
\texttt{J165548.2-440348} & S3152 & 1 & \texttt{G341.912-00.456} & 1 & 253.9505 & $-$44.0633 & 0.7 & 22 & 13.88 & 1.68 & - & - & - & -\\%
\texttt{J165252.6-435422} & S3559\_N2 & 1 & \texttt{G341.701+00.050} & 1 & 253.2187 & $-$43.9075 & 5.3 & 59 & 268.62 & 27.15 & - & - & - & -\\%
\texttt{J165242.4-442641} & S3599 & 1 & \texttt{G341.271-00.265} & 0 & 253.1793 & $-$44.4404 & 18.5 & 73 & 31.53 & 5.27 & - & - & - & -\\%
\texttt{J165226.3-435840} & S3627 & 1 & \texttt{G341.598+00.068} & 0 & 253.1086 & $-$43.9757 & 7.9 & 78 & 9.56 & 1.67 & - & - & - & -\\%
\texttt{J165056.2-440930} & S3840 & 1 & \texttt{G341.286+00.159} & 1 & 252.7356 & $-$44.1591 & 5.0 & 86 & 3.02 & 0.82 & - & - & - & -\\%
\texttt{J165032.8-435407} & S3889 & 1 & \texttt{G341.438+00.383} & 0 & 252.6299 & $-$43.8984 & 27.9 & 84 & 5.79 & 1.42 & - & - & - & -\\%
\texttt{J165026.2-441730} & S3914 & 1 & \texttt{G341.121+00.141} & 1 & 252.6067 & $-$44.2968 & 20.5 & 44 & 4.12 & 1.40 & - & - & - & -\\%
\texttt{J165001.1-440502} & S3967 & 1 & \texttt{G341.238+00.335} & 1 & 252.5043 & $-$44.0834 & 1.5 & 72 & 5.61 & 1.42 & - & - & - & -\\%
\texttt{J170001.0-420837} & S2525\_N1 & 1 & \texttt{G343.912+00.116} & 1 & 255.0404 & $-$42.1413 & 9.0 & 91 & 291.54 & 30.48 & - & - & - & -\\%
\texttt{J165949.8-424710} & S2587 & 1 & \texttt{G343.367-00.236} & 0 & 254.9599 & $-$42.7873 & 10.2 & 61 & 18.65 & 3.48 & - & - & - & -\\%
\texttt{J165921.0-423239} & S2622\_N1 & 1 & \texttt{G343.503-00.016} & 0 & 254.8390 & $-$42.5444 & 6.0 & 32 & 380.56 & 44.50 & - & - & - & -\\%
\texttt{J165904.1-424139} & S2675\_N3 & 1 & \texttt{G343.353-00.068} & 0 & 254.7673 & $-$42.6938 & 1.8 & 29 & 207.69 & 21.21 & - & - & - & -\\%
\texttt{J165751.7-423517} & S2813 & 1 & \texttt{G343.297+00.173} & 0 & 254.4632 & $-$42.5881 & 8.1 & 30 & 31.25 & 3.38 & - & - & - & -\\%
\texttt{J165753.1-423543} & S2813 & 2 & \texttt{G343.293+00.162} & 0 & 254.4715 & $-$42.5981 & 10.7 & 17 & 22.22 & 2.88 & - & - & - & -\\%
\texttt{J165748.5-430427} & S2843 & 1 & \texttt{G342.916-00.120} & 1 & 254.4538 & $-$43.0693 & 17.8 & 82 & 6.68 & 1.02 & - & - & - & -\\%
\texttt{J165706.3-430519} & S2949 & 1 & \texttt{G342.819-00.029} & 0 & 254.2736 & $-$43.0885 & 9.3 & 26 & 2.99 & 0.63 & - & - & - & -\\%
\texttt{J165603.0-430443} & S3091 & 1 & \texttt{G342.705+00.126} & 0 & 254.0104 & $-$43.0806 & 10.4 & 49 & 22.89 & 2.42 & - & - & - & -\\%
\texttt{J165600.7-430418} & S3091 & 2 & \texttt{G342.701+00.134} & 1 & 253.9973 & $-$43.0781 & 30.4 & 30 & 5.09 & 1.00 & - & - & - & -\\%
\texttt{J165509.8-432135} & S3214\_N1 & 1 & \texttt{G342.388+00.074} & 1 & 253.7920 & $-$43.3594 & 3.6 & 68 & 78.20 & 8.35 & - & - & - & -\\%
\texttt{J170413.5-421953} & S2030 & 1 & \texttt{G344.224-00.594} & 1 & 256.0567 & $-$42.3291 & 7.8 & 90 & 758.09 & 76.40 & - & - & - & -\\%
\texttt{J171141.3-411858} & S993 & 1 & \texttt{G345.864-01.103} & 0 & 257.9216 & $-$41.3206 & 16.7 & 53 & 13.35 & 2.00 & - & - & - & -\\%
\texttt{J171028.6-395312} & S1063 & 1 & \texttt{G346.884-00.069} & 0 & 257.6188 & $-$39.8865 & 1.0 & 47 & 6.99 & 1.79 & - & - & - & -\\%
\texttt{J170941.6-400727} & S1168 & 1 & \texttt{G346.603-00.088} & 1 & 257.4208 & $-$40.1239 & 9.9 & 47 & 3.97 & 1.31 & - & - & - & -\\%
\texttt{J170945.1-404741} & S1194 & 1 & \texttt{G346.071-00.498} & 0 & 257.4369 & $-$40.7948 & 4.4 & 99 & 2.28 & 0.93 & - & - & - & -\\%
\texttt{J170931.9-403326} & S1210 & 1 & \texttt{G346.233-00.322} & 0 & 257.3775 & $-$40.5599 & 21.3 & 54 & 14.41 & 2.72 & - & - & - & -\\%
\texttt{J170909.4-400817} & S1238 & 1 & \texttt{G346.529-00.013} & 1 & 257.2843 & $-$40.1386 & 16.8 & 79 & 86.47 & 10.68 & - & - & - & -\\%
\texttt{J170937.8-413211} & S1247 & 1 & \texttt{G345.462-00.920} & 0 & 257.4073 & $-$41.5359 & 1.9 & 34 & 34.19 & 19.50 & - & - & - & -\\%
\texttt{J170836.1-400345} & S1318 & 1 & \texttt{G346.530+00.115} & 1 & 257.1509 & $-$40.0614 & 3.8 & 41 & 34.99 & 6.15 & - & - & - & -\\%
\texttt{J170755.3-401820} & S1419 & 1 & \texttt{G346.255+00.073} & 1 & 256.9785 & $-$40.3063 & 8.2 & 26 & 3.06 & 1.41 & - & - & - & -\\%
\texttt{J170754.3-403135} & S1426 & 1 & \texttt{G346.078-00.057} & 1 & 256.9770 & $-$40.5265 & 3.1 & 60 & 218.96 & 23.01 & - & - & - & -\\%
\texttt{J170752.4-402226} & S1427 & 1 & \texttt{G346.196+00.040} & 1 & 256.9679 & $-$40.3739 & 2.1 & 30 & 9.85 & 1.23 & - & - & - & -\\%
\texttt{J170734.0-400302} & S1450 & 1 & \texttt{G346.423+00.279} & 0 & 256.8956 & $-$40.0490 & 14.6 & 32 & 21.29 & 2.46 & - & - & - & -\\%
\texttt{J170741.7-403125} & S1455\_N1 & 1 & \texttt{G346.056-00.020}$^{i}$ & 1 & 256.9207 & $-$40.5219 & 11.8 & 68 & 365.30 & 38.39 & - & - & - & -\\%
\texttt{J170730.2-395651} & S1457 & 1 & \texttt{G346.496+00.352} & 0 & 256.8754 & $-$39.9462 & 5.1 & 35 & 23.53 & 2.51 & - & - & - & -\\%
\texttt{J170746.3-404635} & S1458 & 1 & \texttt{G345.863-00.188} & 1 & 256.9455 & $-$40.7771 & 10.1 & 27 & 6.78 & 2.40 & - & - & - & -\\%
\texttt{J170744.3-404120} & S1462 & 1 & \texttt{G345.927-00.129} & 1 & 256.9338 & $-$40.6905 & 6.4 & 36 & 13.44 & 1.90 & - & - & - & -\\%
\texttt{J170729.6-403432} & S1483 & 1 & \texttt{G345.997-00.024} & 0 & 256.8773 & $-$40.5710 & 21.3 & 55 & 22.00 & 3.19 & - & - & - & -\\%
\texttt{J170720.3-402938} & S1505 & 1 & \texttt{G346.039+00.048} & 1 & 256.8358 & $-$40.4947 & 5.1 & 30 & 6.29 & 1.45 & - & - & - & -\\%
\texttt{J170701.7-401029} & S1525 & 1 & \texttt{G346.259+00.288} & 1 & 256.7564 & $-$40.1741 & 3.5 & 38 & 28.73 & 3.31 & - & - & - & -\\%
\texttt{J170704.5-403711} & S1549 & 1 & \texttt{G345.909+00.013} & 0 & 256.7688 & $-$40.6191 & 2.2 & 30 & 18.21 & 2.82 & - & - & - & -\\%
\texttt{J170623.5-405015} & S1633 & 1 & \texttt{G345.656-00.016} & 1 & 256.5995 & $-$40.8392 & 8.6 & 48 & 86.01 & 13.50 & - & - & - & -\\%
\texttt{J170608.1-405742} & S1674\_N1 & 1 & \texttt{G345.529-00.049} & 0 & 256.5320 & $-$40.9599 & 8.6 & 47 & 550.79 & 62.82 & - & - & - & -\\%
\texttt{J170511.9-405058} & S1826 & 1 & \texttt{G345.510+00.160} & 0 & 256.2975 & $-$40.8493 & 7.0 & 32 & 11.10 & 1.96 & - & - & - & -\\%
\texttt{J170428.1-404620} & S1914 & 1 & \texttt{G345.486+00.315} & 1 & 256.1143 & $-$40.7739 & 11.6 & 40 & 661.47 & 66.82 & - & - & - & -\\%
\texttt{J170431.9-404635} & S1914 & 2 & \texttt{G345.493+00.303} & 1 & 256.1325 & $-$40.7756 & 3.4 & 57 & 136.27 & 15.22 & - & - & - & -\\%
\texttt{J171124.4-392953} & S945 & 1 & \texttt{G347.305+00.016} & 0 & 257.8527 & $-$39.4964 & 6.9 & 46 & 31.92 & 5.44 & - & - & - & -\\%
\texttt{J170916.8-395013} & S1205 & 1 & \texttt{G346.791+00.143} & 0 & 257.3247 & $-$39.8353 & 17.4 & 57 & 12.13 & 3.29 & - & - & - & -\\%
\texttt{J170911.7-393901} & S1211 & 1 & \texttt{G346.932+00.270} & 0 & 257.3003 & $-$39.6458 & 17.3 & 81 & 14.41 & 2.63 & - & - & - & -\\%
\texttt{J171632.4-392718} & S324 & 1 & \texttt{G347.921-00.763}$^{j}$ & 1 & 259.1392 & $-$39.4536 & 15.4 & 69 & 148.69 & 15.27 & - & - & $-$0.74$^{k}$ & 0.10\\%
\texttt{J171432.7-391514} & S532 & 1 & \texttt{G347.854-00.335} & 0 & 258.6375 & $-$39.2584 & 16.9 & 92 & 40.45 & 5.54  & - & - & - & -\\%
\texttt{J171431.1-391400} & S536 & 1 & \texttt{G347.872-00.314} & 0 & 258.6301 & $-$39.2321 & 4.5 & 92 & 26.61 & 3.32 & - & - & - & -\\%
\texttt{J171205.8-385547} & S841 & 1 & \texttt{G347.838+00.250} & 0 & 258.0162 & $-$38.9286 & 28.7 & 59 & 41.02 & 5.38 & - & - & - & -\\%
\hline%
\end{longtable}

\begin{itemize}
\scriptsize
\item[$^a$] ATCA source has two components. Their fluxes were summed up to compute the spectral index.
\item[$^{b}$] Spectral index value obtained with noisy ATCA subband measurements.
\item[$^{c}$] Spectral index value not reported as not reliable (e.g. high fit $\widetilde{\chi}^{2}$ due to noisy ATCA subband measurements)
\item[$^{d}$] ASKAP flux potentially unreliable. Spectral index estimated using ATCA data only.
\item[$^{e}$] This ASKAP source includes flux from a nearby more compact \hii{} region (\texttt{G344.989-00.269}) that cannot be deblended. Both sources are observed in ATCA. No ATCA fluxes are reported for this source.
\item[$^{f}$] This ASKAP source includes flux from a nearby point-source (visible in ATCA) that cannot be deblended. Matching source in ATCA data has a relatively extended morphology that cannot be properly fitted with a gaussian model.
\item[$^{g}$] Matching source in ATCA data has a relatively extended morphology that cannot be properly fitted with a gaussian model.
\item[$^{h}$] This source matches also with a confirmed star (\texttt{IRAS 16472-4401}).
\item[$^{i}$] This source matches also with a confirmed star (\texttt{2MASS J17074166-4031240}).
\item[$^{j}$] This source matches also with a confirmed star (\texttt{Cl* NGC 6318 PCA 7229}).
\item[$^{k}$] Spectral index obtained using MGPS and NVSS data.
\end{itemize}

\end{center}
\end{landscape}



\end{document}